\documentclass[12pt]{article}
\usepackage{epsfig,cite,amsmath,axodraw,hhline}

\input paperdef

\evensidemargin \oddsidemargin
\makeatletter
\renewcommand{\section}{\@startsection {section}{1}{\z@}%
                           {-3.5ex \@plus -1ex \@minus -.2ex}%
                           {2.3ex \@plus.2ex}%
                           {\mathversion{bold}\normalfont\Large\bfseries}}
\renewcommand{\subsection}{\@startsection{subsection}{2}{\z@}%
                           {-3.25ex\@plus -1ex \@minus -.2ex}%
                           {1.5ex \@plus .2ex}%
                           {\mathversion{bold}\normalfont\large\bfseries}}
\renewcommand{\subsubsection}{\@startsection{subsubsection}{3}{\z@}%
                           {-3.25ex\@plus -1ex \@minus -.2ex}%
                           {1.5ex \@plus .2ex}%
                           {\mathversion{bold}\normalfont\normalsize\bfseries}}
\makeatother

\allowdisplaybreaks

\begin{document}
\thispagestyle{empty}

\def\thefootnote{\fnsymbol{footnote}}

\begin{flushright}
DCPT/07/80\\
IPPP/07/40 \\
arXiv:0708.3052 [hep-ph]
\end{flushright}

\vspace{0cm}

\begin{center}

{\large\sc 
{\bf Studying the MSSM Higgs sector by forward proton tagging\\[.5em]
     at the LHC}}

\vspace{0.0cm}

\vspace{1cm}

{\sc 
S.~Heinemeyer$^{1}$%
\footnote{email: Sven.Heinemeyer@cern.ch}%
, V.A.~Khoze$^{2, 3}$%
\footnote{email: V.A.Khoze@durham.ac.uk}%
, M.G.~Ryskin$^{2, 3}$%
\footnote{email: Ryskin@MR11084.spb.edu}%
, W.J.~Stirling$^{2, 4}$%
\footnote{email: W.J.Stirling@durham.ac.uk}%
,\\[.3em] 
M.~Tasevsky$^{5, 6}$%
\footnote{email: Marek.Tasevsky@cern.ch}%
~and G.~Weiglein$^{2}$%
\footnote{email: Georg.Weiglein@durham.ac.uk}
}

\vspace*{0.5cm}

{\sl
$^1$Instituto de Fisica de Cantabria (CSIC-UC), 
Santander, Spain

\vspace*{0.25cm} 

$^2$IPPP, Department of Physics, Durham University, 
Durham DH1 3LE, U.K.

\vspace*{0.25cm} 

$^3$Petersburg Nuclear Physics Institute, Gatchina, 
St.~Petersburg, 188300, Russia

\vspace*{0.25cm} 

$^4$Department of Mathematical Sciences,
Durham University, DH1 3LE, U.K.

\vspace*{0.25cm}

$^5$Institute of Physics, 
18221 Prague 8, Czech Republic

\vspace*{0.25cm}

$^6$University of Antwerpen, Physics Department, 
B-2610 Antwerpen, Belgium%
\footnote{former address}

}

\end{center}

\vspace*{0.2cm}
\begin{abstract}
We show that the use of forward proton detectors at the LHC installed  
at 220~m and 420~m distance around ATLAS and / or CMS
can provide important information on the Higgs sector of the MSSM.
We analyse central exclusive production of
the neutral $\cp$-even Higgs bosons $h$ and $H$ and their decays into
bottom quarks, $\tau$ leptons and $W$ bosons in different MSSM benchmark
scenarios. Using plausible
estimates for the achievable experimental efficiencies and the relevant
background processes, we find that the prospective sensitivity of
the diffractive Higgs production will allow to probe interesting
regions of the $\MA$--$\tb$ parameter plane of the MSSM. 
Central exclusive production of the $\cp$-even Higgs bosons of the
MSSM may provide a unique opportunity to access the bottom Yukawa
couplings of the Higgs bosons up to masses of $\MH \lsim 250 \gev$.
We also discuss the 
prospects for identifying the $\cp$-odd Higgs boson, $A$, in diffractive
processes at the LHC.
\end{abstract}

\def\thefootnote{\arabic{footnote}}
\setcounter{page}{0}
\setcounter{footnote}{0}

\newpage

%%%%%%%%%%%%%%%%%%%%%%%%%%%%%%%%%%%%%%%%%%%%%%%%%%%%%%%%%%%%%%%%%%%%%%%%%%%%%%%
%%%%%%%%%%%%%%%%%%%%%%%%%%%%%%%%%%%%%%%%%%%%%%%%%%%%%%%%%%%%%%%%%%%%%%%%%%%%%%%

\section{Introduction}

Searches for Higgs bosons and the study of their properties are among
the primary goals of the Large Hadron Collider (LHC) at CERN.
For the Higgs boson of the Standard Model (SM)
the discovery is, in principle, guaranteed for any 
mass~\cite{atlastdr,atlasrev,cms,CMS-TDR}.
Various extended models predict a large diversity of Higgs-like
bosons with different masses, couplings and  $\cp$-parities. 
The most elaborate extension of the SM up to now is the Minimal 
Supersymmetric Standard Model (MSSM)~\cite{susy}, in which there are
three neutral  
($h$, $H$ and $A$) and two charged ($H^+,H^-$) Higgs bosons. At lowest
order the Higgs sector of the MSSM is $\cp$-conserving, with the
$\cp$-even states $h$ and $H$ ($\Mh< \MH$) and the $\cp$-odd state $A$.
The Higgs sector of the MSSM is affected by large higher-order
corrections (see for example\ \citere{reviews} for recent reviews), which
have to be taken into account for reliable phenomenological predictions.

Within the MSSM, the LHC will be able to observe all the Higgs states 
of the model over a significant part of the MSSM parameter space. There
exists an important parameter region, however, where the LHC will
detect only one of the MSSM Higgs bosons with SM-like properties. 
Revealing that a detected new state is indeed a Higgs boson and
distinguishing the Higgs boson(s) of the SM or the MSSM from the states
of extended Higgs theories will be non-trivial. This goal will require
a comprehensive programme of precision Higgs measurements. In
particular, it will be of utmost importance to determine the spin and 
$\cp$ properties of a new state and to measure precisely its mass, width
and couplings.

While ultimately the cleaner experimental environment of electron--positron 
collisions will be required to assemble a comprehensive phenomenological
profile of the Higgs
sector~\cite{teslatdr,orangebook,acfarep,Snowmass05Higgs}, 
it will be highly important to fully exploit the experimental
capabilities of the LHC. The ``standard'' LHC production channels are 
gluon fusion, weak boson fusion and associated production with heavy
quarks or vector bosons. The accuracy in determining the mass of the new
particle via these channels will depend on whether the $H \to \ga\ga$ or
$H \to ZZ \to 4 \mu$ channels will be accessible. The observation of the 
new state in different channels will
provide valuable information on its
couplings~\cite{HcoupLHCSMold,HcoupLHCSMMD,HcoupLHCSM} and
will also enable initial studies of further
properties~\cite{LHCfurtherHiggs}. 

\begin{figure}
\begin{center}
\includegraphics[height=5.5cm]{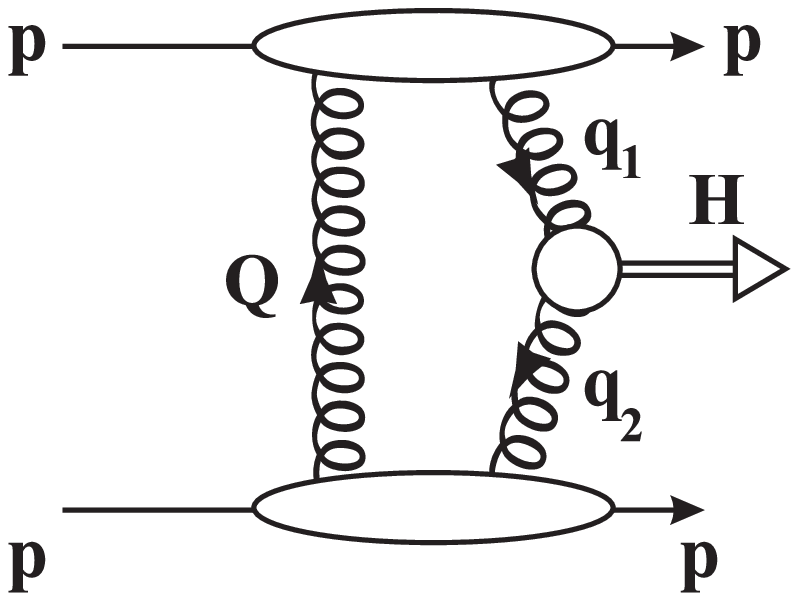}
\caption{Schematic diagram for central exclusive diffractive (CED)
Higgs production at the LHC,
$pp \to p \oplus H \oplus p$.}  
\label{fig:H}
\end{center}
\end{figure}

There has been a great deal of attention devoted recently to
the possibility of complementing the standard LHC physics menu by 
adding forward proton detectors to the CMS and ATLAS experiments 
(see for example \citeres{KMR,KMRProsp,DKMOR,
cox1,JE,LOI,bh75,CR,KPR,ar,totem,CMS-Totem,RP220,clp} and references therein).
The use of forward proton tagging  
would provide an exceptionally clean environment to search for
new phenomena at the LHC and to identify their nature. Of particular
interest in this context is ``central exclusive diffractive'' 
(CED) Higgs-boson production $pp\to p \oplus H \oplus p$, where the 
$\oplus$ signs are used to denote the presence of large rapidity 
gaps\footnote{
We focus here on neutral Higgs-boson production. 
Charged Higgs bosons can also be exclusively produced in
$pp$ collisions predominantly via the photon fusion mechanism 
$\ga\ga \to H^+H^-$.
}.
~In these exclusive processes there is no
hadronic activity between the outgoing protons and the decay products
of the central system. 
The predictions for exclusive production are obtained by calculating
 the diagram of \reffi{fig:H} 
 using techniques developed in \citeres{KMR,KMRProsp,jeff}. 
%
%%%%%%%%%%%%%%%%%%%%%%% F I G U R E %%%%%%%%%%%%%%%%%%%%%%%%%%%%%%%%%%%%%%%%%%%
%%%%%%%%%%%%%%%%%%%%%%% F I G U R E %%%%%%%%%%%%%%%%%%%%%%%%%%%%%%%%%%%%%%%%%%%

There are several major reasons why central exclusive diffractive
(CED) production is so attractive for Higgs boson studies.
First, if the outgoing protons remain intact and scatter through small
angles then, to a very good approximation, the primary active di-gluon
system obeys a $J_z=0$, $\cp$-even selection rule
\cite{Liverpool,KMRmm}. Here $J_z$ is the projection of the total
angular momentum along the proton beam axis. This selection rule
readily permits a clean determination of the quantum numbers of the
observed Higgs resonance which  will be dominantly produced in a
scalar state. Furthermore, because the process is exclusive, the energy
loss of the outgoing protons is directly related to the mass of the
central system, allowing a potentially excellent mass resolution,
irrespective of the decay mode of the produced
particle\footnote{
Current studies suggest~\cite{LOI,RO,cox1} that the
missing mass resolution will be of order $1\%$ for Higgs
masses above $120 \gev$, assuming both protons are detected at 420~m
from the interaction point.
}.
Another important feature of the CED process is that it may enable to 
a signal-to-background ratio of order 1 (or even 
better) to be achieved~\cite{DKMOR,cox1}. A particular advantage of forward
proton tagging is that it would allow  all the main
Higgs-boson decay modes, $b \bar b$,  $WW$  and $\tau\tau$, to be observed in
this channel. It may in this way provide a unique
possibility to study the Higgs coupling to bottom quarks, which may be 
difficult
to access in other search channels at the LHC~\cite{CMS-TDR} despite the
fact that 
$H \to b \bar b$ is by far the dominant decay mode for a light SM-like
Higgs boson.

Within the MSSM, CED Higgs-boson production can be even more important
than in the SM. The coupling of the lightest MSSM Higgs boson to bottom
quarks and $\tau$~leptons can be strongly enhanced for large values of
$\tb$, the ratio of the vacuum expectation values of the two Higgs
doublets in the MSSM Higgs sector, and
relatively small values of the $\cp$-odd Higgs-boson mass,
$\MA$. As a consequence, in this parameter region the
expected Higgs signal-to-background ratios in the $b \bar b$ channel
are much larger than for the corresponding SM process~\cite{KKMRext} 
(see also \citere{bclpr}). 
It is interesting to note in this context that
in some MSSM scenarios CED production would provide the possibility 
for lineshape analyses (to discriminate between different Higgs-boson
signals)~\cite{KKMRext,JE} and offer a way 
for direct observation of a $\cp$-violating signal in the Higgs
sector~\cite{KMRCP,JE}. 
In a situation where the total width is larger than the mass resolution,
the CED process may provide a unique opportunity to  
measure the total width.

The lightest MSSM Higgs boson becomes SM-like for larger
values of $\MA$. For $\MH \approx \MA \gsim 2 \MW$ the lightest MSSM
Higgs boson couples to gauge bosons with about SM strength, while the
heavy MSSM Higgs bosons, $H$ and $A$, decouple from the gauge bosons. 
The search for heavy MSSM Higgs bosons therefore differs significantly
from the case of a heavy SM-like Higgs boson. While for a SM-like Higgs
boson the weak-boson fusion channel is a promising production process
and the decay $H \to ZZ \to 4 \mbox{ leptons}$ is the ``gold-plated'' 
search channel~\cite{atlastdr,CMS-TDR}, none of these channels (nor
Higgs boson decay into $W$~bosons) can be
used in the search for heavy MSSM Higgs bosons. This leads to the result that
in a significant part of the MSSM parameter space, the
well-known ``LHC wedge region''~\cite{atlastdr,CMS-TDR,higgscms},
the heavy MSSM Higgs bosons escape detection at the LHC.

In CED the heavy $\cp$-even MSSM Higgs boson $H$ can be produced and its
decay into $b \bar b$ can be utilised. While in the SM the 
${\rm BR}(H \to b \bar b)$ is strongly suppressed for $\MH \gsim 2 \MW$
because of the dominant decay into gauge bosons, in the MSSM 
$H \to b \bar b$ remains by far the dominant decay mode also for larger
masses, as long as no decays into supersymmetric particles (or lighter
Higgs bosons) are open. CED Higgs boson production with decay 
to $b \bar b$
is therefore important over a much larger mass range than in the
SM. In this paper we study the prospects for the channels 
$h, H \to b \bar b$ and $h, H \to \tau^+\tau^-$ in CED production and
discuss the discovery reach of the various channels in the
$\MA$--$\tb$ parameter plane.
We also analyse the channels $h, H \toWW$ and compare them with
the SM case. 

CED production of the $\cp$-odd Higgs boson is less promising than
production of the $\cp$-even state because this mode is strongly suppressed
by the $\cp$-even selection rule~\cite{Liverpool,KMRmm}.
We therefore investigate the prospects of this channel in a less exclusive
reaction and discuss the possibilities for distinguishing $A$ and $H$.

The outline of the paper is as follows. The Higgs sector of the MSSM
is briefly described in \refse{sec:Higgsbench}, where we also define the
benchmark scenarios used later for the numerical analysis. In
\refse{sec:sigmaprod} we review the relevant Higgs-boson production
cross sections, while in \refse{sec:backgrounds} the various backgrounds
are discussed.
The experimental aspects of CED production of $\cp$-even MSSM Higgs
bosons are discussed in \refse{sec:cedprodhH}.
Our numerical
analysis of the CED production processes of the $\cp$-even MSSM Higgs bosons 
is presented in \refse{sec:discovery}. 
In \refse{sec:cedA} we comment on the
observability of the $\cp$-odd MSSM Higgs boson.
Our conclusions can be found in \refse{sec:conclusions}.

%%%%%%%%%%%%%%%%%%%%%%%%%%%%%%%%%%%%%%%%%%%%%%%%%%%%%%%%%%%%%%%%%%%%%%%%%%%%%%%
%%%%%%%%%%%%%%%%%%%%%%%%%%%%%%%%%%%%%%%%%%%%%%%%%%%%%%%%%%%%%%%%%%%%%%%%%%%%%%%

\section{The MSSM Higgs sector: notations and benchmark 
     scenarios}
\label{sec:Higgsbench}

\subsection{Tree-level structure}

Unlike in the SM, in the MSSM {\it two} Higgs doublets
are required.
At the tree level, the Higgs sector can be described with the help of two  
independent parameters, usually chosen as 
the ratio of the two
vacuum expectation values,  
$\tb \equiv v_2/v_1$, and $\MA$, the mass of the $\cp$-odd $A$ boson.
Diagonalisation of the bilinear part of the Higgs potential,
i.e.\ the Higgs mass matrices, is performed via rotations
by angles $\al$ for the $\cp$-even part and 
$\be$ for the $\cp$-odd and the charged part.
The angle $\al$ is thereby determined through
\begin{equation}
\tan 2\al = \tan 2\be \; \frac{\MA^2 + \MZ^2}{\MA^2 - \MZ^2} ;
\qquad  -\frac{\pi}{2} < \al < 0~.
\label{alphaborn}
\end{equation}
One obtains five physical states, the neutral $\cp$-even Higgs bosons
$h, H$, the $\cp$-odd Higgs boson $A$, and the charged Higgs bosons
$H^\pm$. Furthermore there are three unphysical Goldstone boson states,
$G^0, G^\pm$.
At lowest order, the Higgs-boson masses can be expressed
in terms of $\MZ, \MW$, and $\MA$,
\begin{eqnarray}
\label{mlh}
 M_{h,H}^2 &=& \edz \KKL \MA^2 + \MZ^2  \mp 
          \sqrt{(\MA^2 + \MZ^2)^2 - 4 \MA^2\MZ^2 \CQZb} \KKR, \non \\
\label{mhp}
\MHp^2 &=& \MA^2 + \MW^2 .
\end{eqnarray}

In the decoupling limit, which is typically reached for $\MA \gsim 150 \gev$
(depending on $\tb$), the heavy MSSM
Higgs bosons are nearly degenerate in mass, 
$\MA \approx \MH \approx \MHp$.  The
couplings of the neutral Higgs bosons to SM gauge bosons are
proportional to
\begin{equation}
VVh \sim \Sba\,, VVH \sim \Cba\,,
\qquad (V = Z, W^\pm)~, 
\end{equation}
while the coupling $VVA \equiv 0$ at tree level.
In the decoupling limit one finds $\be - \al \to \pi/2$, 
i.e.\ $\Sba \to 1$, $\Cba \to 0$.
Consequently, for $\MA \gsim 150 \gev$ one finds the
following decay patterns for the neutral MSSM Higgs bosons at tree-level:
\begin{itemize}
\item[$h$:]
the light $\cp$-even Higgs boson has SM-like decays. Due to its upper mass
limit of $\Mh \lsim 130$ $\gev$~\cite{feynhiggs,mhiggsAEC} (see below), a
non-negligible decay to gauge bosons only occurs in a limited window
of $\Mh$ values close to this upper limit.
\item[$H,A$:]
compared to a SM Higgs boson with mass $\MHSM \gsim 150 \gev$, which 
would decay predominantly into SM gauge bosons, 
the decays of $H, A$ to SM gauge bosons are strongly suppressed. In turn, the
branching ratios to $b \bar b$ and $\tau^+\tau^-$ are much larger in
this mass range compared to the SM case. As a rule of
thumb, $\br(H,A \to b \bar b) \approx 90\%$ and 
$\br(H,A \to \tau^+\tau^-) \approx 10\%$, if SUSY particles (such as
charginos and neutralinos) are too heavy to be produced in the decays of
$H$ and $A$.
\end{itemize}

%%%%%%%%%%%%%%%%%%%%%%%%%%%%%%%%%%%%%%%%%%%%%%%%%%%%%%%%%%%%%%%%%%%%%%%%%%%%%%%
%%%%%%%%%%%%%%%%%%%%%%%%%%%%%%%%%%%%%%%%%%%%%%%%%%%%%%%%%%%%%%%%%%%%%%%%%%%%%%%

\subsection{Higher-order corrections}
\label{subsec:HO}

Higher-order corrections in the MSSM Higgs sector are 
in general quite large. In particular, higher-order corrections 
give rise to an upward shift of the upper bound on the light $\cp$-even 
Higgs-boson mass from the tree-level value, $\Mh \leq \MZ$, to about
$\Mh \lsim 130 \gev$~\cite{mhiggslong,mhiggsAEC}. Besides the impact on
the masses, large higher-order corrections also affect the Higgs-boson
couplings.  For the evaluation of the theoretical predictions for the  
relevant observables in the MSSM Higgs sector we use the code
\fhtt~\cite{feynhiggs,mhiggslong,mhiggsAEC,feynhiggs2.5}.

\medskip
For simplicity, in the following we confine our attention to the
$\cp$-conserving case, i.e.\ we do not consider $\cp$-violating complex
phases. However the analysis in this paper can easily be extended to the case
of non-vanishing complex phases.
In order to fix our notation, we list the conventions for the inputs
from the scalar top and scalar bottom sector of the MSSM:
the mass matrices in the basis of the current eigenstates 
$\tilde t_L, \tilde t_R$ and
$\tilde b_L, \tilde b_R$ are given by
\begin{eqnarray}
\label{stopmassmatrix}
{\cal M}^2_{\tilde t} &=&
  \left( \begin{array}{cc}
  \MSQ^2 + \mt^2 + \cos 2\beta 
                           (\frac{1}{2} - \frac{2}{3} \sw^2) \MZ^2 &
      \mt \Xt \\
      \mt \Xt &
      \MstR^2 + \mt^2 + \frac{2}{3} \cos 2\beta \sw^2 \MZ^2 
  \end{array} \right), \\
&& \nonumber \\
\label{sbotmassmatrix}
{\cal M}^2_{\tilde b} &=&
  \left( \begin{array}{cc} \MSQ^2 + \mb^2 + \cos 2\beta 
                               (-\frac{1}{2} + \frac{1}{3} \sw^2) \MZ^2 &
      \mb \Xb \\
      \mb \Xb &
      \MsbR^2 + \mb^2 - \frac{1}{3} \cos 2\beta \sw^2 \MZ^2 
  \end{array} \right),
\end{eqnarray}
where 
\begin{equation}
\mt \Xt = \mt (\At - \mu \cot\beta) , \quad
\mb \Xb = \mb\, (\Ab - \mu \tb) .
\label{eq:mtlr}
\end{equation}
Here $\At$ denotes the trilinear Higgs--stop coupling, $\Ab$ denotes the
Higgs--sbottom coupling, and $\mu$ is the higgsino mass parameter.
As an abbreviation we will use
\BE
\msusy \equiv \MSQ = \MstR = \MsbR~.
\EE

\medskip
The relation between the bottom-quark mass and the Yukawa coupling
$h_b$, which also controls the interaction between the Higgs fields and
the sbottom quarks, reads at lowest order $\mb =h_b v_1$. 
This relation is affected at \onel\ order by large radiative
corrections \cite{deltamb1,deltamb2,deltamb2b,deltamb3},
proportional to $h_b v_2$,  
in general giving rise to $\tb$-enhanced contributions.
These terms proportional to $v_2$ are generated either by
gluino--sbottom \onel\ diagrams (resulting in \order{\alb\als}
corrections, $\alb \equiv h_b^2/(4\pi)$),
%\mla
or by chargino--stop loops (giving 
\order{\alb\alt} corrections, $\alt \equiv h_t^2/(4\pi)$ with $h_t$ being
the top Yukawa coupling). Because the $\tb$-enhanced
contributions can be numerically relevant, an accurate prediction
of $h_b$ from the experimental value of the bottom mass requires a
resummation of such effects to all orders in the perturbative
expansion, as described in \citeres{deltamb2,deltamb2b}.

The leading effects are included in the effective Lagrangian
formalism developed in \citere{deltamb2}.
Numerically this is by far the dominant part of the
contributions from the sbottom sector (see also
\citeres{mhiggsEP4,mhiggsEP4b,mhiggsFD2}). 
The effective Lagrangian is given by
\BEA
\cL = \frac{g}{2\MW} \frac{\mbms}{1 + \db} \Bigg[ 
&& \tb\; A \, i \, \bar b \ga_5 b 
   + \wz \, V_{tb} \, \tb \; H^+ \bar{t}_L b_R \non \\
&+& \KL \frac{\Sa}{\Cb} - \db \frac{\Ca}{\Sbe} \KR h \bar{b}_L b_R 
                                                               \non \\
&-& \KL \frac{\Ca}{\Cb} + \db \frac{\Sa}{\Sbe} \KR H \bar{b}_L b_R
    \Bigg] + {\rm h.c.}~.
\label{effL}
\EEA
Here $\mbms$ denotes the running bottom quark mass including SM QCD
corrections. The prefactor $1/(1 + \db)$ in \refeq{effL} arises from the
resummation of the leading corrections to all orders. 
The additional terms $\sim \db$ in the $h\bar b b$ and $H\bar b b$
couplings arise from the mixing and coupling of the ``other'' Higgs
boson, $H$ and $h$ respectively, to the $b$~quarks.

As explained above, 
the function $\db$ consists of two main contributions, 
an \order{\als} correction from a
sbottom--gluino loop and an \order{\alt} correction
from a stop--higgsino loop. The explicit
form of $\db$ in the limit of $\msusy \gg \mt$ and $\tb \gg 1$
reads~\cite{deltamb1}
\BE
\db = \frac{2\als}{3\,\pi} \, \mgl \, \mu \, \tb \,
                    \times \, I(\msbe, \msbz, \mgl) +
      \frac{\alt}{4\,\pi} \, \At \, \mu \, \tb \,
                    \times \, I(\mste, \mstz, \mu) ~.
\label{def:dmb}
\end{equation}
The function $I$ is given by
\BEA
I(a, b, c) &=& \ed{(a^2 - b^2)(b^2 - c^2)(a^2 - c^2)} \,
               \KL a^2 b^2 \log\frac{a^2}{b^2} +
                   b^2 c^2 \log\frac{b^2}{c^2} +
                   c^2 a^2 \log\frac{c^2}{a^2} \KR \\
 &\sim& \ed{\mbox{max}(a^2, b^2, c^2)} ~. \non
\EEA
The sign of $\db$ is governed by the sign of the parameter $\mu$ (and
for the second term of \refeq{def:dmb} also by the sign of $\At$).
As a consequence of \refeq{effL}, positive values of $\db$ lead to a
suppression of the bottom Yukawa coupling.
On the other hand, for negative values of 
$\db$ the bottom Yukawa coupling
may be strongly enhanced (and can even acquire 
non-perturbative values when $\db \to -1$).
The CED channel, $pp \to p \oplus H \oplus p$ with $H \to b \bar b$,
receives important contributions from the $\db$ corrections via the
bottom Yukawa coupling. 
We will in the following discuss the impact of the $\db$ corrections 
by considering different values of the parameter 
$\mu$.

\medskip
Another important source of higher-order corrections are
Higgs-propagator corrections. They affect the Higgs-boson masses (as
discussed above) and all Higgs-boson couplings. In the coupling of the
light $\cp$-even Higgs boson to bottom quarks, for example, the
Higgs-propagator corrections lead, to a good approximation, to the 
replacement~\cite{hff}
\BE
h b \bar b \sim y_b \frac{\Sa}{\Cb} \; \to \;
                y_b \frac{\Saeff}{\Cb}~,
\EE
where $\aeff$ contains the contributions from the Higgs-boson
propagator corrections~\cite{hff}. For certain parts of the MSSM
parameter space, up to $\MA \lsim 350 \gev$, 
it is possible that $\aeff \to 0$, and thus the coupling of the light Higgs
boson to bottom quarks becomes tiny~\cite{hff}. An example of a scenario
where this is realised is
the ``small~$\aeff$'' scenario, as discussed in the next subsection.

%%%%%%%%%%%%%%%%%%%%%%%%%%%%%%%%%%%%%%%%%%%%%%%%%%%%%%%%%%%%%%%%%%%%%%%%%%%%%%%
%%%%%%%%%%%%%%%%%%%%%%%%%%%%%%%%%%%%%%%%%%%%%%%%%%%%%%%%%%%%%%%%%%%%%%%%%%%%%%%

\subsection{Benchmark scenarios -- bounds from Higgs-boson searches}
\label{subsec:benchmarks}

Due to the large number of MSSM parameters, a number of benchmark 
scenarios~\cite{benchmark2,benchmark3}
have been used for the interpretation of MSSM Higgs
boson searches at LEP~\cite{LEPHiggsSM,LEPHiggsMSSM} and at the
Tevatron~\cite{D0bounds,D0bounds2,CDFbounds,CDFbounds2,Tevcharged}.

Since at tree level the Higgs sector of the MSSM is 
governed by two parameters (in
addition to $\MZ$ and the SM gauge couplings), the definition of the
benchmarks is such that the two tree-level parameters, $\MA$ and $\tb$,
are varied while the values of all other parameters are fixed at certain 
benchmark settings. 
From the most commonly used benchmark scenarios for the
$\cp$-conserving MSSM from \citeres{benchmark2,benchmark3} we list here
the three scenarios that are relevant for our analysis.

%%%%%%%%%%%%%%%%%%%%%%%%%%%%%%%%%%%%%%%%%%%%%%%%%%%%%%%%%%%%%%
%%%%%%%%%%%%%%%%%%%%%%%%%%%%%%%%%%%%%%%%%%%%%%%%%%%%%%%%%%%%%%

\begin{itemize}

\item
\ul{The $\Mhmax$ scenario:}\\[.5em]
In this scenario the parameters are chosen such that the mass of the
light $\cp$-even Higgs boson acquires its maximum possible values 
as a function of $\tb$ (for fixed $M_{\rm SUSY}$, $\mt$ 
and $\MA$ set to its maximum value, $\MA = 1$~TeV).
This was used in particular to obtain conservative $\tb$ 
exclusion bounds~\cite{tbexcl} at LEP~\cite{LEPHiggsMSSM}.
The parameters are%
\footnote{
Using instead the current experimental central top quark mass value of 
$\mt = 170.9 \gev$~\cite{mt1709} would have only a minor impact on our
analysis. 
}%
:
\begin{eqnarray}
&& \mt = 172.7 {\rm ~GeV}, \quad \msusy = 1 {\rm ~TeV}, \quad
\mu = 200 {\rm ~GeV}, \quad M_2 = 200 {\rm ~GeV}, \nonumber \\
\label{mhmax}
&& \Xt = 2\, \msusy  \quad
   \Ab = \At, \quad m_{\tilde g} = 0.8\,\msusy~.
\end{eqnarray}

\item
\ul{The no-mixing scenario:}\\[.5em]
This benchmark scenario is the same as the $\Mhmax$ scenario, but with
vanishing mixing in the $\tilde t$~sector and with a higher SUSY mass
scale (the latter has been chosen to avoid conflict with the exclusion
bounds from the LEP Higgs searches~\cite{LEPHiggsMSSM,LEPHiggsSM}),
\begin{eqnarray}
&& \mt = 172.7 {\rm ~GeV}, \quad \msusy = 2 {\rm ~TeV}, \quad
\mu = 200 {\rm ~GeV}, \quad M_2 = 200 {\rm ~GeV}, \nonumber \\
\label{nomix}
&& \Xt = 0  \; \quad
   \Ab = \At, \quad m_{\tilde g} = 0.8\,\msusy~.
\end{eqnarray}

\item
\ul{The small-$\alpha_{\rm eff}$ scenario:}\\[.5em]
As explained above, the decays $h \to b \bar b$ (and also 
$h \to \tau^+\tau^-$) can be strongly affected by corrections entering
via the effective mixing angle $\aeff$. 
If $\alpha_{\rm eff}$ is 
small, these two decay channels can be strongly suppressed in the MSSM
due to the additional factor $-\sin\alpha_{\rm eff}/\cos\beta$ 
compared to the SM coupling. 
Such a suppression occurs for large $\tb$ and not too large $M_A$
for the following parameters: 
\begin{eqnarray}
&& \mt = 172.7 {\rm ~GeV}, \quad \msusy = 800 {\rm ~GeV}, \quad
\mu = 2.5 \, \msusy, \quad M_2 = 500 {\rm ~GeV}, \nonumber \\
\label{smallaeff}
&& \Xt = -1100 {\rm ~GeV}, \quad
   \Ab = \At, \quad m_{\tilde g} = 500 {\rm ~GeV}~.
\end{eqnarray}
\end{itemize}

As discussed above, in order to study the impact of potentially large
corrections in the $b/\tilde{b}$ sector it is useful to vary the absolute 
value and sign of the parameter $\mu$. 
%\mla
This has led to the definition of an extension of the $\Mhmax$ and 
the no-mixing scenario by the following values of $\mu$~\cite{benchmark3}
\begin{equation}
\mu = \pm 200, \pm 500, \pm 1000 {\rm ~GeV} ~,
\label{eq:musuggest}
\end{equation}
allowing both an enhancement and a suppression of the bottom Yukawa
coupling and taking into account the limits from direct searches for
charginos at LEP.

The other MSSM parameters that are not specified above have only a minor
impact on MSSM Higgs-boson phenomenology. In our numerical analysis
below we fix them such that all soft SUSY-breaking parameters in the
diagonal entries of the sfermion mass matrices are set to $\msusy$, and
the trilinear couplings for all sfermions are set to $\At$.

For the exclusion bounds from the LEP Higgs searches the 
channel $e^+e^- \to Z^* \to Z h, H$ played a major role. We will
indicate the bounds obtained from this channel (for fixed $\msusy$ and
$\mt$)~\cite{LEPHiggsMSSM} in our figures below. As expected, the bounds
are weaker in the $\Mhmax$ scenario compared to the no-mixing scenario.
Limits from Run~II of the Tevatron have been published for the
following channels~\cite{D0bounds,D0bounds2,CDFbounds,CDFbounds2,Tevcharged}
($\phi = h, H, A$):
\BEA
\label{scen1}
    && p \bar p \to b \bar b \phi,  \phi \to b \bar b 
       ~(\mbox{with one additional tagged } b \mbox{ jet}) , \\[.3em] 
\label{scen2}
       && p \bar p \to \phi \to \tau^+\tau^-  
        ~(\mbox{inclusive}) , \\[.3em]
\label{scen3}
     &&p \bar p \to t \bar t \to H^\pm W^\mp \, b \bar b,
     H^{\pm} \to \tau \nu_{\tau} ~.
\EEA
While these limits begin to probe the region of small $\MA$ and large 
$\tb$ that is of particular interest for the CED Higgs production
analyses performed in this paper, the parameter region with 
$\MA \gsim 100 \gev$ and $\tb \lsim 50$ currently remains unaffected by the
Tevatron exclusion bounds.

%%%%%%%%%%%%%%%%%%%%%%%%%%%%%%%%%%%%%%%%%%%%%%%%%%%%%%%%%%%%%%%%%%%%%%%%%%%%%%%
%%%%%%%%%%%%%%%%%%%%%%%%%%%%%%%%%%%%%%%%%%%%%%%%%%%%%%%%%%%%%%%%%%%%%%%%%%%%%%%

\section{Cross sections for CED Higgs production in the MSSM}
\label{sec:sigmaprod}

In what follows we use the formalism of
\citeres{KMR,KMRProsp,KMRmm}
to obtain the cross sections for CED production of Higgs bosons,
similarly to \citere{KKMRext}. The amplitudes,
corresponding to the diagram of \reffi{fig:H}, are calculated 
using perturbative QCD techniques~\cite{KMR,KMRProsp}. 
There is also the possibility of soft rescattering in which particles
from the underlying events populate the gaps. Accounting for these
absorptive effects leads to a rather small probability for the
survival of the rapidity gaps, the so-called ``survival
factor" $S^2$~\cite{KMRsoft,maor}. A typical value for $S^2$ is about 
$S^2 = 0.025$, determined from summation of multi-pomeron
amplitudes~\cite{KMRsoft}.

For the purposes of this paper it is sufficient to use simple approximate 
formulae for the Higgs signal and background cross sections as a function of
mass, 
derived in \citeres{KKMRcentr,KKMRext}. At LHC energies these formulae 
approximate the full results of \citere{KKMRext} with an accuracy
of better than 10\% for Higgs masses in the region 
$50 \gev \lsim \Mh, \MH \lsim 350 \gev$.
For a more detailed analysis the formalism of \citeres{KMR,KMRProsp,KMRmm}
should be applied and appropriate MC programs, such as ExHuME (see 
\citeres{exhume,bhmps}), together with ATLAS or CMS simulation programs 
should be used to account for detector effects.

Here and in what follows we evaluate the cross sections $\si^{\rm excl}$ for
CED production of $h, H$ using the simplified formula
%\mda
\begin{equation}
\si^{\rm excl} \, \br^{\rm MSSM} =3 \, {\rm fb} \left(\frac{136}{16+M}\right)^{3.3}
  \left(\frac{120}M\right)^3 
  \frac{\Ga(h/H \to gg)}{0.25\mev} \,\br^{\rm MSSM},
\label{eq1}
\end{equation}
%\mua
where the gluonic partial width $\Ga(h/H\to gg)$ and the 
branching ratios for the various channels in the MSSM,
$\br^{\rm MSSM}$, are calculated using the
program \fh~\cite{feynhiggs,mhiggslong,mhiggsAEC,feynhiggs2.5}. 
The mass $M$ (in GeV) represents the Higgs-boson mass, i.e.\ 
$M = \Mh$ for $h$ and $M = \MH$ for $H$.
The factor $(136/(16+M))^{3.3}$ accounts for the mass dependence of the
effective ``exclusive" $gg^{PP}$ luminosity, see 
\citeres{KKMRcentr,KKMRext}.
The normalization is fixed  at $M=120 \gev$, where in accordance with 
\citere{KKMRext} we obtain $\si^{\rm excl} =3$~fb with the width  
$\Ga(\HSM \to gg)=0.25 \mev$.
In \citere{KKMRext} various uncertainties in the prediction of the
CED cross sections were evaluated. Combining different sources of
possible uncertainties in the  gluon-gluon  $gg^{PP}$
luminosity, it was found that the normalisation of
3~fb may be uncertain to a factor of about 2.5.
One particular source of uncertainty is the sensitivity to  the gluon
distribution, which on its own generates a factor of 1.5~uncertainty.
We anticipate that an accurate calibration of the  $gg^{PP}$ luminosity
at the LHC will be obtained by measuring exclusive high $E_T$
dijet (or exclusive $\ga\ga$) production, see \citeres{KMRmm,gam}.
Note also that the existing CDF data on measurements of such events
at the Tevatron~\cite{dino} are in reasonable quantitative agreement
with the expectations~\cite{bh75} using exactly the  same
formalism \cite{KMRProsp,KMRmm} that we are using to obtain the cross sections
for CED  Higgs-boson production%
\footnote{Additional uncertainties in the production cross sections of
up to $\sim 20\%$ could arise for large $\tb$ due to the imperfect
inclusion of NNLO QCD corrections.}%
.

Eq.~(\ref{eq1}) yields a total number of signal events of about~100 for a SM
Higgs boson with $\MHSM = 120 \gev$ with an integrated luminosity of 
60~fb$^{-1}$ if only the
forward detector acceptances are accounted for and no cuts and
efficiences in the central detector are imposed.

In the case of pseudoscalar $A$ production, which is
suppressed by the $\cp$-even selection rule (see \citeres{Liverpool,KMRmm}),
the signal cross section is given approximately by
%\mda
\begin{equation}
\si^{\rm excl} \, \br^{\rm MSSM} =3 \, {\rm fb} 
 \KL 0.0071 \; \frac{136^{3.3}}{(0.5+M)^{3.45}} \KR \;
  \left(\frac{120}M\right)^3 
  \frac{\Ga(A \to gg)}{0.25\mev} \, \br^{\rm MSSM} ,
\label{eq2}
\end{equation}
%\mua
with $M = \MA$. Again, 
the gluonic partial width $\Ga(A\to gg)$ and $A$ decay 
branching ratios are evaluated with the
program \fh~\cite{feynhiggs,mhiggslong,mhiggsAEC,feynhiggs2.5}.

%%%%%%%%%%%%%%%%%%%%%%%%%%%%%%%%%%%%%%%%%%%%%%%%%%%%%%%%%%%%%%%%%%%%%%%%%%%%%%%
%%%%%%%%%%%%%%%%%%%%%%%%%%%%%%%%%%%%%%%%%%%%%%%%%%%%%%%%%%%%%%%%%%%%%%%%%%%%%%%

\section{Background processes}
\label{sec:backgrounds}

\subsection{Summary of the backgrounds to the 
                     $p \oplus (h, H\to b\bar b) \oplus p$ signal} 
\label{sec:bgbb}

As we have already discussed, the advantage of the 
$p \oplus (h,H \to b\bar b) \oplus p$ 
signal is that there exists a $J_z=0$ selection rule, which requires the
leading order $gg^{PP}\to b\bar b$ QCD background subprocess to vanish in
the limit of massless quarks and forward outgoing protons.
However, there are still various sources of potentially important
background contributions, for more details see
\citeres{DKMOR,bh75,krs2}. 

As discussed in \citere{krs2} it is convenient to consider
separately the  quark helicity conserving (QHC) and the quark helicity
non-conserving (QHNC) background amplitudes. These amplitudes do not
interfere, and their contributions can be treated independently which,
in particular, avoids double-counting. 

There are four main types of background subprocess contributions.

\begin{itemize}
\item[(i)] 
The prolific (LO)
$gg^{PP}\to gg$ subprocess can mimic $b\bar b$ production since one may
misidentify the outgoing gluons as $b$ and $\bar{b}$ jets.
This contribution was first evaluated in \citere{DKMOR}
assuming a 1\% probability of misidentification per jet and
applying a $60^\circ<\theta<120^\circ$ jet cut.

\item[(ii)]
An admixture of $|J_z|=2$ production, arising from non-forward going
protons, which contributes to the (QHC) LO $gg^{PP}\to b\bar b$ background. 
 
\item[(iii)] 
Since the $b$-quarks have non-zero mass there is a contribution to the
$J_z=0$ (QHNC) cross section of order $\mb^2/E_T^2$. It is this term
which currently raises the main concern from the theory side.
The problem is that the result is strongly affected by the 
(uncomfortably large) higher-order QCD effects which can dominate over 
the Born-level prediction, see \citeres{fkm,krs2}. 
In particular, the one-loop double logarithmic 
contribution is larger than the Born term, and the final result
becomes strongly dependent on the NNLO  effects as well as
on the scale $\mu$ at which the QCD coupling $\al_S$
is evaluated, and on the running $b$~quark mass.  
Although no complete result for these higher-order 
effects to the $gg^{PP}\to b\bar b$ process currently exists,
an estimate has been given in \citere{krs2} based on a seemingly
plausible hypothesis that the NNLO effects can be incorporated
in a way similar to the simpler process $\ga\ga \to q\bar q g$,
where the full one-loop result is known.
However without the complete result for the higher-order 
radiative corrections corresponding to the 
$gg^{PP}\to b\bar b$ amplitude it is impossible to make a firm prediction.

It is our understanding that the estimate of this contribution, given in
\citere{krs2}, should be valid to an  
accuracy not better than a factor of 2-4 (or maybe even worse), and
this exclusive piece currently represents the main source of 
theoretical uncertainty in the predictions for the overall
background coming from single proton-proton collisions. 
A welcome feature of this contribution is that it
decreases with 
increasing $E_T$ much faster than the other background terms, see below
and \citeres{KKMRext, KMRCP}. 

\item[(iv)] 
Finally, there is the possibility of NLO $gg^{PP}\to b\bar b g$
(dominantly QHC) background contributions, which for hard
gluon radiation at large angles do not obey the selection rules, 
see \citeres{BKSO,DKMOR, krs2}.
Of course the extra gluon can in principle be observed
experimentally (as an additional jet) and the contribution of such
background events reduced. 
However, there are important exceptions which are discussed in detail
in \citeres{DKMOR,krs2}.

\begin{itemize}
\item[(a)] 
The extra gluon may go unobserved in the direction of
a forward proton. This background may be reduced by
requiring the approximate equality $M_{\rm missing} = M_{b\bar b}$.
The degree of this reduction  depends on the mass resolution
in the proton detector and jet energy resolution in the central detector.
The calculations in \citere{kmrN} show that
for the $b\bar b$ mass interval $\De M_{bb}=0.2 M_{b\bar b}$ 
and a mass window of $\Delta M=5 \gev$ 
($\Delta M$ indicates the mass window
over which the signal and background are collected)
the background arising from these
forward going gluons (with $k_T<5 \gev$) does not exceed  
5\% of the SM Higgs signal and, therefore, can be safely neglected.

\item[(b)]
The remaining danger is large-angle hard gluon emission which is
collinear with either the $b$ or $\bar{b}$ jet, and, therefore,
unobservable. 
A general study of a background coming from three-jet  $b \bar b g$ production
has been performed in \citere{krs2}. This background source results
in a sizable contribution and is therefore taken into account in the final
background formula, see \refeq{eq:backbb} below. 
\end{itemize}
\end{itemize}

In principle, we also have to consider the higher-order contributions
corresponding to the radiation of additional (soft) gluons.
These terms have to be taken into account together with
the virtual loop radiative corrections to the 
lowest-order cross section discussed in \citeres{KMRProsp,krs2}.
However, numerically, these contributions do not signficantly increase
the radiative background, see \citere{kmrN}. It should be noted
that the effect of gluon emission off the screening (labeled ``Q" in
\reffi{fig:H}) is also numerically small \cite{myths}.

There are also some other potentially worrying background sources,
which after a thorough investigation~\cite{krs2,kmrN} have
not been included in the final expression for the  $b \bar b$
%\mla
background in \refeq{eq:backbb} below. This is either because their
contributions are numerically small from the very beginning,
or because they can be reduced to an acceptable level 
by straightforward experimental cuts.
First, there is the NNLO QHC contribution to the
exclusive process which comes from the one-loop box diagrams.
This piece is not mass-suppressed and is potentially important 
especially for large $M_H$. As a consequence 
of rotational invariance (see \citere{BKSO}) 
it has a different angular dependence than the other contributions.
In particular, this term vanishes at $\theta=\pi/2$ and reaches its
maximum at $\theta=\pi/4$. 
However, numerically, this contribution is comparatively small.

\bigskip
Second, a potential background source can arise from  
the interaction of two soft Pomerons. This can result in
the two main event categories:
\begin{itemize}
\item[(a)] central Higgs-boson production accompanied by
two (or more) additional gluon jets,

\item[(b)] production of a high $E_T$ $b\bar b$-pair accompanied by gluon
jets.
\end{itemize}
In these cases the Higgs boson or the $b\bar b$ pair are produced in the
collision of two gluons (from the Pomeron wave functions) via the hard
subprocesses ($gg\to H$ or $gg\to b\bar b$) similarly to the usual
inelastic event.
In both processes the mass, $M_{bb}$, of the central  $b\bar b$ system
(resulting either from the Higgs decay or from the QCD background)
is not equal to the `Pomeron-Pomeron' mass $M_{PP}=M_{\rm missing}$, which is
measured with good accuracy by the forward proton detectors.
As discussed in \citere{DKMOR}, the suppression of such backgrounds
is controlled by the requirement that $|M_{\rm missing}-M_{bb}|$ should
lie within the $\Delta M_{bb}$ mass interval.
Unfortunately, the currently expected accuracy of the mass measurement
in the central detector is not as good as anticipated in \citere{DKMOR},
and more detailed studies are therefore needed in order to 
evaluate these backgrounds more carefully, see \citere{kmrN}. 
To retain sufficient Higgs signal statistics
we take the mass interval to be twice the mass
resolution in the central detector, $\De M_{bb}\simeq 24 \gev$.
Already this requirement imposes a strong
restriction on the diffractive parton distribution function (DPDF):
the longitudinal momentum fraction $\be$
of the Pomeron should be comparatively large, $\be > 0.6 - 0.7$.
At the same time, at the large scales ($\mu \sim M_H/2$) relevant for
Higgs production, the diffractive gluon distribution vanishes rapidly 
for $\be \to 1$. As a result, the expected background cross sections appear
to be quite small. Using the MRW2006 DPDFs~\cite{MRW}
and imposing a cut $60^\circ < \theta < 120^\circ$ on the final-state jets,
we arrive at the following conclusion: 
for $\De M_{bb}\simeq 24 \gev$ the $gb\bar bg$ and $gHg$ contributions 
are each less than about  6\% 
of the SM Higgs 
signal\footnote{It should be noted
that since the diffractive gluon density $g^D(\beta,x_{\rm Pom},\mu)$
(in our case $x_{\rm Pom}=\xi$) vanishes at least as 
$g^D(\be,x_{\rm Pom},\mu) \propto (1-\be)$ for $\be \to 1$, 
the $gHg$ and $gb\bar bg$ contributions fall with decreasing $\De M_{bb}$ 
as $(\De M_{bb}/M_{bb})^k$ with $k>4$. 
Two powers come from the phase space limitation,
and two powers come from the behaviour of the diffractive gluons ($g^D$) 
at large $\be$.
}.
 Moreover, a sizable fraction of such  events
can be further rejected by observing the extra (gluon) jets in the
detectors.

It should be noted that one could try to generate events produced in the
collision of  two soft Pomerons
using the POMWIG Monte Carlo~\cite{Pomwig} (modulo a proper account of
the soft survival factor $S^2$), see for instance \citere{CMS-Totem}. 
Unfortunately, the published version of  this generator 
uses the old H1 DPDFs (at the scale $\mu^2 \sim 75 \gev^2$),
which are almost flat for $\be > 0.5$.
At present, the large $\be$ behaviour of the DPDFs is not sufficiently 
well constrained by the inclusive diffractive DIS 
data.
In the new H1 QCD analysis \cite{H1new} a good description 
of the diffractive DIS data is obtained by two fits:
H1 2006 DPDF Fit~A  with essentially flat gluons and
H1 2006 DPDF Fit~B, where the diffractive gluon density 
rapidly decreases for $\be > 0.5$. However, it is only 
Fit~B that is consistent with the HERA data 
on diffractive charm and dijet production. At the same time,
this fit is quite close to the  MRW2006~\cite{MRW} results,
described in terms of the LO perturbative QCD calculations.
This provides a justification for our choice of the large 
$\be$ behaviour of the DPDFs, and we therefore conclude that
the background arising from soft Pomeron--Pomeron interactions
can safely be neglected.

Finally, note that the exclusive Higgs signal may be smeared
by the additional contribution coming from the CED $H+n\cdot g$
production process. However this piece is suppressed by colour
constraints, since the $t$-channel two-gluon exchange across the gap region
should be colourless. In particular, because of this there is no single
gluon radiation. The first non-zero contribution starts from
$n=2$, but this is additionally colour-suppressed since the pair of
gluons should form a colour singlet. Next, the mass of the $g H g$ system
measured by forward protons is larger than that of the Higgs boson,
and we have to impose the mass matching condition discussed in 
item~(iv) above. 
Furthermore, those gluons with not too small transverse momenta, 
e.g.\ $k_T > 5 \gev$, can be detected and therefore such events can
be rejected. Numerically, this background appears to be small 
(about 15\% of the SM Higgs signal \cite{kmrN}) 
and, again, it is not included in the final formula \refeq{eq:backbb} below.
Therefore, the remaining significant background contributions come from 
exclusive dijet production, listed in items (i-iii, iv(b)) above%
\footnote{
We assume that multi-jet events can be rejected with the help of
the central detector or one of the  forward detectors.
}%
. 

Within the accuracy of the existing calculations \cite{KMRmm,DKMOR,krs2},
the overall background to the $0^+$ Higgs signal in the 
$b \bar b$ mode can be approximated by the following
formula
\BE
\frac{{\rm d}\si^B}{{\rm d} M} \approx 0.5 \, {\rm fb/GeV} \left[
0.92\left(\frac{120}{M}\right)^6 +
\frac{1}{2} \left(\frac{120}{M}\right)^8
                                  \right],
\label{eq:backbb}
\EE
where the first term in the square brackets corresponds to
the processes listed in items (i-ii, iv (b)) above, while the last term 
arises from the mass-suppressed term described in item (iii).

It is worth mentioning that in the derivation of \refeq{eq:backbb}  
a $b$-jet cut ($60^\circ < \theta < 120^\circ$ in the $b \bar b$ rest
frame) was applied to 
suppress the collinear singularities. In order to use the same
efficiencies listed in \refta{expeff} below for both the signal and
background, we fix the normalization in \refeq{eq:backbb} as if the
background had the same (flat) distribution in $\cos\theta$ 
as the Higgs decay signal\footnote{
This approximation proves to be sufficient within the given
limited angular interval  for $\theta$ ($60^\circ < \theta < 120^\circ$).
}.%
Correspondingly, 
the value of $\si^B$ is normalised such that the contribution coming
from the $60^\circ < \theta < 120^\circ$ interval becomes equal to the
genuine QCD cross section integrated over this $\theta$ interval.

It should be noted that in comparison with the background studies in
\citere{DKMOR}, in \refeq{eq:backbb} we use a higher value of the
probability $P_{g/b}$ for misidentifying a gluon as a $b$-jet,
$P_{g/b}=1.3 \%$, corresponding to the values used in ATLAS
studies~\cite{atlastdr}\footnote{
One may expect that in the particularly clean environment of the CED
events, the value of $P_{g/b}$ could be reduced.
}.%

Recall that currently the most serious background
is caused
by  exclusive $b\bar b$ QCD production, which
has {\it exactly\/} the same characteristics as the $h/H\to b\bar b$
signal. The last term in \refeq{eq:backbb} shows the lowest-order result. 
As discussed above,
the  higher-order $\als$ corrections can be quite large.
At the moment, only double-logarithmic effects are known,
see \citere{krs2} and references therein.
It would therefore be desirable to calculate, at least, 
the complete one-loop expression for the $gg^{PP}\to b\bar b$ background.

Finally it is worth noticing that we use \refeq{eq:backbb} 
only for the numerical evaluations of the background contribution, and
no detector simulation has been performed. In a more complete treatment,
involving a detector simulation, optimisation procedures could be applied 
which would potentially further reduce the effects of backgrounds.

%%%%%%%%%%%%%%%%%%%%%%%%%%%%%%%%%%%%%%%%%%%%%%%%%%%%%%%%%%%%%%%%%%%%%%%%%%%%%%%

\subsection{The backgrounds to the 
                    $p \oplus (h, H\to \tau^+\tau^-)\oplus p$ signal}
\label{sec:bgtautau}

Although the $\tau^+\tau^-$ signal has the advantage that there is
practically no irreducible QCD background, there are still
other sources of background events, see for example \citere{KMRCP}.
First, exclusive $\tau^+\tau^-$ events may be produced by $\ga\ga$
fusion. Secondly, there may be a contribution caused by exclusive
production of a pair of high $E_T$ ($\sim M/3$) gluons being
misidentified as a $\tau^+ \tau^-$-pair. In more detail:

\begin{itemize}
\item[(a)] 
The cross section for the QED production mechanism is appreciable,
and is enhanced by two large logarithms arising from the integration 
over the transverse momenta of the exchanged photons.
The corresponding luminosity factor can be calculated
with sufficient accuracy using the equivalent photon approximation,
see for example\ \citeres{KMRProsp,KMRphot}. To suppress the contribution
caused by a collinear singularity in the $\ga\ga \to \tau^+ \tau^-$
cross section, we impose a cut $60^\circ < \theta < 120^\circ$
on the polar angle $\theta$ in the $\tau^+ \tau^-$ rest frame. As in the
$b\bar b$ case, this results in a reduction of the Higgs signal by a
factor of 2.
Neglecting the $\tau$-lepton  mass and following the standard 
calculational procedure we obtain for the LHC energy $\sqrt{s}=14 \tev$
\BE
\label{eq:t1}
\si(pp\to p\ \oplus\ \tau^+ \tau^-\ \oplus \ p)\ =\ 20\,\mbox{fb}
 \frac{2\De M}M
\left(\frac{120 \gev}{M}\right)^2.
\EE
In \refeq{eq:t1} we account for the gap survival factor $S^2\simeq 0.9$,
caused by the `soft' rescatterings (see \citeres{KMRsoft,KMRphot}), and
for reference purposes we use the cross section for
exclusive production of a $\tau^+ \tau^-$-pair with the invariant 
mass $M=120 \gev$.
For the mass window $\De M= 5\gev$ at $M=120 \gev$
the expected QED background is about 2~fb. 
Since the dominant contribution to the  photon--photon luminosity
comes  from very small transverse momenta of the exchanged photons,
to suppress this background we can select events with comparatively
large $p_T$ of the outgoing protons.
For example, if $p_T > 200 \mev$, the QED background is diminished
by a factor of more than 70, while the Higgs signal is reduced by about
 40\%.\footnote{
It should be noted that
since larger photon transverse momenta correspond to
smaller impact parameters, the soft survival factor in the photon fusion
process at $p_T > 200 \mev$ falls  to $S^2 \simeq 0.5$.
}%

\item[(b)] 
The prolific gluon dijets may mimic $\tau^+ \tau^-$ production
since one may misidentify the outgoing gluons as $\tau$'s. To
evaluate this contribution we use the technique of \citere{KMRProsp}
to calculate the effective exclusive $gg^{PP}$ luminosity and
the Born $gg^{PP}\to gg$ hard cross section. Here  $gg^{PP}$ indicates
that the `active' gluons, which interact to form the system with mass~$M$,
originate from colourless $t$-channel exchanges, see \reffi{fig:H}.
Imposing the same $\theta$ cut, $60^\circ < \theta < 120^\circ$, we
obtain 
\BE
\label{eq;t2}
\si(pp\to p\ \oplus\ gg\ \oplus \ p)\ =\ 45\,\mbox{pb}\
\frac{2\De M}M
\left(\frac{120 \gev}{M}\right)^5~.
\EE
To suppress this QCD background down to the level of 1~fb
at $M=120 \gev$, the probability that a gluon is misidentified as a $\tau$,
$P_{g/\tau}$, must be less than 1/50 for {\it each} high $E_T$ jet%
\footnote{
In \citere{misiden}, for the inclusive case, the probability
$P_{g/\tau}$ was evaluated as 1/500. It therefore seems quite realistic 
to expect that the probability $P_{g/\tau}<1/50$ can be achieved in the
much cleaner environment of a CED event.
}%
.
\end{itemize}

%%%%%%%%%%%%%%%%%%%%%%%%%%%%%%%%%%%%%%%%%%%%%%%%%%%%%%%%%%%%%%%%%%%%%%%%%%%%%%%

\subsection{The backgrounds to the $p \oplus (h,H\toWW) \oplus p$ signal}
\label{sec:bgWW}

The $\WW$ decay mode of the Higgs 
is less challenging than the $b\bar b$ channel from a purely
experimental perspective. As discussed in detail in
\citeres{cox2,krs1}, another attractive feature of the $\WW$ channel is
that in this case there is no
relatively large  continuum background process, such as 
central exclusive $b \bar b$ production in the case of $H \to b \bar b$. 
It should be recalled that the latter strongly relies on the
experimental missing mass resolution and misidentification probability
$P_{g/b}$ being sufficiently well under control 
to provide adequate background suppression.

Events with two $W$ bosons in the final state fall into 3 broad
categories depending on the decay modes of the $W$'s: fully-hadronic,
semi-leptonic and fully-leptonic. Experimentally, events in which
at least one of the $W$'s decays in either the electron or muon channel 
are by far the simplest, and the analysis 
(see \citeres{cox2,krs1}) focuses on the semi- and fully-leptonic
modes, which constitute about half the signal. 

The main exclusive backgrounds in the case of the $WW$ channel
can be divided into two broad groups: 
\begin{enumerate}
\item 
Central production of a $WW$ pair, $pp\to p \oplus (\WW) \oplus p$, from
the process $\ga\ga \toWW$.
\item 
The $W$-strahlung process  $pp\to p \oplus Wjj \oplus p$
arising from the $gg^{PP}\to Wq \bar q$ subprocess, where the 
(hadronically decaying) $W^*$
is `faked' by the two quarks.
\end{enumerate}
As shown in \citere{cox2}, over a wide region of Higgs masses
the photon-photon backgrounds can be strongly suppressed
if one requires that the final leptons and jets are central
and impose cuts on the transverse momenta of the protons in the taggers.
At the same time, these cuts do not significantly affect the signal. 

The potentially problematic background arises from the second category
above, i.e.\ from the QCD $W$-strahlung processes, which have been studied in
detail in \citere{krs1}. Following this analysis, above the $WW$
threshold the background should not be a problem, since
its contribution can be suppressed extremely effectively
by requiring the invariant mass of the di-quark system to be close to
$\MW$. 

For $\MH < 2 \MW$, there are two distinct scenarios in which either the
$W$ or the $W^*$ decays leptonically. In the latter case the QCD
background has been shown to be very small. For the former, without
additional cuts on the final state, the calculated $gg$ background is of
the same order as the SM Higgs signal. Though a full MC simulation
would be required to assess the effectiveness of the
further optimisation approaches, we 
expect that procedures such as cuts
on the final state particles and azimuthal correlations
between jets, should enable this background to be 
significantly reduced further without dramatically affecting the signal.

To summarise the current understanding, further optimisation efforts are
required in order to reduce the QCD background contribution, 
arising in the semi-leptonic case when the off-shell $W$ boson decays
hadronically.  
For the fully leptonic decay modes, and for semi-leptonic 
decays in which the on-mass-shell $W$ boson decays hadronically, 
the signal-to-background ratio for a SM-type Higgs boson should be much
greater than unity. 
As explained above, an important property of the $WW$ channel is that in
this case the suppression of the dominant backgrounds does not rely primarily
on the high precision of the missing mass resolution.

%%%%%%%%%%%%%%%%%%%%%%%%%%%%%%%%%%%%%%%%%%%%%%%%%%%%%%%%%%%%%%%%%%%%%%%%%%%%%%%
%%%%%%%%%%%%%%%%%%%%%%%%%%%%%%%%%%%%%%%%%%%%%%%%%%%%%%%%%%%%%%%%%%%%%%%%%%%%%%%

\section{CED production of the  $\cp$-even Higgs bosons 
  $h$ and $H$: experimental aspects} 
\label{sec:cedprodhH}

In the following we will discuss the prospective experimental
efficiencies for the various CED channels. 
As discussed above, in the MSSM the Higgs-boson phenomenology is very
different from the SM case. 
The light $\cp$-even MSSM Higgs boson cannot be heavier than about
130~GeV~\cite{mhiggslong,mhiggsAEC}. For large 
values of $\tb$ and not too large $\MA$, its couplings
to bottom quarks and $\tau$~leptons can be strongly
enhanced compared to the SM case. In certain parameter regions, on the
other hand, also a strong suppression of the couplings of the 
light $\cp$-even MSSM Higgs boson to down-type fermions is possible, leading
in this case to an enhancement of $\br(h \toWW)$. 
Within the SM with increasing Higgs-boson mass the branching ratios of the 
Higgs boson into $b \bar b$ and $\tau^+\tau^-$ fall very rapidly due to
the rise of the branching ratios into the $\WW$ and $ZZ^{(*)}$. In the MSSM,
on the other hand, the  $H$~boson decouples from the gauge bosons if its
mass is much higher than the upper bound on the mass of the light Higgs,
and the $A$~boson has zero couplings to the SM gauge
bosons at tree-level. Consequently, the decays of $H,A$ 
into $b \bar b$ and $\tau^+\tau^-$ remain dominant as long as no decay
channels into supersymmetric particles (or light Higgs bosons) are open.
Therefore in the mass region above about $130 \gev$ 
the decay channels $H \to b \bar b, \tau^+\tau^-$ are
much more important in the MSSM as compared to the SM case. The
couplings of $H,A$ to down-type fermions receive an additional
enhancement in the large $\tb$ region. On the other
hand, the decay of the neutral $\cp$-even MSSM Higgs bosons into $W$ bosons, 
$h, H \toWW$ is significant only for relatively low masses, 
i.e.\ $\Mh, \MH \lsim 130 \gev$.

A detailed investigation of the Higgs sector is one of the central
physics targets of the recent proposal~\cite{FP420} to add 
proton tagger (Roman Pot (RP)) detectors positioned at a distance 
$\pm 420$~m from the interaction 
points of the ATLAS and CMS experiments at the LHC.
At nominal LHC optics this will allow coverage in the proton
fractional momentum loss $\xi$ in the range 0.002--0.02,
with an acceptance of around 30\% for a centrally produced system
with a mass around $120 \gev$.
A combination with the forseen proton detectors at 
$\pm 220$~m~\cite{totem,RP220} would 
significantly increase the physics reach of forward studies enlarging
the $\xi$ range up to 0.2.
This would be especially beneficial because of
the acceptance for higher mass states and improvements in the triggering,
see the discussion below and \citere{CMS-Totem}.

Experimentally, for increasing mass of the centrally produced system, 
the detector characteristics improve
and the overall environment becomes cleaner.
When the mass increases, the overall proton tagger
acceptance (which includes all the combinations of 420~m and 220~m
taggers) rises, also the trigger and $b$-tagging efficiencies 
increase, and the Higgs-boson mass resolution can improve. 
The signal-to-background ratio should also increase since it
behaves roughly as $M^3/\Delta M$, where as before 
$\Delta M$ is the mass window
over which the signal and background is collected, see
\refeqs{eq1}, (\ref{eq2}) and (\ref{eq:backbb}).

While the total RP acceptance increases with rising mass, 
the acceptance of the 420~$+$~420~m (referred to
below as ``420'') configuration starts to decrease around a mass of 
$90 \gev$, so that the 
fraction of the CED signal events detected at this configuration
becomes marginal for masses above $200 \gev$. In general, the low mass
signal comprises events detected at 220~$+$~420~m or 420~$+$~220~m
(referred to below as ``combined'' configuration) and in the 420 
configuration,
while the high-mass signal consists mainly of events detected at 
220~$+$~220~m (denoted as ``220'') and in the combined configuration.
For $M_H \gsim 250 \gev$ only the 220 RP configuration is relevant.

The mass resolution, $\delta M$, however, depends on the RP
configuration. At $M=120 \gev$ in the 420 case a mass resolution of
$\de M = 1.9 \gev$ is achievable,
whereas the resolution becomes 2--3 times worse for the combined or 220
configurations. 
Because of worsening mass resolution, in order to collect a sufficient
part of the signal we have to apply a wider mass window than in the
420 case. By enlarging the mass window, the collected signal
increases as a consequence of
the convolution of Gaussian and Breit-Wigner distributions
(as will be explained below), while the background
contribution is directly proportional to the width of the mass window.
At some point, the statistical significance stops rising and starts to
decrease. Therefore, it becomes essential to find the 
optimal mass window widths for all different RP configurations.
The procedure we used to find the optimum mass window is described below.

We now briefly describe the main selection criteria for the three basic
decay channels: 

\begin{itemize}

\item $h,H \to b \bar b$: 
In the current study the signal comprises two main event categories: (a) two
$b$-tagged jets and (b) two jets with at least one $b$-hadron decaying 
into a muon. The
signature for (a) consists of three main ingredients: ($i$) two
scattered protons, one in each arm of the RPs, ($ii$) two well-collimated
$b$-tagged jets (back-to back in the azimuthal angle $\phi$,
$2.85<|\phi_1-\phi_2|<3.43$) with rapidities $|\eta_{\rm jet}|<2.5$
in the central detector, and ($iii$) consistent values of the whole
4-momentum (mass, rapidity and transverse momentum) of the central mass
system as evaluated from the two tagged protons (``missing mass'') and
as determined from the two jets.  In particular the following cuts were
imposed on the longitudinal momentum fractions $x^+,\, x^-$
and the ratio of the dijet mass $M_{bb}$ measured in the central
detector to the missing mass $M_{\rm missing}$, given by the RPs
(see \citere{CMS-Totem}):
\BEA
|x^k_{bb}-x^k_{\rm missing}|/x^k_{\rm missing} &<& 0.3 
\mbox{ (k = $+, -$ denote the light cone components of the } \non \\ 
&& \mbox{\hspace{30mm} momenta in opposite beam directions)} \non \\
\label{eq:bbcuts}
0.85<M_{bb}/M_{\rm missing} &<& 1.15 
                 \mbox{ (for the 420 configuration)} \\
0.8<M_{bb}/M_{\rm missing} &<& 1.2 
                 \mbox{ (for the combined and 220 RP configurations)} \non
\EEA
The other event category (b) with muons in the final state yields about 10\%
of the whole signal sample. 
The signal is selected in the same way as in the event category (a)
except that the jets are not required to be b-tagged and in addition, the 
jet plus muon trigger conditions are applied (a jet with $E_{\rm T} >40 \gev$
and at least one muon with $E_{\rm T} >3 \gev$).
Further details on the selection cuts can be found
in \citere{CMS-Totem}\footnote{
It would be beneficial to include electrons originated from the
$b$-decays; at the moment this is still under discussion.}.%

\item $h,H \to \tau^+\tau^-$: 
This channel has a dominant dijet signature, but also 
sizable branching ratios of $\tau \to l\nu\bar{\nu}$.
These features render this channel similar
to the $h,H \to b \bar b$ channel. Therefore, we conservatively assume the
same selection efficiencies as in the case of decays to $b \bar b$.

\item $h,H \toWW$:
The experimental feasibility of the mode $h, H \toWW$ has been
studied in detail in \citeres{krs1,cox2}. Triggering on this channel is
not a problem, since the final state is rich in high-$p_{\rm T}$
leptons. Efficiencies of about 20\% can be achieved if the standard
leptonic and di-leptonic trigger thresholds are applied. It was
demonstrated in \citeres{krs1,cox2} that there would be
a detectable signal with small and controllable background
for the CED production of a SM-like Higgs boson in the mass interval
between $140 \gev$ and $200 \gev$.
Unfortunately, based on the standard lepton Level~1 trigger thresholds and
Roman Pot acceptances (see \refta{expeff} and \citere{cox2}), 
the yield for a $120 \gev$ SM Higgs boson with an integrated luminosity of 
$\cL = 60~{\rm fb}^{-1}$ is only about 3~events
\footnote{
The yield would rise if the lepton Level~1 leptonic trigger thresholds
could be reduced~\cite{cox2}.}%
.~The prospects become better for higher
luminosity. Furthermore,
as we will discuss below, the potential of the $h,H \toWW$ mode may improve
in favourable regions of the MSSM parameter space when the decay 
$h \to b \bar b$ is suppressed~\cite{benchmark2}. 

\end{itemize}

For the purpose of this paper we define the overall efficiency, $\epsilon$,
which incorporates the RP and central detector acceptances, experimental cuts,
efficiencies of the central detector and Level~1 triggers. 
In the $h,H \to b\bar b$ case it also accounts
for the $b$-tagging efficiency and for the cut on the rapidities of the
central jets $|\eta_{\rm jet\,1}-\eta_{\rm jet\,2}|<1.1$%
\footnote{
In the configuration with two forward outgoing protons this is
equivalent to the cut $60^\circ<\theta <120^\circ$ on the polar angle of
$b$-jets in the $b\bar{b}$ rest frame, required to reduce the QCD background
(see the discussion in \refse{sec:bgbb}).}%
.~The same angular (rapidity) cut is imposed in the $\tau\tau$ channel
in order to suppress the QED background. The parameter $\epsilon$
relates the cross section that can actually
be detected, $\si_{\rm det}$, to the theoretical cross
section $\si_{\rm th}$ (calculated using \refeqs{eq1}, 
(\ref{eq2}) and (\ref{eq:backbb})), which would correspond to the case
that no experimental cuts are applied and no efficiency losses occur.
Accordingly,
\BE
\si_{\rm det} = \epsilon \cdot \si_{\rm th} .
\label{eq:redfac}
\EE

The background processes relevant for the different CED channels have
been discussed in \refses{sec:bgbb}--\ref{sec:bgWW}.
To retain the signal at the Level~1 trigger, 
the following trigger conditions can be used: 

\begin{enumerate}

\item 
{\em Single-sided 220~m RP and at least two jets, each with 
{$E_{\rm T}>40 \gev$}, measured in the central detector}.\\
As was shown in \citere{Diffrtrig}, for this trigger a tolerable Level~1
bandwidth of 1~kHz per experiment may be kept only up to luminosities of
about
${\cal L} \sim 2 \times 10^{33} \, {\rm cm}^{-2} \, {\rm s}^{-1}$
due to pile-up background.
Requiring the single-sided 220~m RP condition implies that the symmetric
420 RP configuration may only be possible for the events where the
acceptances of 220~m and 420~m RPs at the same side overlap. In such
events (amounting roughly to 40\% of the total event yield for the 
$b \bar b$ decay mode at a mass of $120 \gev$ in the CMS--TOTEM system)
the information from 420~m is going to be used due to a better resolution on 
the fractional momentum loss $\xi$ of the proton.
In the rest of the signal retained by this Level~1 trigger condition, 
the combined RP configuration will be present resulting in a 2--3 times
worse mass resolution than for the 420 case (see above).

The efficiency of this trigger to retain that part of the signal which
passes the cuts on the two jets and
does not include selected muons rises from about 60\% at $M = 120 \gev$
 to 100\% at $M = 200 \gev$.

\item 
{\em A jet with {$E_{\rm T}>40 \gev$} and at least one muon with
{$E_{\rm T}>3 \gev$}, both measured in the central detector}.\\
This trigger retains that part of the signal that comes from the decays
of $b$-hadrons or $\tau$~leptons into muons. Therefore it also increases
the fraction of the 420~RP configuration in the total signal sample. The
saving efficiency for the signal events of the above type that pass the
cuts is 100\% for all masses.

\item 
{\em At least two jets each with {$E_{\rm T}>90 \gev$} measured in the
central detector}.\\
This is the lowest $E_{\rm T}$ threshold for any dijet trigger designed at
ATLAS or CMS. Obviously, all the signal with masses well above $200 \gev$ 
(and passing the cuts)
may be retained at Level~1 just by this trigger.

\item 
{\em Leptonic triggers, requiring electrons or muons in the central
detector}.\\
This trigger serves mainly to retain events with $W$~bosons decaying
leptonically or semi-leptonically. 
In CMS, the $E_{\rm T}$ thresholds are 29, 10, 14 and $3 \gev$ for the
single-electron, double-electron, single-muon and double-muon triggers,
respectively. Corresponding values for ATLAS are 25, 15, 20 and 
$10 \gev$, but the trigger efficiencies are similar.

\end{enumerate}

For the process $h,H \to b \bar b$, a combination of the triggers~1
and~2 allows 
the retention of about 65\% of the signal events passing the relevant cuts 
at $M = 120 \gev$ and up to 100\% at 
$M = 200 \gev$, while at masses well above $200 \gev$ the trigger~3
retains the whole signal sample selected by the cuts. 
On the other hand, the signal for the process $h,H \toWW$ (with 
the leptonic decay of at least one $W$) is collected
by applying the cuts corresponding to the trigger definition only,
without any additional selections.

%%%%%%%%%%%%%%%%%%%%% T A B L E %%%%%%%%%%%%%%%%%%%%%%%%%%%%%%%%%%%%%%%%%%%%%%%
\begin{table}[htb!]
\renewcommand{\arraystretch}{1.2}
\begin{center}
\begin{tabular}{||c||c|c|c|c|c|c|c||}
\hhline{|t:========:t|}
%\hline\hline
$M$[GeV] & $A_{420}$ & $A_{\rm comb}$ & $A_{220}$ &
$\epsilon_{420}$($b \bar b$) & $\epsilon_{\rm comb}$($b \bar b$) &
$\epsilon_{220}$($b \bar b$) & $\epsilon$($\WW$) \\ \hline\hline
100          & 37  & 13 & 0   & 1.2    & 0.8  & 0   &  -   \\ \hline
120          & 31  & 25 & 0   & 1.7    & 2.5  & 0   & 11.7 \\ \hline
140          & 25  & 37 & 0   & 1.6    & 5.1  & 0   & 14.6 \\ \hline
160          & 19  & 49 & 0   & 1.5    & 7.6  & 0   & 20.4 \\ \hline
180          & 14  & 60 & 0   & 1.2    & 9.6  & 0   & 21.5 \\ \hline
200          & 9   & 69 & 0   & 0.4    & 11.0 & 0   & 24.2 \\ \hline
300          & 0   & 76 & 13  & 0      & 12.5 & 2.0 &  -   \\
\hhline{|b:========:b|}
%\hline\hline
\end{tabular}
\end{center}
\renewcommand{\arraystretch}{1}
\caption {RP acceptances ($A$) for the 420, combined and 220 RP
configurations, and total experimental selection efficiencies for the
signal decay channels $h,H \to b \bar b$ and $h,H \toWW$, evaluated 
under the assumption that the whole signal is collected at a given mass,
see also \citere{CMS-Totem}. All numbers are given in \%. They have been 
obtained with ExHuME~1.3~\cite{exhume}. 
Conservatively, $\epsilon(h,H \to \tau^+\tau^-)$ is 
set equal to $\epsilon(h,H \to b\bar b)$ (with $\epsilon$
defined in \refeq{eq:redfac}).}
\label{expeff} 
\end{table}
%%%%%%%%%%%%%%%%%%%%% T A B L E %%%%%%%%%%%%%%%%%%%%%%%%%%%%%%%%%%%%%%%%%%%%%%%

\refta{expeff} shows RP acceptances as a function of the Higgs-boson
mass for all possible RP configurations. We also show the signal selection
efficiencies that have been used in our evaluation of the statistical
significances. Following \citere{CMS-TDR}, the statistical significance 
has been calculated as a
probability from the Poisson distribution with mean equal to the number
of background events to observe a number of events equal or greater than
a sum of signal and background events, converted to an equivalent number
of sigmas of Gaussian distribution \cite{NarBit}. 
The number of signal events at a given mass has been obtained by scaling the
signal cross section from \refeq{eq1} or (\ref{eq2}) by the signal
selection efficiencies from \refta{expeff}. The same procedure has been
applied to obtain the number of background events with the cross section
calculated using \refeq{eq:backbb}. In the calculation of the
significance, we considered the number of signal and background
events without systematic uncertainties.
As mentioned above, both the signal and the background consist of two
contributions corresponding to the two RP configurations, namely 420 and
combined up to masses of $200 \gev$, and 220 and combined for masses above
$200 \gev$.

In the case of large $\tb$ the total width ($\Ga_{\rm tot}$) of the
Higgs bosons may become rather large. Therefore, in what follows we
choose the optimal mass window
\begin{equation}
\De M(M, \tb) = 2 \sqrt{{\delta M}^2(M)+\Ga_{\rm tot}^2(M, \tb)} , 
\label{eq:masswindow}
\end{equation}
where the mass resolution $\delta M$ was taken from the RP studies made in
\citere{RO}. The values of the mass window
range from 3 to 20$\gev$ for the 420 case and 
between 11 and 22$\gev$ for the combined or 220 RP configurations. The lower 
boundaries correspond to low values of $\Ga_{\rm tot}$, while the larger 
boundaries correspond to larger values of $\Ga_{\rm tot}$.

In our analysis we use the 
optimum mass window as the region over which the signal as
well as
background cross sections are integrated to obtain the numbers of signal ($S$) 
and background ($B$) events:
\BEA
  S &=& \si^{S}(M)\,{\cal L}\,
  \KKL    {\cal I}_{420}(M, \tb)\,\epsilon_{420}(M)
        + {\cal I}_{\rm comb}(M, \tb)\,\epsilon_{\rm comb}(M) \KKR~, \\[.3em]
B &=& \frac{{\rm d}\si^B}{{\rm d} M}{\cal L}\,
  \KKL     \De M_{420}(M, \tb) \epsilon_{420}(M)
         + \De M_{\rm comb}(M, \tb) \epsilon_{\rm comb}(M)\KKR~,
\label{B}
\EEA
where $\si^{S}$ and ${\rm d}\si^{B}/{\rm d}M$ are the cross sections 
calculated using 
\refeqs{eq1}, (\ref{eq2}) and (\ref{eq:backbb})%
\footnote{
When calculating $B$,
we actually integrated over the mass window from $M-\Delta M/2$ up to
$M+\Delta M/2$ explicitly. We obtained the same result as using eq.~(\ref{B})
which is due to the fact that the effect of the first
derivative with respect to $M$ is cancelled by the integration over
the symmetric interval, and only the second derivative may contribute.}%
.~${\cal I}_{420}$ and ${\cal I}_{\rm comb}$ are the integrals of convolutions 
of Gaussian and Breit-Wigner mass distributions over a given mass window for 
the 420 and combined RP configuration, respectively,
and $\De M_{420}$ and $\De M_{\rm comb}$ are the corresponding mass
windows. Since the total width, which depends on $\tb$, enters 
$\De M_{420}$ and $\De M_{\rm comb}$, our background rate given in
\refeq{B} formally depends on $\tb$. It should be noted that in a
situation where the total width is larger than the mass resolution, the
CED process may provide a unique opportunity to actually 
{\em measure \/}  the total width. This information may be very
important for distinguishing an MSSM Higgs boson from a SM-type Higgs.

Because of the acceptance,
the 220 RP configuration only contributes for the highest mass
value given in \refta{expeff}, i.e. $M = 300 \gev$. 
Because of similar
mass resolutions for the combined and 220 RP configurations, in this
mass region contributions of
both the RP configurations are summed and denoted by ``combined''.
When
calculating the integrals ${\cal I}$, we account for the mass dependence of the
hard subprocess, the experimental efficiencies (from Table~\ref{expeff}) and
the effective exclusive $gg^{PP}$ luminosity. For this particular choice of
the optimum mass window, the values of ${\cal I}_{420}$ and 
${\cal I}_{\rm comb}$ are shown in Table~\ref{masswind}. Evidently they are
close to 0.67 over the whole studied range of the values of the mass and 
the total width. As a systematic cross-check, we have also 
tried another option for the optimum mass window (arising from a more
complete calculation), namely 
$\De M = \sqrt{{(2.7\,\delta M)}^2+(1.5\Ga_{\rm tot})^2}$, and observed very
similar results as for the previous option.

%%%%%%%%%%%%%%%%%%%%% T A B L E %%%%%%%%%%%%%%%%%%%%%%%%%%%%%%%%%%%%%%%%%%%%%%%
\begin{table}[htb!]
\renewcommand{\arraystretch}{1.2}
\begin{center}
\begin{tabular}{||c||c|c|c|c|c||}
\hhline{|t:======:t|}
%\hline\hline
$M$[GeV] & $\Ga_{\rm tot}=$0.2 & 1 & 5 & 10 & 20 [GeV] \\
\hline\hline
120      & 65/67  & 62/66 & 69/66 & 71/71 & 75/80 \\ \hline
140      & 64/67  & 61/65 & 69/63 & 69/65 & 70/67 \\ \hline
160      & 62/67  & 62/65 & 69/62 & 70/66 & 71/68 \\ \hline
180      & 62/67  & 62/64 & 67/64 & 67/67 & 68/69 \\ \hline
200      & 61/67  & 59/64 & 62/63 & 65/67 & 83/69 \\ \hline
300      & --/68  & --/65 & --/63 & --/67 & --/70   \\
\hhline{|b:======:b|}
%\hline\hline
\end{tabular}
\end{center}
\renewcommand{\arraystretch}{1}
\caption{The values of the integrals ${\cal I}_{420}$ / ${\cal I}_{\rm comb}$ 
[in \%] for the 420 and combined  RP configurations.}
\label{masswind} 
\end{table}
%%%%%%%%%%%%%%%%%%%%% T A B L E %%%%%%%%%%%%%%%%%%%%%%%%%%%%%%%%%%%%%%%%%%%%%%%

For the $h, H \to \tau^+\tau^-$ sample it may be possible
to exploit the full 420 RP information in the off-line analysis.
Using information on the event topology and other characteristics of the
$h, H \to \tau^+\tau^-$ sample, it may be possible to avoid the
single-sided 220~m RP condition for the $\tau^+\tau^-$ final
state. However, no detailed studies of the trigger for such events have
been performed so far~\cite{mg}.

{}From the discussion above, it is evident that it would be beneficial if
the 420~m RPs could be incorporated into the Level~1
trigger. Based on the currently foreseen hardware this seems not
to be feasible, given the scope of the Level~1 trigger buffer and the time
that the signal needs to travel from the 420~m RP detector. An
increase of the Level~1 trigger latency would allow one to use the
420~m RP information for triggering at Level~1~\cite{ma}. Such a
setup could be very advantageous for suppressing pile-up and non-pile-up
backgrounds.
It is worth noting in this context that an increase of the Level~1
trigger latency, as needed for triggering on the 420~m RP information, has been
discussed in the context of the LHC luminosity upgrade
(SLHC)~\cite{SLHC}. 

In the above discussion of the experimental prospects for CED Higgs
production we have particularly emphasised the importance of
triggering in the central detector on leptons arising from $b$, $\tau$ or
$W$ decays.  From the experimental point of view, in order to further 
enhance the physics potential of the CED Higgs processes 
an improvement of the signal selection efficiency by lowering the lepton 
thresholds would therefore be beneficial.

Bearing in mind the remaining theoretical uncertainties on the signal
rates (e.g.\ from
higher-order QCD effects), the results for the signal rates in this
study represent a conservative evaluation.
A possible enhancement of the signal rates by a factor of about 2
would still be compatible with our estimates of the theoretical
uncertainties. Furthermore, on the experimental side a gain in
sensitivity might be expected from improvements and optimisation of the
event selection procedures, triggers and various efficiencies. 
Besides the above-mentioned possibility of including the 420 RP into
the Level~1 trigger, one could envisage improvements from lowering the
lepton $p_{\rm T}$ thresholds in the existing Level~1 triggers,
from considering, possibly, an electron trigger and from improving the 
selection procedure for the $b \bar b$ and $\tau\tau$ signals. 
In order to illustrate the physics gain that could be expected if such a
more optimistic scenario were to be realised, we will include in the
following a
discussion of the cross sections and experimental efficiencies outlined
above with scenarios where the event rates are higher by a factor 
of~2 (due to improvements on the experimental side and possibly higher
signal rates, denoted by ``eff$\times 2$'').

Finally, it is important to recall that the price to pay for the increase
of the LHC luminosity is a rise in the average number, $N$, of soft
proton--proton interactions per bunch crossing. Thus, at an instantaneous
luminosity of $2 \times 10^{33} {\rm cm}^{-2} {\rm s}^{-1}$ $N =7$, while at
$10^{34} {\rm cm}^{-2} {\rm s}^{-1}$ $N=35$. In such a situation
each hard scale central event%
\footnote{Here ``hard scale event" refers to an
event with a large $E_{\rm T}$ jet (or lepton), or where a 
particle with a large mass (for example, a $W$-boson) is produced.}%
~is accompanied by a luminosity-dependent number of minimum bias events,
which dominantly are of soft origin. A certain fraction of these pile-up
events contains protons within the acceptances of the RPs.
In particular, from an evaluation based on \citere{kmr3P} it follows that
about 2--3\% (0.75--1\%) of the minimum bias events can be detected in
the RPs at 
220~m (420~m) on one side. When the same RP acceptances are applied,
these numbers appear to be in a reasonable agreement with the results
of \citere{CMS-Totem}, where Phojet~\cite{phojet} was used to generate
minimum bias events. 
The pile-up events could potentially endanger the prospects of CED
studies at high luminosities, since they can be overlaid with hard scale
inclusive non-diffractive events.
None of these events would separately survive the signal selection cuts,
but once two single diffractive events, each with a proton within the RP
acceptance, are overlaid with a hard scale central signal, this mixture
of three events may perfectly fake the signal.
The problem is that the rates of inclusive processes are so huge
compared to the CED signal that this situation may occur quite
frequently. As a result, event pile-up is (currently) considered to be one of
the most important background sources in the planned high instantaneous
luminosity runs. 

The pile-up issue is currently under very intensive study within ATLAS
and CMS (for a detailed discussion see \citere{CMS-Totem}).
Although at first sight the issue of pile-up backgrounds
does not look favourable, the situation is far
from being hopeless. Apart from concentrating on the 420~m RP
configuration, which enables a narrow mass window to be imposed, there
are other possible leverages that could bring the pile-up problem under
control. 

In the FP420 project~\cite{FP420} there are prospects for installing fast
timing detectors with an expected vertex resolution of better than 3\,mm.
Such a precision would enable one to determine whether the protons seen in
the RPs came from the same vertex as the hard scale central
signal. Preliminary Monte Carlo studies indicate that, with the nominal
LHC running conditions, a rejection of a factor of about 40 should be
possible~\cite{RP220}.  

The pile-up effect can further be reduced by exploiting precise
vertex detectors with vertex resolution of better than 100\,$\mu$m: events
with more than one vertex in a few mm window in the beam direction given
by the timing detectors (presumably from pile-up) can then be
rejected~\cite{CMS-Totem}. 
Another possibility to reduce the effect of pile-up is to exploit
different track multiplicity properties of signal and pile-up events. 
A cut reflecting this difference may provide another rejection factor of
up to 100~\cite{Nchcut}. Here, however, further studies have to be performed,
 in particular with full detector simulations.

Finally, matching the whole 4-momentum (not just the mass, but also
the rapidity and transverse momenta) of the central heavy system (dijet,
$\WW$-pair) measured in the central detector to the corresponding values coming
from the forward protons taggers will further suppress the pile-up
background.

It should be noted that the situation concerning pile-up backgrounds 
is more favourable in the $\WW$ channel compared to the channel with a
$b \bar b$ pair in the central detector.
Indeed,
in the SM case the signals expected in the $b\bar b$ and $\WW$ channels
are of the same order of magnitude. The lower probability for the light
Higgs to decay to $\WW$ is partly compensated by a better efficiency of
the $W$ signal selection. On the other hand, the probability of $W$
production in the pile-up event is lower than the
probability of high $E_{\rm T}$ $b$-jet production. In fact the suppression
caused by the large mass ($\MW \sim 80  \gev$) of a $W$~boson is slightly
stronger than that caused by the large $E_{\rm T}\sim 40$-$50 \gev$ of a
$b$-quark jet. Besides this, the electroweak coupling, which controls
the $W$ production rate, is smaller than the QCD  coupling $\als$. As a
result, the signal-to-background ratio $S/B_{\rm pile-up}$, 
where $B_{\rm pile-up}$ is the 
the pile-up background, is expected to be higher by around two orders of
magnitude for the $\WW$ channel as compared to the $b\bar b$ case.

Concerning triggering at Level~1, one may hope that the pile-up
background contaminating the Level~1 trigger output rate may be managed 
even at instantaneous 
luminosities higher than $2 \times 10^{33} {\rm cm}^{-2} {\rm s}^{-1}$ 
(which was found in
initial studies~\cite{Diffrtrig} to be the highest luminosity
compatible with a tolerable Level~1 bandwidth of 1~kHz per experiment).
In particular, if a Higgs boson has already been detected in the
standard (non-diffractive) search channels, the knowledge about its mass 
can be embedded into the proposed diffraction Level~1 trigger. 
Thus this information can be used to significantly suppress the
pile-up background (by restricting $\xi$ values accepted at Level~1).
Another option would be to
incorporate the information from fast timing detectors into the Level~1
trigger logic. 

In conclusion, we believe that it is conceivable that with intensive studies
ATLAS and CMS will succeed in bringing the effect of pile-up down to a
tolerable level even at the highest luminosities which will be 
delivered after several years of LHC running.
In other words, even at the highest instantaneous luminosities 
it may be possible to select and study exclusive (CED) events in the
presence of pile-up interactions.

Based on the above discussion, in our numerical analysis below we
consider four scenarios for the achievable luminosities and the
experimental conditions for CED processes at the LHC:

\newcommand{\sixoo}{60 \ifb}
\newcommand{\sixooeff}{60 \ifb\,eff$\times2$}
\newcommand{\sixooo}{600 \ifb}
\newcommand{\sixoooeff}{600 \ifb\,eff$\times2$}

\begin{itemize}
\item \underline{\sixoo :}\\
An integrated LHC luminosity of ${\cal L} = 2 \times 30~{\rm fb}^{-1}$,
corresponding roughly to three years of running at an instantaneous
luminosity
${\cal L} \sim 10^{33} \, {\rm cm}^{-2} \, {\rm s}^{-1}$ by both ATLAS
and CMS. With such a luminosity the effect of pile-up is not negligible
but can be safely kept under control. The signal selection efficiencies
are based on \refta{expeff} (and are correspondingly reduced by
taking the optimal mass windows into account).

\bigskip
\item \underline{\sixooeff :}\\
The same integrated LHC luminosity as in the above scenario but with 
event rates that are higher by a factor of 2 (see the discussion of possible
improvements and theoretical uncertainties above). 

\item \underline{\sixooo :}\\
An integrated LHC luminosity of ${\cal L} = 2 \times 300~{\rm fb}^{-1}$
and the same efficiency factors as in the scenario with
${\cal L} = 60~{\rm fb}^{-1}$.
This corresponds roughly to three years of running at an instantaneous
luminosity $\cL \approx 10^{34}\,{\rm cm}^{-2} \, {\rm s}^{-1}$ by both
ATLAS and CMS.

\item \underline{\sixoooeff :}\\ 
The same integrated LHC luminosity as in the scenario 
with ${\cal L} = 2 \times 300~{\rm fb}^{-1}$
but with 
event rates that are higher by a factor of 2.
\end{itemize}

%%%%%%%%%%%%%%%%%%%%%%%%%%%%%%%%%%%%%%%%%%%%%%%%%%%%%%%%%%%%%%%%%%%%%%%%%%%%%%%
%%%%%%%%%%%%%%%%%%%%%%%%%%%%%%%%%%%%%%%%%%%%%%%%%%%%%%%%%%%%%%%%%%%%%%%%%%%%%%%

\section{Discovery reach for neutral $\cp$-even Higgs bosons 
in the MSSM in CED production}
\label{sec:discovery}

In this section we discuss the prospects for observing the neutral
$\cp$-even MSSM Higgs bosons in CED production. We display our results
in terms of the lowest-order parameters of the MSSM Higgs sector, 
$\MA$ and $\tb$, for the benchmark scenarios described in
\refse{subsec:benchmarks}.
Also shown in the plots as dark shaded (blue) areas are the
parameter regions excluded by the LEP Higgs
searches in the channel 
$e^+e^- \to Z^* \to Z h, H$~\cite{LEPHiggsSM,LEPHiggsMSSM}. 

For each point in the parameter space we have evaluated the relevant Higgs
production cross section, see \refse{sec:sigmaprod}, times
the Higgs branching ratio corresponding to the decay mode under
investigation. The Higgs-boson masses, the decay branching ratios and
the effective couplings for the production cross sections
have been calculated with the program
\fh~\cite{feynhiggs,mhiggslong,mhiggsAEC,feynhiggs2.5}.
The resulting theoretical cross section has been multiplied by the
experimental efficiencies taking into account detector acceptances,
experimental cuts and triggers as discussed in \refse{sec:cedprodhH}.
The backgrounds
have been estimated according to \refse{sec:backgrounds}.

This procedure has been carried out for four different assumptions on the
luminosity scenario, see \refse{sec:cedprodhH}, and the $5\sigma$
discovery contours (and contours for $3\sigma$ significances, see below)
have been obtained as described above.

%%%%%%%%%%%%%%%%%%%%%%%%%%%%%%%%%%%%%%%%%%%%%%%%%%%%%%%%%%%%%%%%%%%%%%%%%%%%%%%
%%%%%%%%%%%%%%%%%%%%%%%%%%%%%%%%%%%%%%%%%%%%%%%%%%%%%%%%%%%%%%%%%%%%%%%%%%%%%%%

\subsection{Prospective sensitivities for CED production of the 
    light $\cp$-even Higgs boson}

%%%%%%%%%%%%%%%%%%%%%%%%%%%%%%%% Begin FIGURE %%%%%%%%%%%%%%%%%%%%%%%%%%%%%%%%%
\begin{figure}[htb!]
\begin{center}
\includegraphics[width=14cm,height=8.8cm]
                {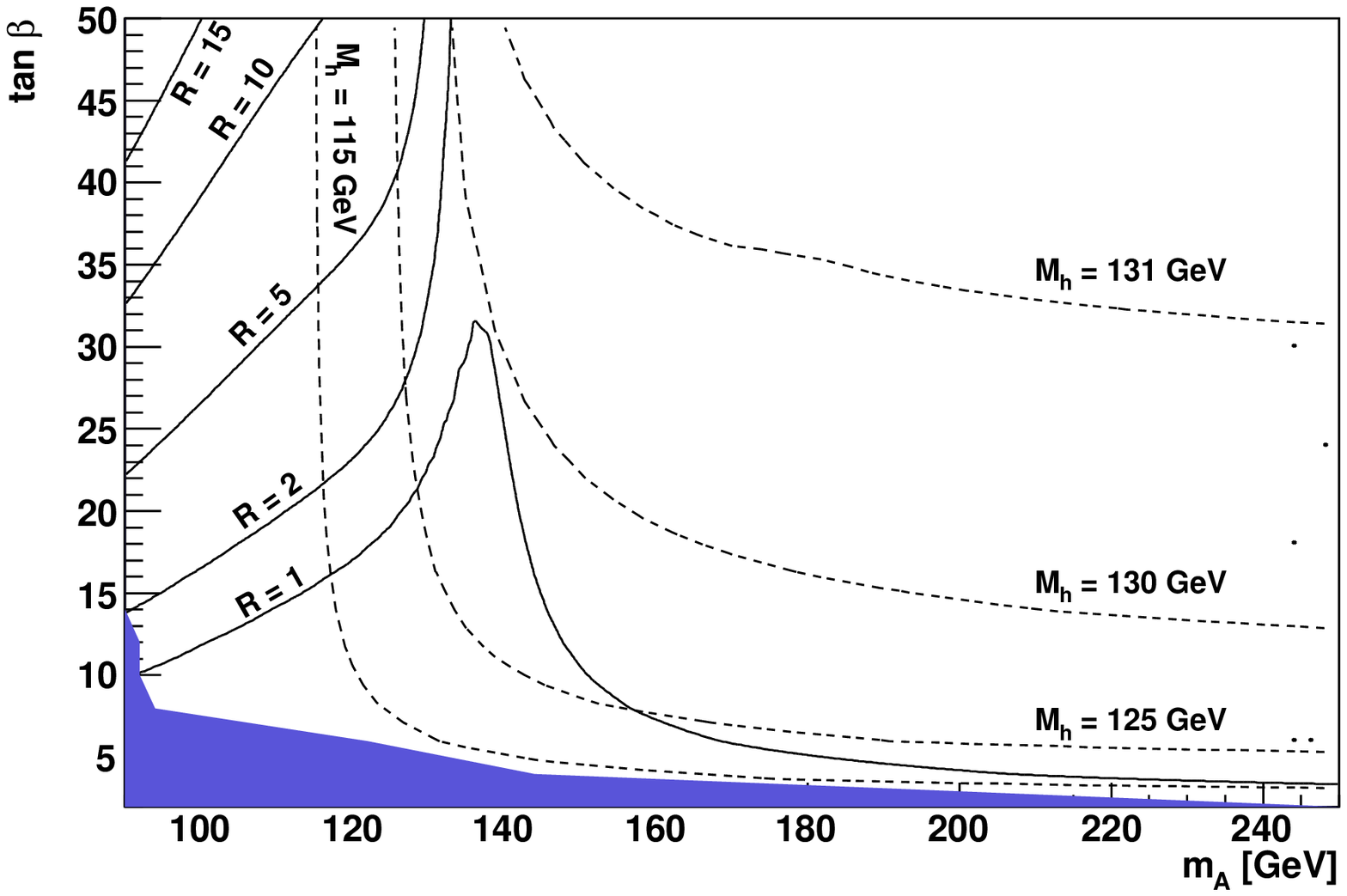}
\includegraphics[width=14cm,height=8.8cm]
                {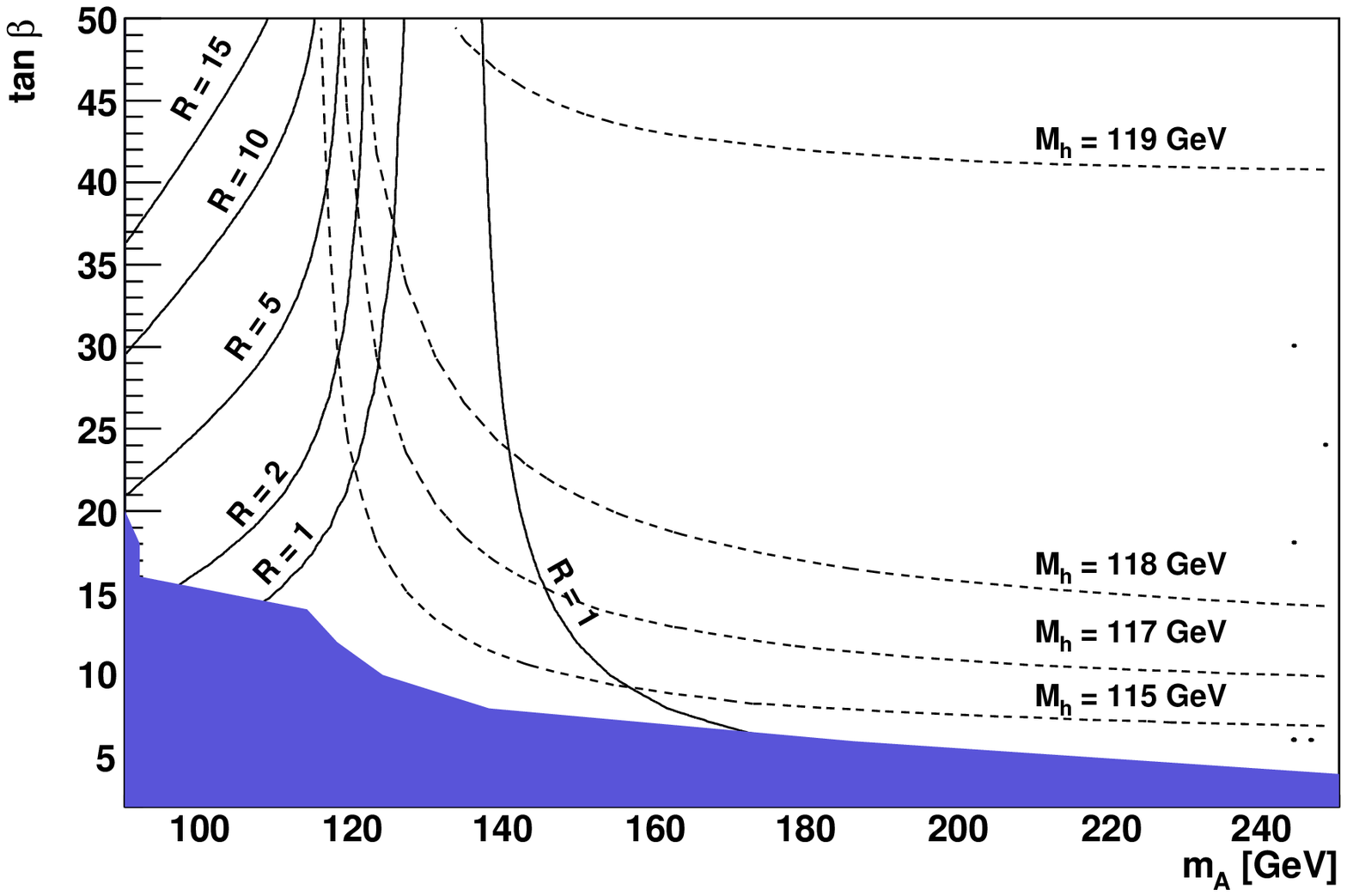} 
\caption{
Contours for the ratio of signal events in the MSSM to those in the SM in 
the $h \to b \bar b$ channel in 
CED production in the $\MA$--$\tb$ plane. The ratio is 
shown in the $\Mhmax$ benchmark scenario (with $\mu = +200 \gev$,
upper plot) and in the no-mixing scenario (with $\mu = +200 \gev$,
lower plot). 
The values of the mass of the light $\cp$-even Higgs boson, $\Mh$, are
indicated by dashed contour lines. The dark shaded (blue)
region corresponds to the parameter region that
is excluded by the LEP Higgs
searches in the channel 
$e^+e^- \to Z^* \to Z h, H$~\cite{LEPHiggsSM,LEPHiggsMSSM}. 
}
\label{fig:hbb-ratio}
\end{center}
\end{figure}
%%%%%%%%%%%%%%%%%%%%%%%%%%%%%%%% End FIGURE %%%%%%%%%%%%%%%%%%%%%%%%%%%%%%%%%%%

We start our analysis with the production of the lighter $\cp$-even Higgs
boson, $h$, and its decay into bottom quarks. As explained in
\refse{sec:Higgsbench}, the $hb\bar b$ coupling can be significantly
enhanced compared to the SM case in the region of relatively small $\MA$
and large $\tb$ (while the lighter $\cp$-even Higgs boson of the MSSM
behaves like the SM Higgs in the decoupling region, $\MA \gg \MZ$).
CED production of the lighter $\cp$-even Higgs boson of the MSSM with
subsequent decay to $b \bar b$ therefore yields a higher event rate in
this parameter region compared to the SM case. This is shown in
\reffi{fig:hbb-ratio} where the ratio of signal events for the MSSM 
to those for the SM (with $\MHSM = \Mh$)
is displayed in the $\MA$--$\tb$ plane for the 
$\Mhmax$ (upper plot) and no-mixing (lower plot) benchmark scenarios as 
specified in \refeqs{mhmax}, (\ref{nomix}). 
For illustration, contour lines for the mass of the lighter
$\cp$-even Higgs boson are also given.
The dark shaded (blue) region indicates the part of the $\MA$--$\tb$
plane that is excluded by the LEP Higgs searches
in the channel 
$e^+e^- \to Z^* \to Z h, H$~\cite{LEPHiggsSM,LEPHiggsMSSM}. 
As discussed above, the exclusion bounds from the Higgs search at the
Tevatron so far do not impose constraints on the parameter region with 
$\MA \gsim 100 \gev$ and $\tb \lsim 50$.

\reffi{fig:hbb-ratio} shows
that the signal rate can be enhanced by up to a factor of $R = 15$ in the
region of large $\tb$ and relatively 
small $\MA$. An enhancement by a factor of $R = 5$
is possible for $\tb$ values down to about $\tb \approx 25$. The
enhancement in the region of large $\tb$ and small $\MA$ is slightly more
pronounced in the no-mixing scenario (lower plot). This behaviour can
easily be understood from the discussion in \refse{subsec:HO}. For the
value of $\mu$ chosen in the $\Mhmax$ and no-mixing scenarios, 
$\mu = +200 \gev$, the higher-order contribution $\db$ is positive and
therefore leads to a (slight) suppression of the bottom Yukawa coupling. 
Since the numerical value of $\db$ in the $\Mhmax$ scenario is larger than in
the no-mixing scenario, as a consequence of the second term in
\refeq{def:dmb}, the $\Mhmax$ scenario yields slightly smaller
enhancement factors compared to the no-mixing scenario in the region of 
large $\tb$ and small $\MA$. In contrast, the differences between
the two scenarios in the parameters of the scalar top sector entering 
via higher-order corrections (affecting in particular the predicted
value of $\Mh$) give rise to a possible sizable
enhancement in the $\Mhmax$ scenario up to higher values of 
$\MA$ compared to the no-mixing scenario. While an enhancement by a
factor of $R = 2$ occurs within the $\Mhmax$ scenario up to 
$\MA \lsim 130 \gev$, the corresponding contour is shifted towards lower
$\MA$ values in the no-mixing scenario by about $\De\MA = 10 \gev$.
For $\MA \gsim 150 \gev$, on the other hand, the signal rates are close
to the SM case in both scenarios over practically the whole range of 
$\tb$ values, as expected from the discussion of the decoupling limit
given above.

%%%%%%%%%%%%%%%%%%%%%%%%%%%%%%%% Begin FIGURE %%%%%%%%%%%%%%%%%%%%%%%%%%%%%%%%%
\begin{figure}[htb!]
\begin{center}
\includegraphics[width=14cm,height=8.8cm]
                {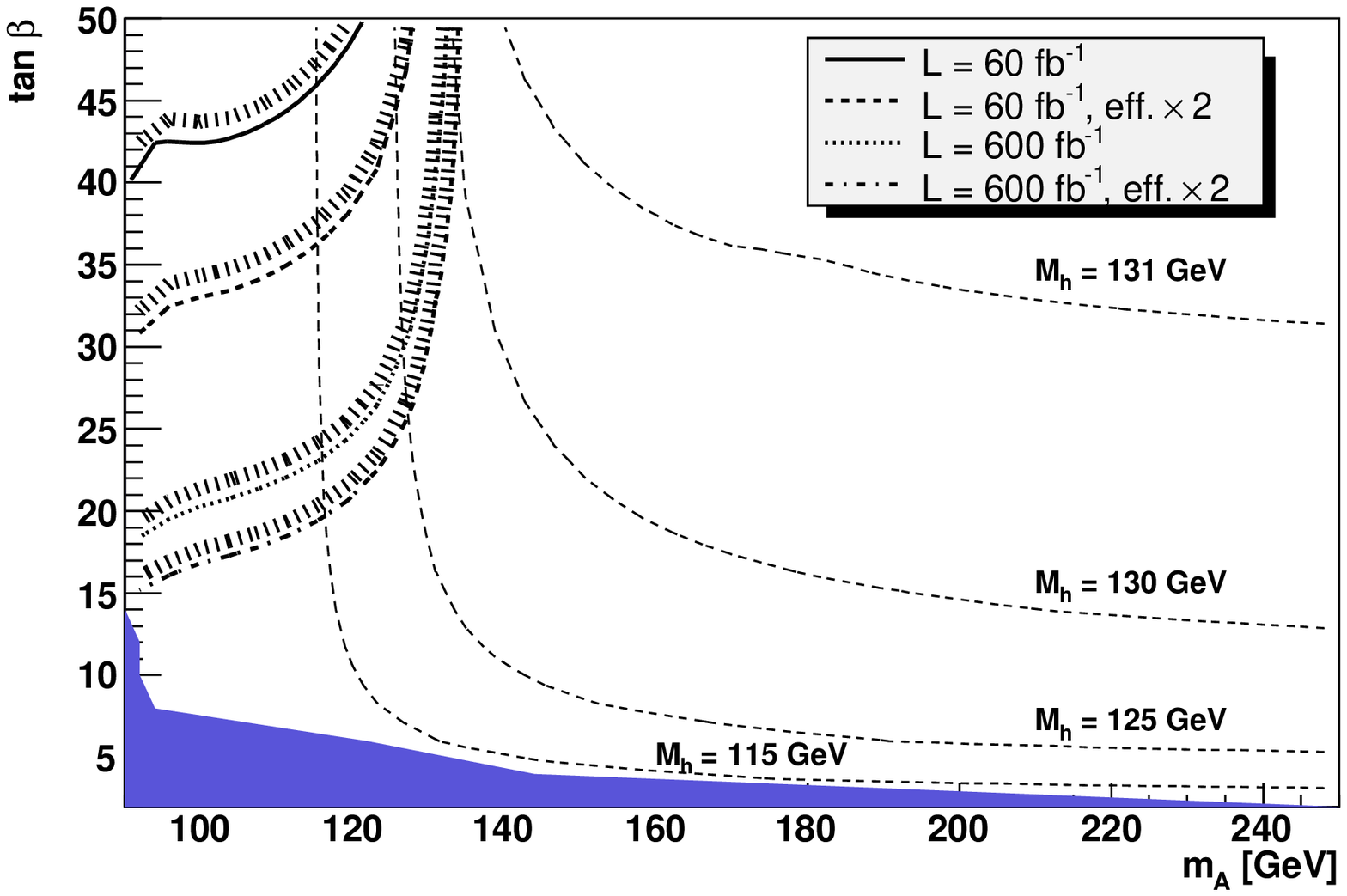}
\includegraphics[width=14cm,height=8.8cm]
                {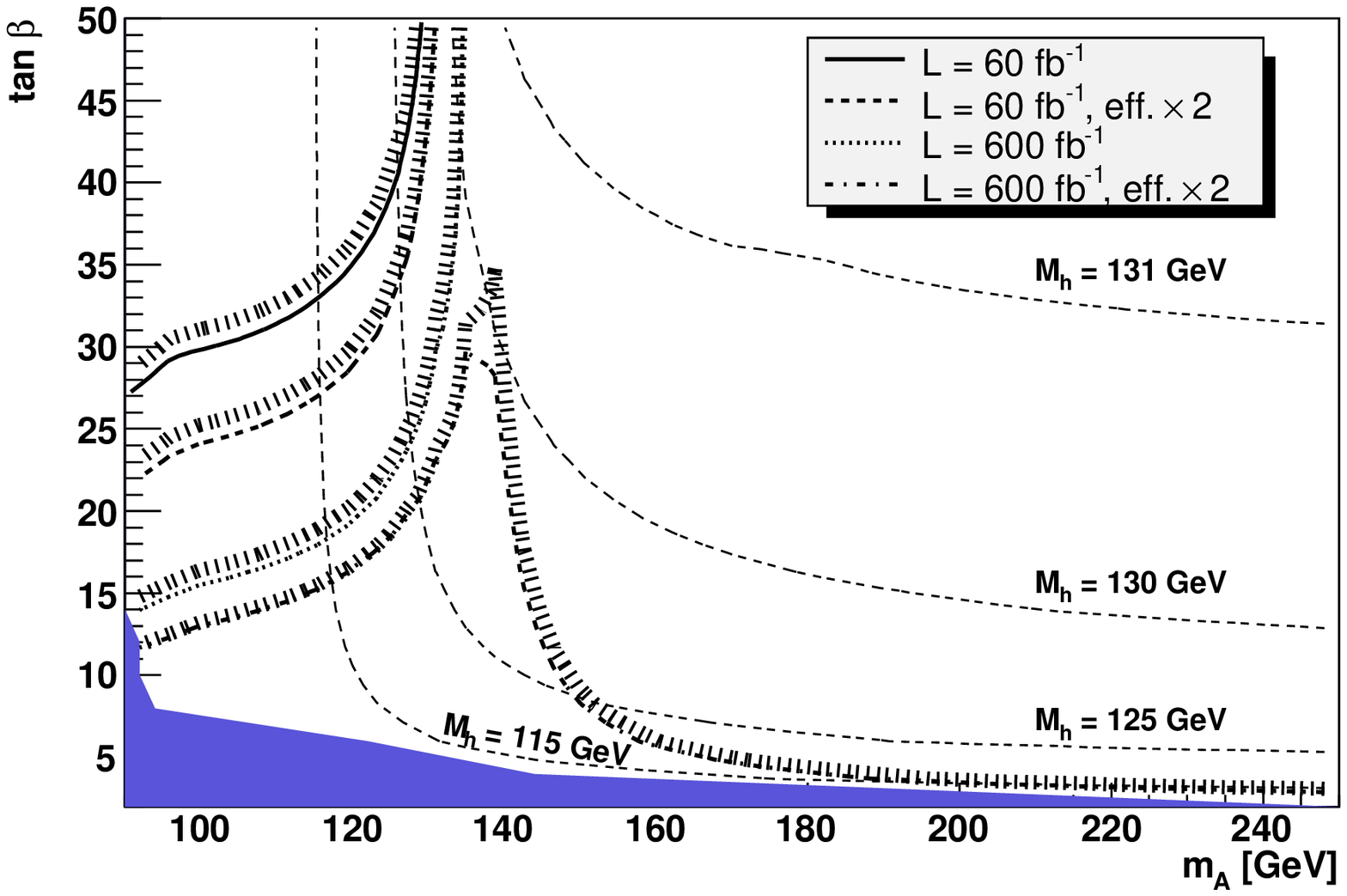}
\caption{
$5 \si$ discovery contours (upper plot) and contours of $3 \si$
statistical significance (lower plot) for the $h \to b \bar b$ channel in
CED production in the $\MA$--$\tb$ plane of the MSSM within the $\Mhmax$
benchmark scenario. The results are shown for assumed effective
luminosities (see text, combining ATLAS and CMS) of \sixoo,
\sixooeff, \sixooo\ and \sixoooeff. 
The values of the mass of the light $\cp$-even Higgs boson, $\Mh$, are
indicated by contour lines. The dark shaded (blue)
region corresponds to the parameter region that
is excluded by the LEP Higgs 
searches in the channel 
$e^+e^- \to Z^* \to Z h, H$~\cite{LEPHiggsSM,LEPHiggsMSSM}. 
}
\label{fig:hbb1}
\end{center}
\end{figure}
%%%%%%%%%%%%%%%%%%%%%%%%%%%%%%%% End FIGURE %%%%%%%%%%%%%%%%%%%%%%%%%%%%%%%%%%%

%%%%%%%%%%%%%%%%%%%%%%%%%%%%%%%% Begin FIGURE %%%%%%%%%%%%%%%%%%%%%%%%%%%%%%%%%
\begin{figure}[htb!]
\begin{center}
\includegraphics[width=14cm,height=8.8cm]
                {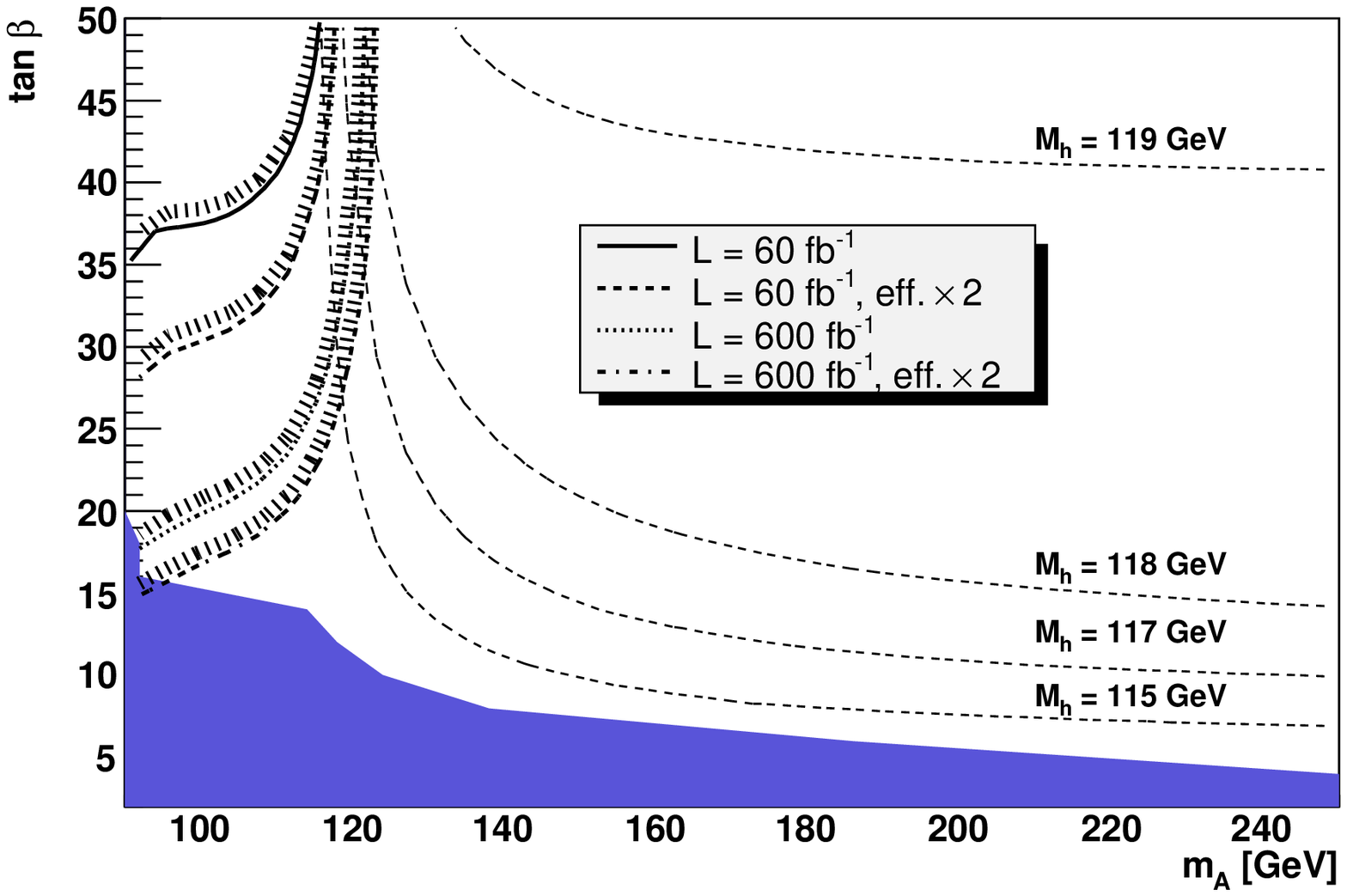}
\includegraphics[width=14cm,height=8.8cm]
                {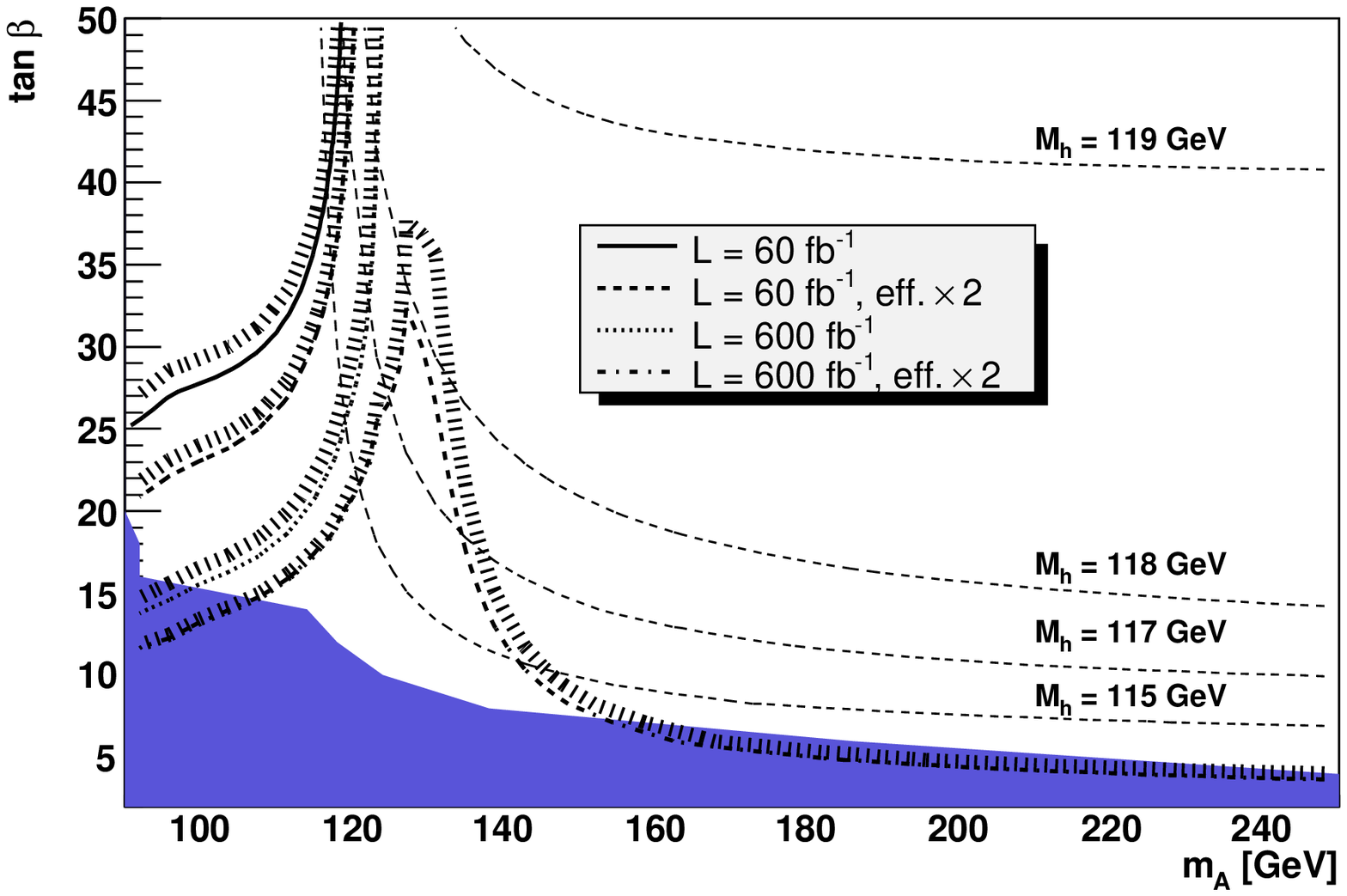}
\caption{
$5 \si$ discovery contours (upper plot) and contours of $3 \si$
statistical significance (lower plot) for the $h \to b \bar b$ channel in
CED production in the $\MA$--$\tb$ plane of the MSSM within the
no-mixing benchmark scenario. The results are shown for assumed effective
luminosities (see text, combining ATLAS and CMS) of \sixoo,
\sixooeff, \sixooo\ and \sixoooeff. 
The values of the mass of the light $\cp$-even Higgs boson, $\Mh$, are
indicated by contour lines. The dark shaded (blue)
region corresponds to the parameter region that
is excluded by the LEP Higgs 
searches in the channel 
$e^+e^- \to Z^* \to Z h, H$~\cite{LEPHiggsSM,LEPHiggsMSSM}. 
}
\label{fig:hbb2}
\end{center}
\end{figure}
%%%%%%%%%%%%%%%%%%%%%%%%%%%%%%%% End FIGURE %%%%%%%%%%%%%%%%%%%%%%%%%%%%%%%%%%%

The $5 \si$ discovery contours in the $\MA$--$\tb$ plane obtained for the 
four luminosity scenarios 
specified above are given for the $\Mhmax$ and no-mixing
benchmark scenarios (i.e.\ the same parameters as in
\reffi{fig:hbb-ratio}) in the upper plots of \reffis{fig:hbb1} 
and~\ref{fig:hbb2}, respectively. The shapes of the $5 \si$ discovery
contours in the two benchmark scenarios follow the patterns discussed in 
\reffi{fig:hbb-ratio}: the region of high $\tb$ and low $\MA$ can be
covered with slightly lower integrated luminosity in the no-mixing
scenario than in the $\Mhmax$ scenario. While in the ``\sixoo''
scenario $\tb$ values down to about $\tb = 40$ can be covered with 
$5 \si$ significance in the low $\MA$ region of the $\Mhmax$ scenario, 
the coverage in the no-mixing scenario extends to about $\tb = 35$ with
the same luminosity (this would improve by $\De\tb \approx 5$--10 in the
more optimistic ``\sixooeff'' scenario).
On the other hand, the coverage in the $\Mhmax$
scenario extends to somewhat higher $\MA$ values. With 600~\ifb the
parameter region with $\MA \lsim 135 (125) \gev$ and $\tb \gsim 20$ can be 
covered in the $\Mhmax$ (no-mixing) scenario at the $5 \si$ level 
(and a slightly
better coverage can be achieved in the more optimistic 
``\sixoooeff'' scenario). In the $\Mhmax$ and no-mixing
scenarios the parameter $\mu$ is set to $\mu = +200\gev$. It follows
from the discussion of the $\db$ corrections given above that the
discovery reach would increase for negative values of $\mu$. We will
give an example below for the case of the heavy $\cp$-even Higgs boson.
As an example of expected event rates, for \sixoo one would expect
after all cuts in the $\Mhmax$ scenario with 
$\tb = 50$ and $\Mh \approx 120 \gev$ about 32 signal events and 
28 background events, while in the no-mixing scenario with
$\tb = 50$ and $\Mh \approx 113 \gev$ one would expect 41 signal events
and 35 background events.

In all our considered luminosity scenarios, a $5 \si$ significance for
the production of the lighter $\cp$-even Higgs boson of the MSSM and 
its decay into bottom quarks is found only in the region of relatively
small $\MA$ and large $\tb$ in \reffis{fig:hbb1} and~\ref{fig:hbb2}.
This is related to the fact that for a SM Higgs in the mass range
corresponding to \reffis{fig:hbb1},~\ref{fig:hbb2} the statistical
significance remains below the $5 \si$ level. Because of the enhanced
signal rate in the respective part of the MSSM parameter space, as
discussed above, a $5 \si$ discovery region occurs in the MSSM while it
is absent in the SM case.

Since the lighter $\cp$-even Higgs boson of the MSSM is likely to be
detectable also in ``conventional'' Higgs search channels at the LHC
(see for example \citeres{atlastdr,CMS-TDR}), it may not be necessary to 
require a statistical significance as high as $5 \si$ 
for the CED channel. For illustration,
we therefore also show the contours of $3 \si$ statistical
significances, given in the lower plots of \reffis{fig:hbb1}
and~\ref{fig:hbb2}. Higgs production in the CED channel will provide
unique information on the Higgs-boson properties. In particular, it will
be important for determining the $\cp$ quantum numbers, for a precise
mass measurement, and it may even allow a direct measurement of the
Higgs-boson width. The CED production process with subsequent decay into
bottom quarks is of particular relevance since this channel may be the
only possibility for directly accessing the $hb \bar b$ coupling,
although the decay into bottom quarks is by far the dominant decay mode
of the lighter $\cp$-even Higgs boson in nearly the whole
parameter space of the MSSM (and it is also the dominant decay of a light 
SM-like Higgs). 
For this reason information on the bottom Yukawa coupling is important for
determining {\em any\/} Higgs-boson coupling at the LHC (rather than
just ratios of couplings), see \citere{HcoupLHCSM}.

\reffis{fig:hbb1},~\ref{fig:hbb2} show that at the $3 \si$ level a
significantly larger part of the $\MA$--$\tb$ plane can be covered
compared to the $5 \si$ discovery contours. In particular, in the
``\sixoooeff'' scenario the coverage in both benchmark
scenarios extends to large $\MA$ values and small values of $\tb$.
With the exception of a small parameter region around 
$\MA \approx 140 (130) \gev$, in the $\Mhmax$ (no-mixing) scenario the
whole $\MA$--$\tb$ plane of the MSSM (and also the case of a light
SM-like Higgs) can be covered with the CED process in this
case. This important result implies that if the CED channel can be utilised 
at high instantaneous
luminosity (which requires in particular that pile-up background
is brought under control, see the discussion in
\refse{sec:cedprodhH}) there is a good chance to detect the 
lighter $\cp$-even Higgs boson of the MSSM in this channel with
subsequent decay into bottom quarks, yielding crucial information on the
properties of the new state.

%%%%%%%%%%%%%%%%%%%%%%%%%%%%%%%% Begin FIGURE %%%%%%%%%%%%%%%%%%%%%%%%%%%%%%%%%
\begin{figure}[htb!]
\begin{center}
\includegraphics[width=14cm,height=8.8cm]
                {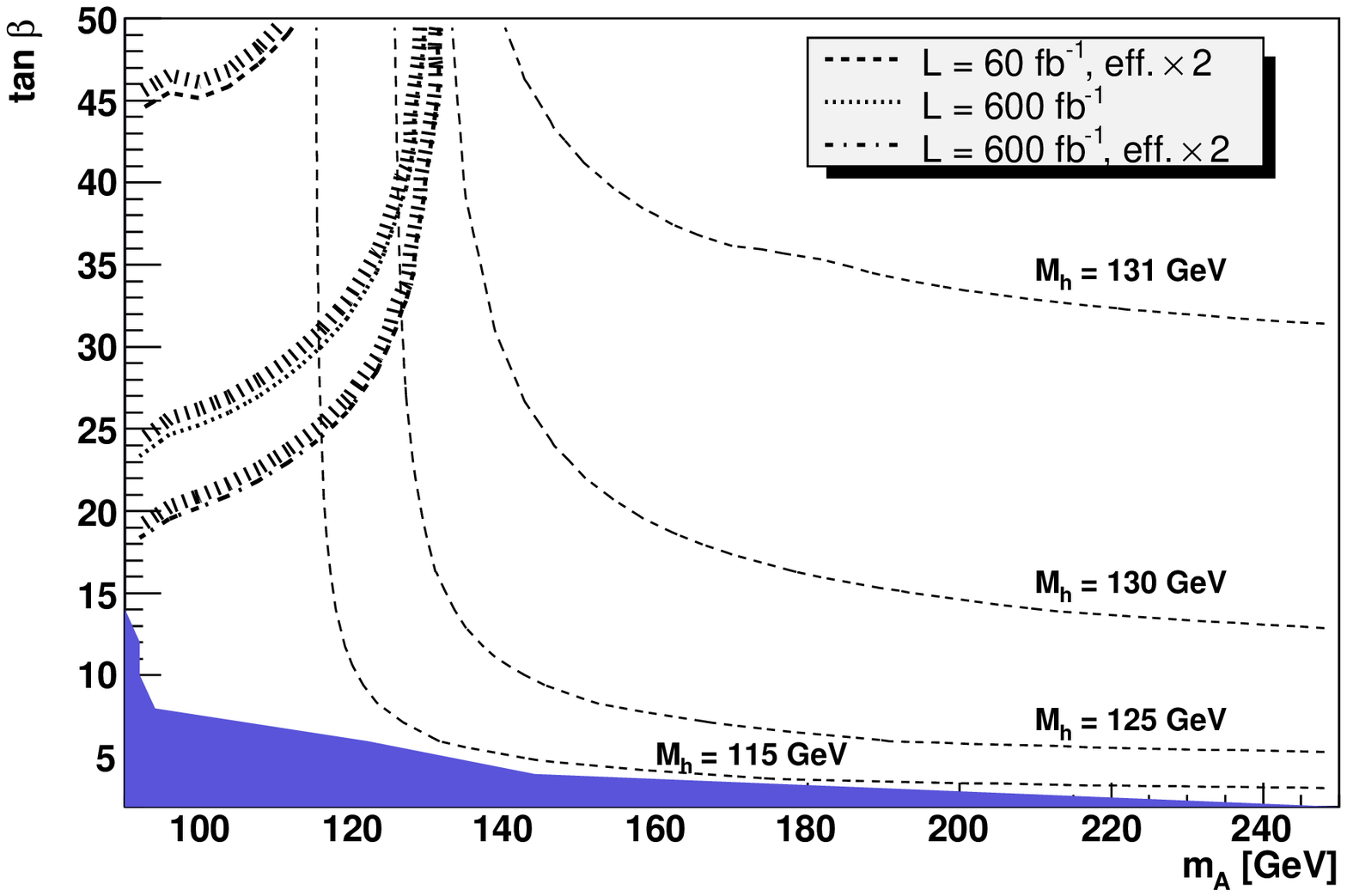}
\includegraphics[width=14cm,height=8.8cm]
                {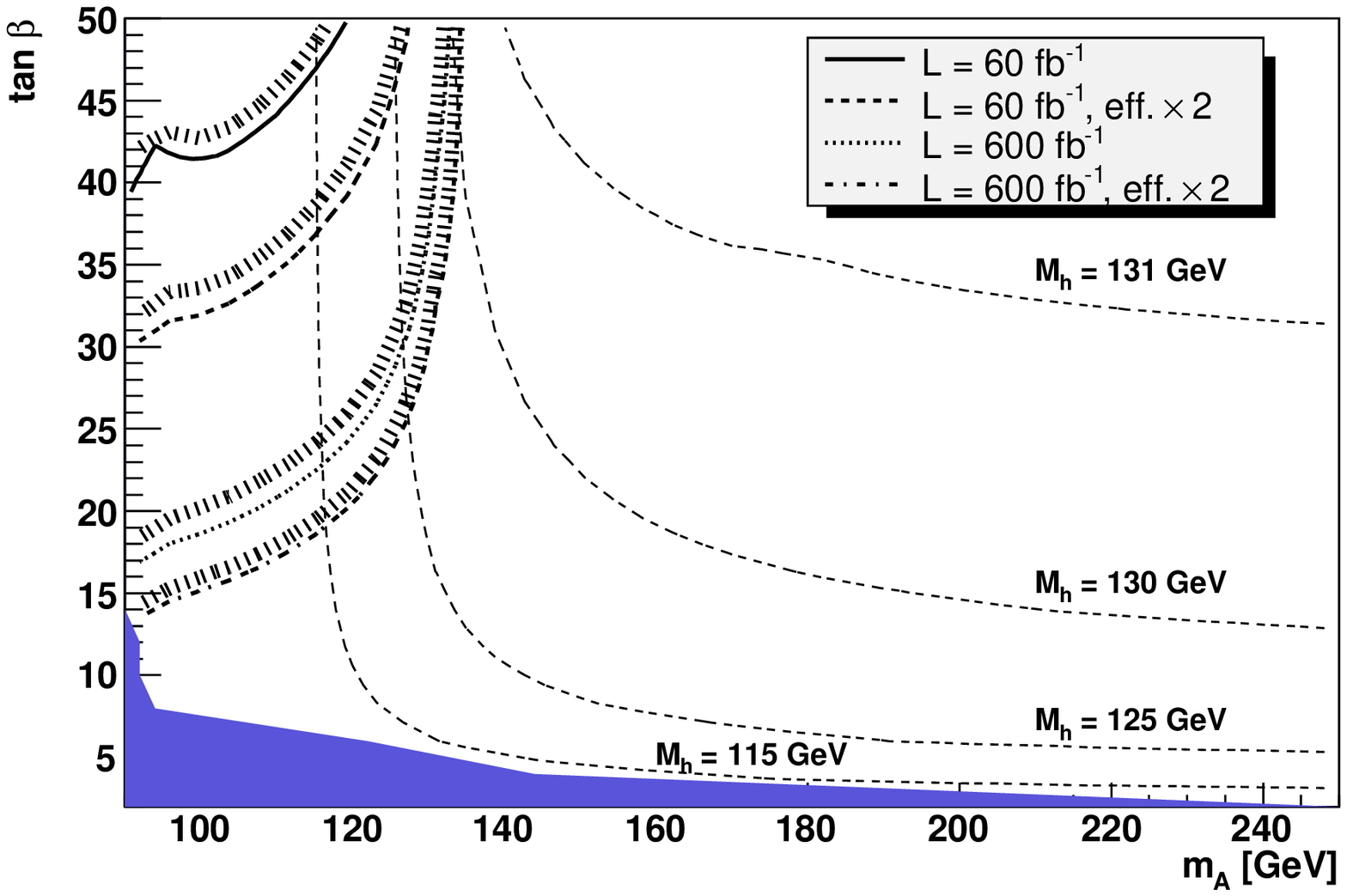}
\caption{
$5 \si$ discovery contours (upper plot) and contours of $3 \si$
statistical significance (lower plot) for the $h \to \tau^+\tau^-$ channel
in
CED production in the $\MA$--$\tb$ plane of the MSSM within the $\Mhmax$
benchmark scenario. The results are shown for assumed effective
luminosities (see text, combining ATLAS and CMS) of \sixoo,
\sixooeff, \sixooo\ and \sixoooeff.
The values of the mass of the light $\cp$-even Higgs boson, $\Mh$, are
indicated by contour lines. The dark shaded (blue)
region corresponds to the parameter region that
is excluded by the LEP Higgs
searches in the channel
$e^+e^- \to Z^* \to Z h, H$~\cite{LEPHiggsSM,LEPHiggsMSSM}.
}
\label{fig:htautau}
\end{center}
\end{figure}
%%%%%%%%%%%%%%%%%%%%%%%%%%%%%%%% End FIGURE %%%%%%%%%%%%%%%%%%%%%%%%%%%%%%%%%%%

In \reffi{fig:htautau} we show the results for $h \to \tau^+\tau^-$ in
the $\Mhmax$ scenario (with $\mu = +200 \gev$).
Despite the fact that this decay is easier to disentangle from the
background than the $h \to b \bar b$ channel, the lower branching ratio 
compared to the decay $h \to b \bar b$ results in a slightly worse coverage 
of the MSSM parameter space in the $\MA$--$\tb$ plane. While in the
``\sixoo'' scenario the $b \bar b$ channel allows the region
of low $\MA$ and high $\tb$ to be probed at the $5 \si$ level (see
\reffi{fig:hbb1}), the statistical significance achievable with the 
$\tau^+\tau^-$ channel at the same luminosity 
remains below $5 \si$ for $\tb < 50$. For the most optimistic luminosity
scenarios the coverage of the $\tau^+\tau^-$ channel is only slightly
worse than that of the $b \bar b$ channel shown in \reffi{fig:hbb1}.
The qualitative features of the no-mixing scenario are similar to those
discussed above for the $h \to b \bar b$ channel, and we do not show it
here. In interpreting the prospects for the $h \to \tau^+\tau^-$ channel
it should be noted that we have conservatively assumed the same
selection efficiencies for this channel as for the 
$h \to b \bar b$ channel. As discussed above, an improved selection
procedure could yield a significant gain for the $\tau^+\tau^-$ 
channel.

%%%%%%%%%%%%%%%%%%%%%%%%%%%%%%%% Begin FIGURE %%%%%%%%%%%%%%%%%%%%%%%%%%%%%%%%%
\begin{figure}[htb!]
\begin{center}
\includegraphics[width=14cm,height=8.8cm]
                {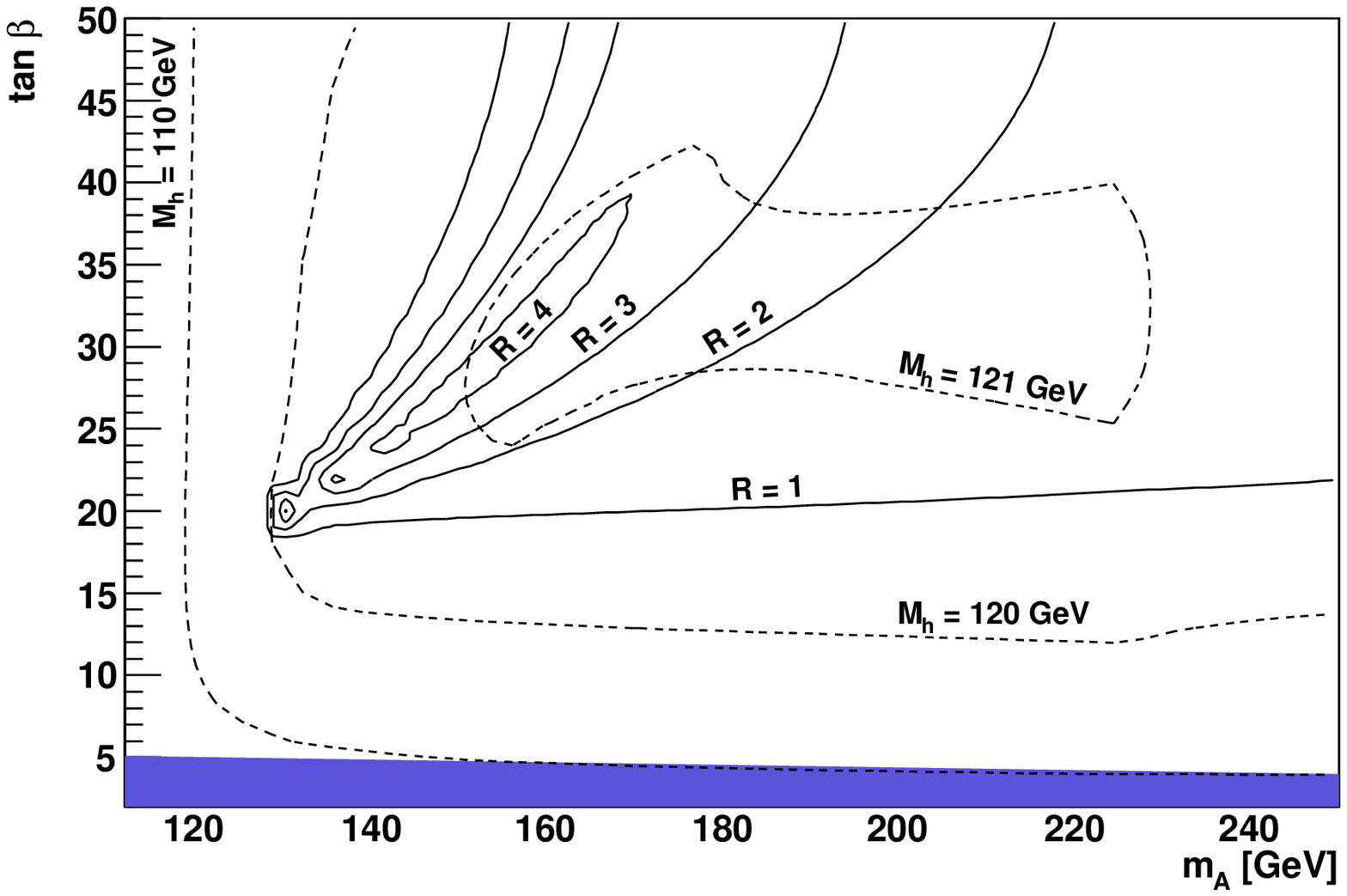}
\caption{
Contours for the ratio of MSSM to SM signal events in 
the $h \toWW$ channel in
CED production in the $\MA$--$\tb$ plane
(with $\MHSM = \Mh$). The ratio is 
shown in the small-$\aeff$ benchmark scenario.
The values of the mass of the light $\cp$-even Higgs boson, $\Mh$, are
indicated by dashed contour lines. The dark shaded (blue)
region corresponds to the parameter region that
is excluded by the LEP Higgs
searches in the channel
$e^+e^- \to Z^* \to Z h, H$~\cite{LEPHiggsSM,LEPHiggsMSSM}.
}
\label{fig:hww-ratio}
\end{center}
\vspace{-1em}
\end{figure}
%%%%%%%%%%%%%%%%%%%%%%%%%%%%%%%% End FIGURE %%%%%%%%%%%%%%%%%%%%%%%%%%%%%%%%%%%

As a further channel for the lighter $\cp$-even Higgs boson of the MSSM
we consider the decay $h \toWW$. 
Since the irreducible background to this
channel has not yet been fully investigated, it would be premature to present
$5\,\si$~discovery regions as for the other channels. Thus we only
present the ratio of signal events in the MSSM to those in the SM
(with $\MHSM = \Mh$). As discussed above, an enhancement of the 
$h b \bar b$ coupling compared to the SM case typically occurs in the 
region of large $\tb$ and small $\MA$, see \reffi{fig:hbb-ratio}. This
in turn leads to a reduction of the branching ratio of $h \toWW$, so
that one would expect that in general the $h \toWW$ channel in the MSSM 
should have a smaller or at most equal event rate compared to
the corresponding process in the SM (with $\MHSM = \Mh$). 
However, as discussed in 
\refse{sec:Higgsbench} in the context of the small-$\aeff$ benchmark
scenario, higher-order corrections in the MSSM can also have 
the opposite effect and yield a significant suppression of the 
$h b \bar b$ coupling, giving rise to an enhancement of $\br(h \toWW)$.
\reffi{fig:hww-ratio} shows 
contours for the ratio of signal events in the MSSM to those in the SM in 
the $h \toWW$ channel within the $\MA$--$\tb$ plane of the small-$\aeff$
benchmark scenario, see \refeq{smallaeff}. For 
$140 \gev \lsim \MA \lsim 170 \gev$ and intermediate $\tb$ an
enhancement of the MSSM rate compared to the SM case of up to a factor
of $R = 4$ is possible. In this region one would expect about 12~signal
events in CED Higgs production with $h \toWW$ for a luminosity of
\sixoo\ within the acceptance of the Roman Pots at 420 and 220~m,
and after applying the standard lepton Level~1 trigger thresholds.

%%%%%%%%%%%%%%%%%%%%%%%%%%%%%%%%%%%%%%%%%%%%%%%%%%%%%%%%%%%%%%%%%%%%%%%%%%%%%%%
%%%%%%%%%%%%%%%%%%%%%%%%%%%%%%%%%%%%%%%%%%%%%%%%%%%%%%%%%%%%%%%%%%%%%%%%%%%%%%%

\subsection{Prospective sensitivities for CED production of the 
    heavy $\cp$-even Higgs boson}

We now turn to the prospects for producing the heavier $\cp$-even Higgs 
boson of the MSSM in CED channels. The discovery reach in the 
``conventional'' search channels at the LHC, in particular 
$b\bar b H/A, H/A \to \tau^+\tau^-$, covers the parameter region of high
$\tb$ and not too large $\MA$~\cite{atlastdr,atlasrev,cms,CMS-TDR}, 
while a ``wedge region''~\cite{atlastdr,CMS-TDR,higgscms} remains
where the heavy MSSM Higgs bosons escape detection at the LHC
(the discovery reach is somewhat extended if decays of the heavy MSSM
Higgs bosons into supersymmetric particles can be
utilised~\cite{atlastdr,CMS-TDR}). CED production of the heavier
$\cp$-even Higgs boson of the MSSM with subsequent decay into bottom
quarks provides a unique opportunity for accessing its bottom Yukawa
coupling in a mass range where for a SM Higgs boson the decay rate into
bottom quarks would be negligibly small. 

%%%%%%%%%%%%%%%%%%%%%%%%%%%%%%%% Begin FIGURE %%%%%%%%%%%%%%%%%%%%%%%%%%%%%%%%%
\begin{figure}[htb!]
\begin{center}
\includegraphics[width=14cm,height=8.8cm]
                {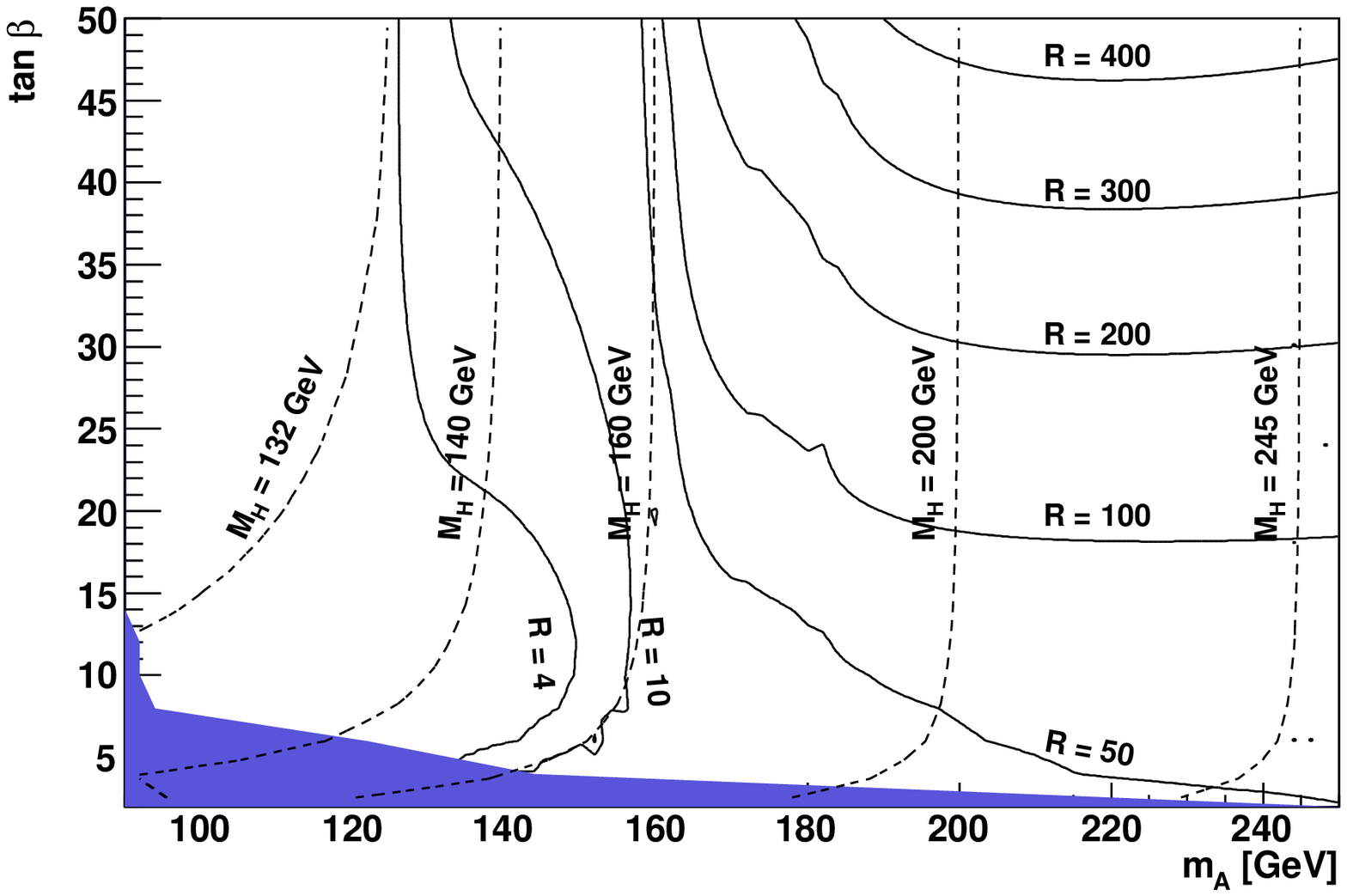}
\includegraphics[width=14cm,height=8.8cm]
                {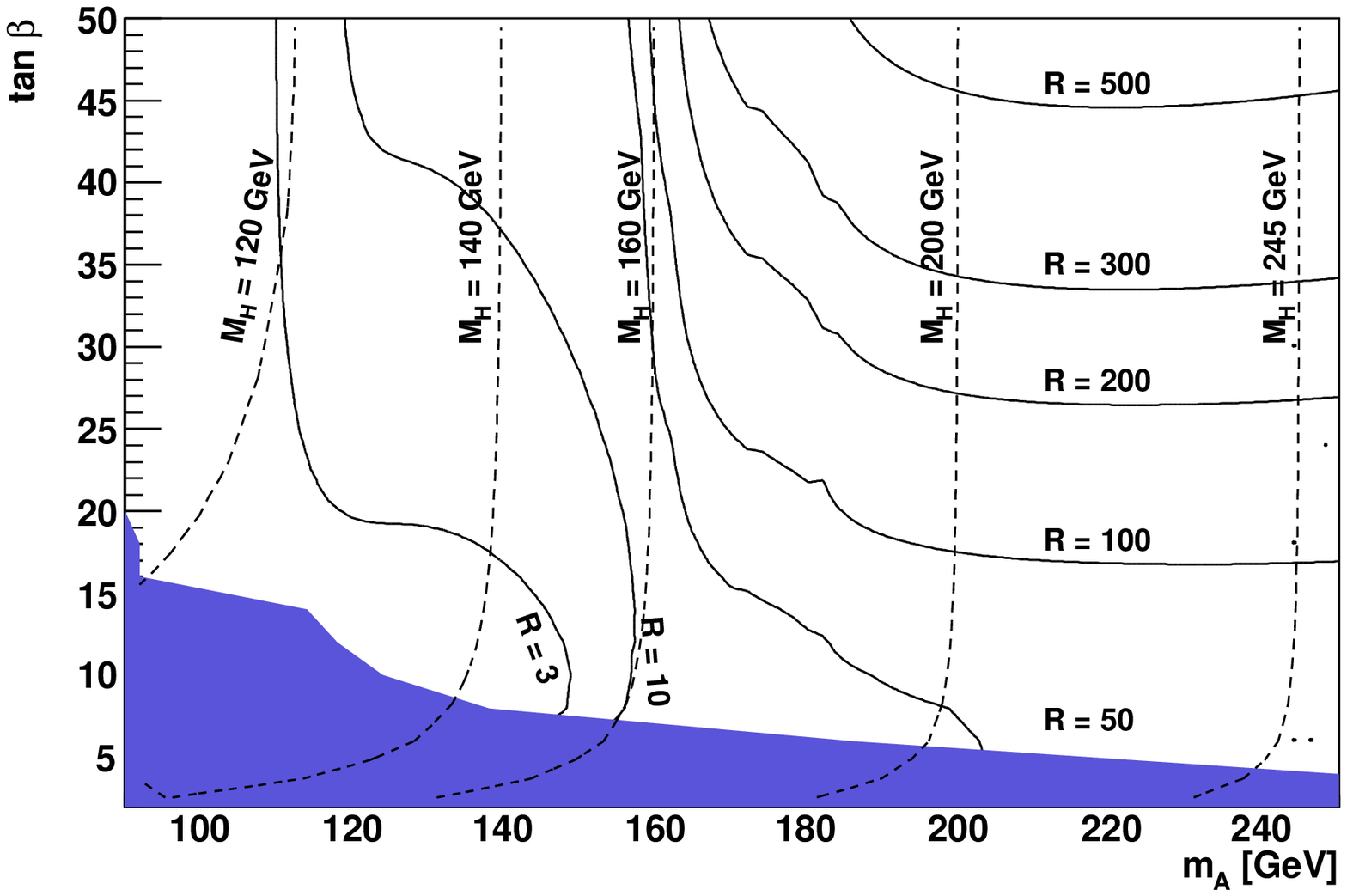} 
\caption{
Contours for the ratio of signal events in the MSSM to those in the SM in
the $H \to b \bar b$ channel in
CED production in the $\MA$--$\tb$ plane. The ratio is
shown in the $\Mhmax$ benchmark scenario (with $\mu = +200 \gev$,
upper plot) and in the no-mixing scenario (with $\mu = +200 \gev$,
lower plot).
The values of the mass of the heavier $\cp$-even Higgs boson, $\MH$, are
indicated by dashed contour lines. The dark shaded (blue)
region corresponds to the parameter region that
is excluded by the LEP Higgs
searches in the channel
$e^+e^- \to Z^* \to Z h, H$~\cite{LEPHiggsSM,LEPHiggsMSSM}.
}
\label{fig:Hbb200-ratio}
\end{center}
\end{figure}
%%%%%%%%%%%%%%%%%%%%%%%%%%%%%%%% End FIGURE %%%%%%%%%%%%%%%%%%%%%%%%%%%%%%%%%%%

%%%%%%%%%%%%%%%%%%%%%%%%%%%%%%%% Begin FIGURE %%%%%%%%%%%%%%%%%%%%%%%%%%%%%%%%%
\begin{figure}[htb!]
\begin{center}
\includegraphics[width=14cm,height=8.8cm]
                {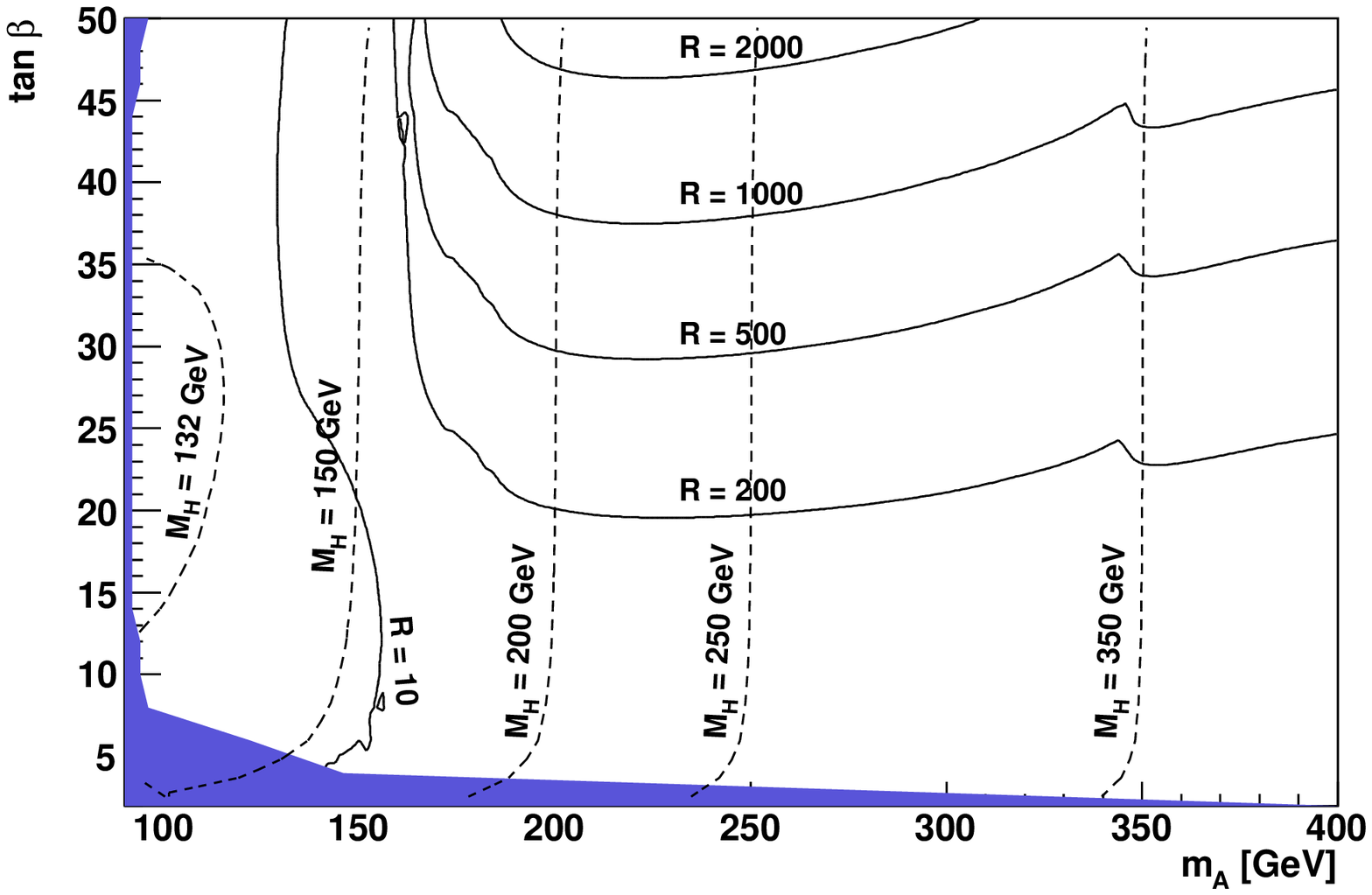}
\includegraphics[width=14cm,height=8.8cm]
                {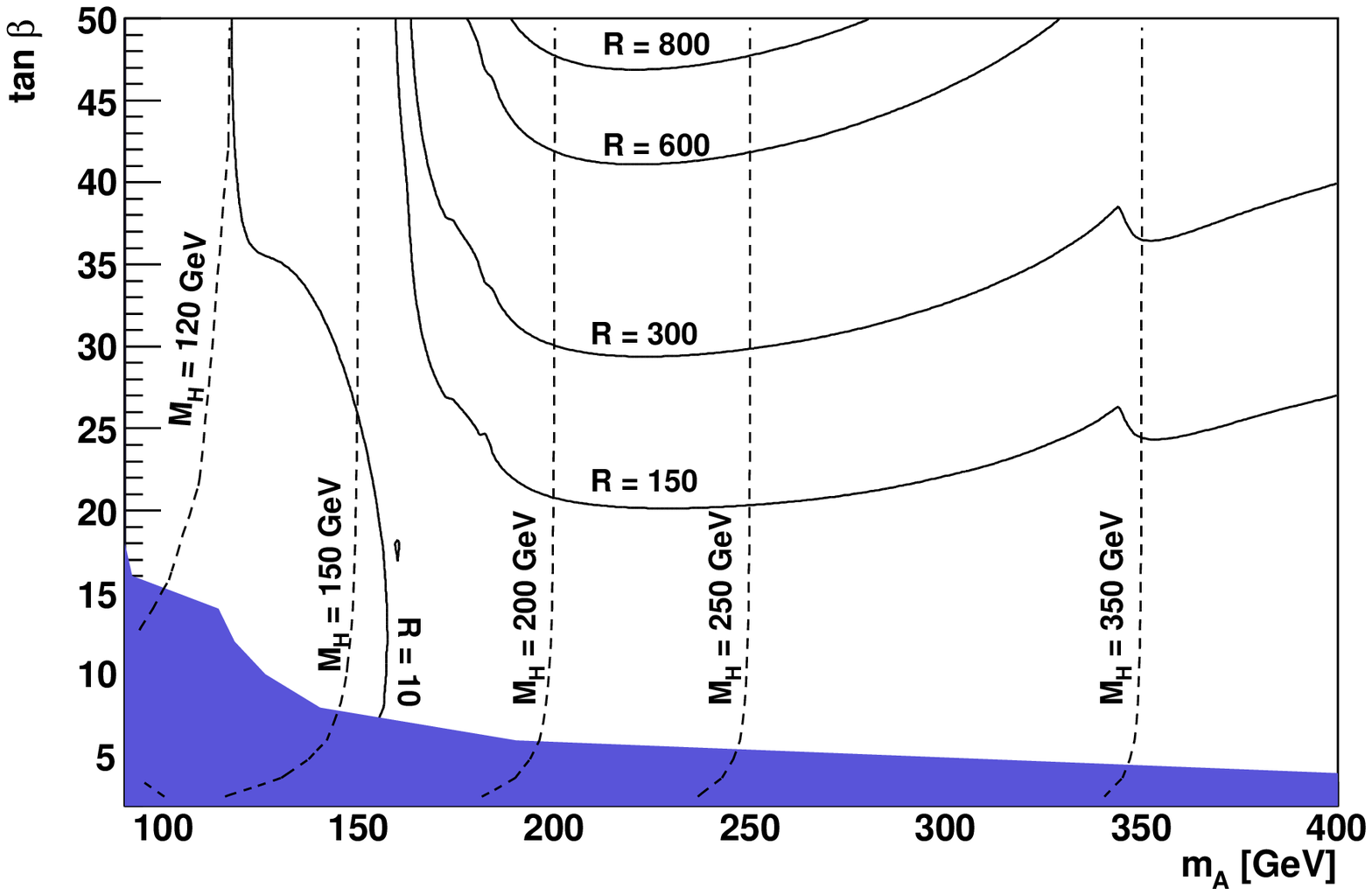} 
\caption{
Contours for the ratio of signal events in the MSSM to those in the SM in
the $H \to b \bar b$ channel in
CED production in the $\MA$--$\tb$ plane. The ratio is
shown in the $\Mhmax$ benchmark scenario (with $\mu = -500 \gev$,
upper plot) and in the no-mixing scenario (with $\mu = -500 \gev$,
lower plot).
The values of the mass of the heavier $\cp$-even Higgs boson, $\MH$, are
indicated by dashed contour lines. The dark shaded (blue)
region corresponds to the parameter region that
is excluded by the LEP Higgs
searches in the channel
$e^+e^- \to Z^* \to Z h, H$~\cite{LEPHiggsSM,LEPHiggsMSSM}.
}
\label{fig:Hbb-500-ratio}
\end{center}
\end{figure}
%%%%%%%%%%%%%%%%%%%%%%%%%%%%%%%% End FIGURE %%%%%%%%%%%%%%%%%%%%%%%%%%%%%%%%%%%

As explained above, the properties of the heavier $\cp$-even Higgs boson
of the MSSM differ very significantly from the ones of a SM Higgs in 
the region where $\MH, \MA \gsim 150 \gev$. While for a SM Higgs the 
$\br(H \to b \bar b)$ is strongly suppressed in this mass region, the 
decay into bottom quarks is the dominant decay mode for the heavier
$\cp$-even MSSM Higgs boson (as long as no decays into supersymmetric
particles or lighter Higgs bosons are open). The drastic difference
between the heavier $\cp$-even MSSM Higgs boson and its SM counterpart
with the same mass is clearly visible in \reffis{fig:Hbb200-ratio},
\ref{fig:Hbb-500-ratio}, 
where the ratio of signal events for the MSSM over
the events for the SM is displayed in the $\MA$--$\tb$ plane for the 
$\Mhmax$ (upper plots) and no-mixing (lower plots) benchmark scenarios 
with $\mu = +200 \gev$ (\reffi{fig:Hbb200-ratio}) and 
$\mu = -500\gev$ (\reffi{fig:Hbb-500-ratio}).
For illustration, contour lines for the mass of the heavier
$\cp$-even Higgs boson are also given. As before,
the dark shaded (blue) region indicates the part of the $\MA$--$\tb$
plane that is excluded by the LEP Higgs searches~\cite{LEPHiggsMSSM}
in the channel 
$e^+e^- \to Z^* \to Z h, H$~\cite{LEPHiggsSM,LEPHiggsMSSM}. 

\reffi{fig:Hbb200-ratio} shows that the MSSM process in the 
$H \to b \bar b$ channel is significantly enhanced compared to the 
SM case essentially everywhere in the unexcluded part of the 
$\MA$--$\tb$ plane. For $\MA \approx 120 \gev$ the rate is enhanced by
about a factor of $R = 3$, while for large $\MA$ we find a huge
enhancement by typically two orders of magnitude. In accordance with our
discussion of the $\db$ corrections given above, the enhancement is 
somewhat more pronounced in the no-mixing scenario, where it reaches a
factor of $R = 500$ in the high-$\tb$ region, while
\reffi{fig:Hbb200-ratio} shows an enhancement of up to $R = 400$ in
the $\Mhmax$ scenario. 

While in the case of positive $\mu$ shown in \reffi{fig:Hbb200-ratio} 
the higher-order contribution $\db$ yields a suppression of the 
bottom Yukawa coupling, the opposite effect occurs if the parameter $\mu$ is
negative. This is illustrated in \reffi{fig:Hbb-500-ratio}, where 
$\mu = -500\gev$ has been chosen. As a consequence, the enhancement
compared to the SM rate is even larger in this case, reaching a factor
of $R = 2000$ in the $\Mhmax$ scenario. (The fact that the enhancement in
the no-mixing scenario with $\mu = -500\gev$ is less pronounced is due
to the numerically smaller value of $\db$ in the no-mixing scenario, as
explained above.)

%%%%%%%%%%%%%%%%%%%%%%%%%%%%%%%% Begin FIGURE %%%%%%%%%%%%%%%%%%%%%%%%%%%%%%%%%
\begin{figure}[htb!]
\begin{center}
\includegraphics[width=14cm,height=8.8cm]
                {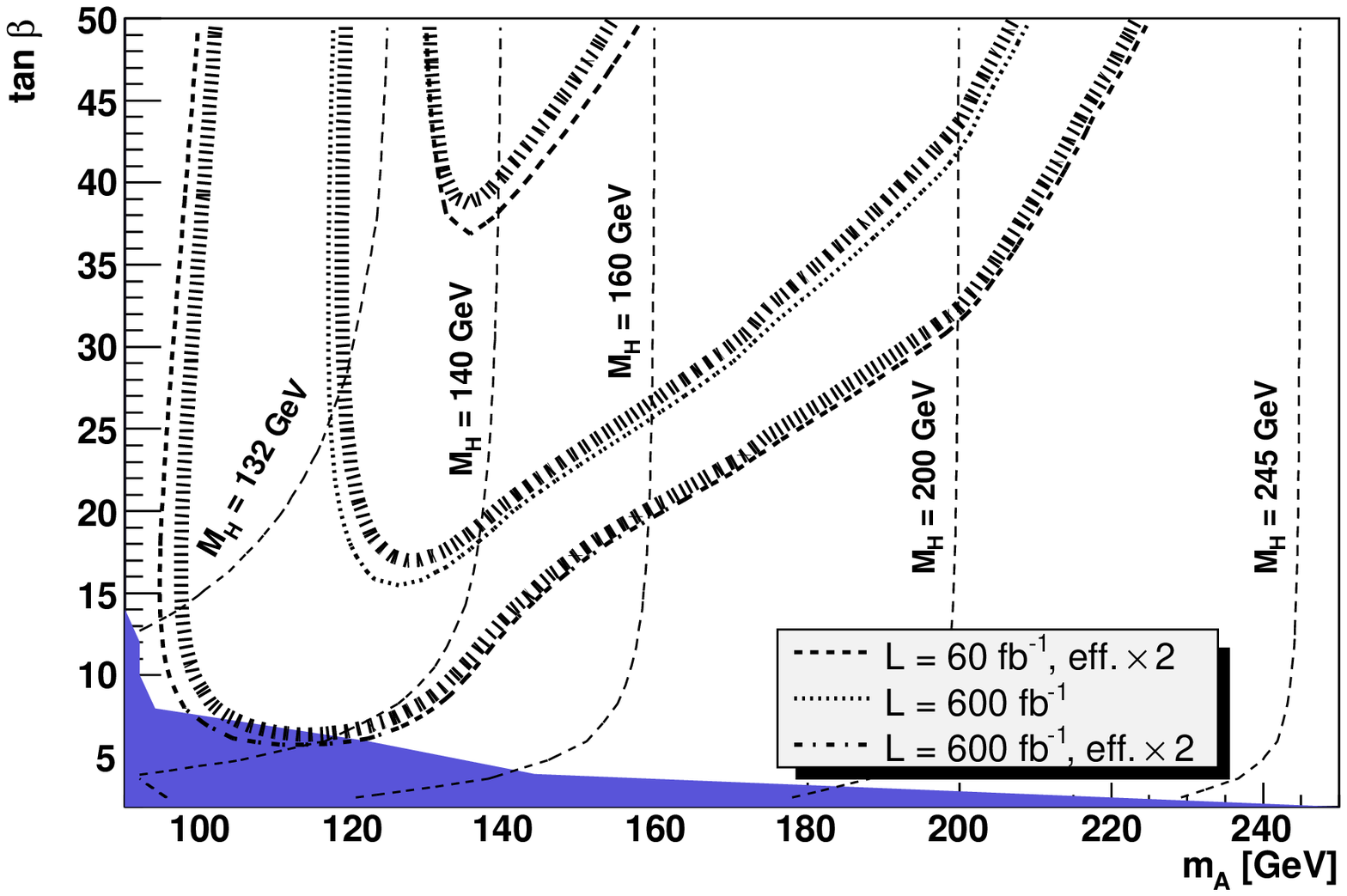}
\includegraphics[width=14cm,height=8.8cm]
                {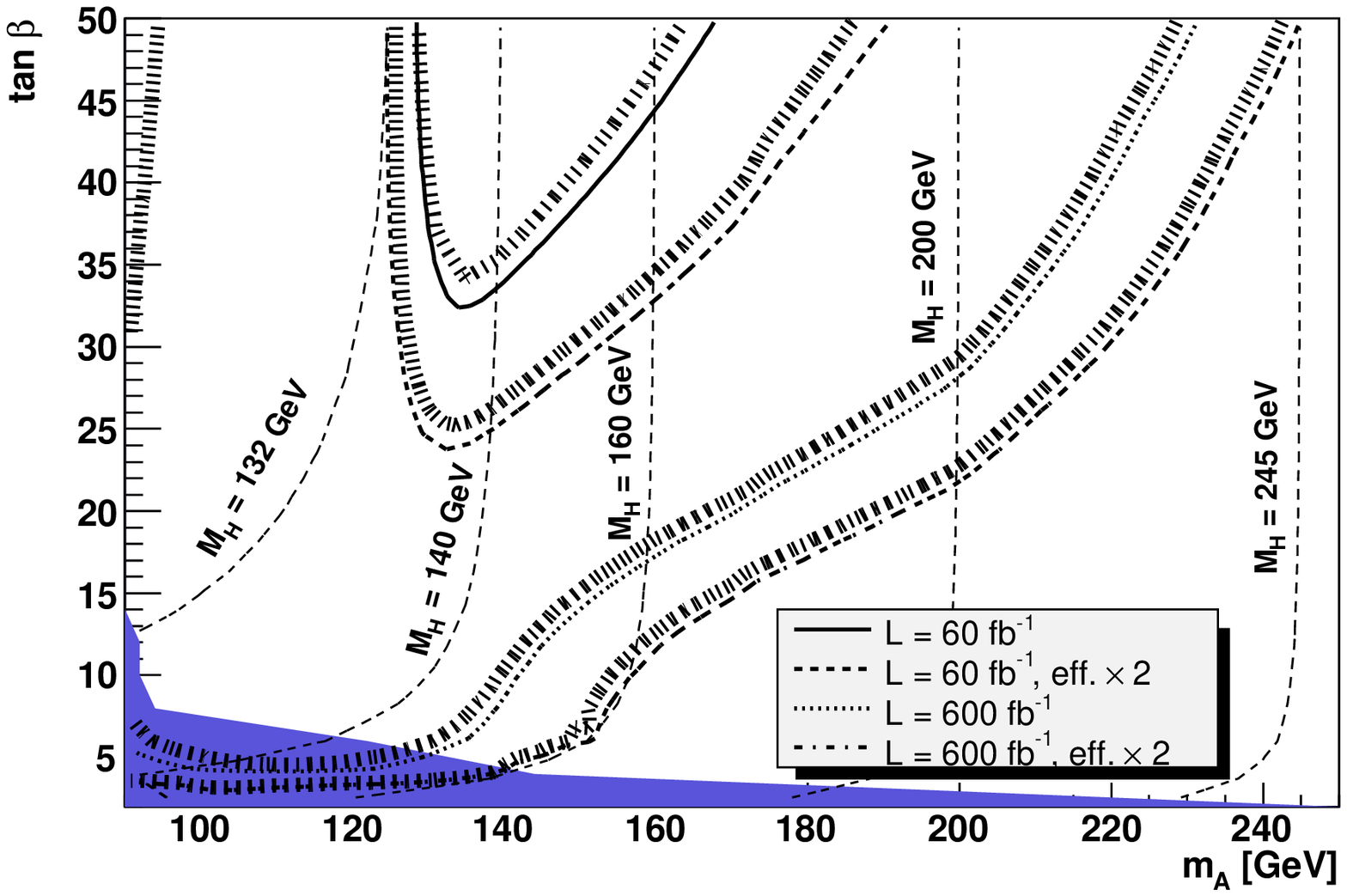}
\caption{
$5 \si$ discovery contours (upper plot) and contours of $3 \si$
statistical significance (lower plot) for the $H \to b \bar b$ channel
in CED production in the $\MA$--$\tb$ plane of the MSSM within the $\Mhmax$
benchmark scenario (with $\mu = +200 \gev$). The results are shown for 
assumed effective luminosities (see text, combining ATLAS and CMS) of \sixoo,
\sixooeff, \sixooo\ and \sixoooeff.
The values of the mass of the heavier $\cp$-even Higgs boson, $\MH$, are
indicated by contour lines. The dark shaded (blue)
region corresponds to the parameter region that
is excluded by the LEP Higgs
searches in the channel
$e^+e^- \to Z^* \to Z h, H$~\cite{LEPHiggsSM,LEPHiggsMSSM}.
}
\label{fig:Hbb200a}
\end{center}
\end{figure}
%%%%%%%%%%%%%%%%%%%%%%%%%%%%%%%% End FIGURE %%%%%%%%%%%%%%%%%%%%%%%%%%%%%%%%%%%

%%%%%%%%%%%%%%%%%%%%%%%%%%%%%%%% Begin FIGURE %%%%%%%%%%%%%%%%%%%%%%%%%%%%%%%%%
\begin{figure}[htb!]
\begin{center}
\includegraphics[width=14cm,height=8.8cm]
                {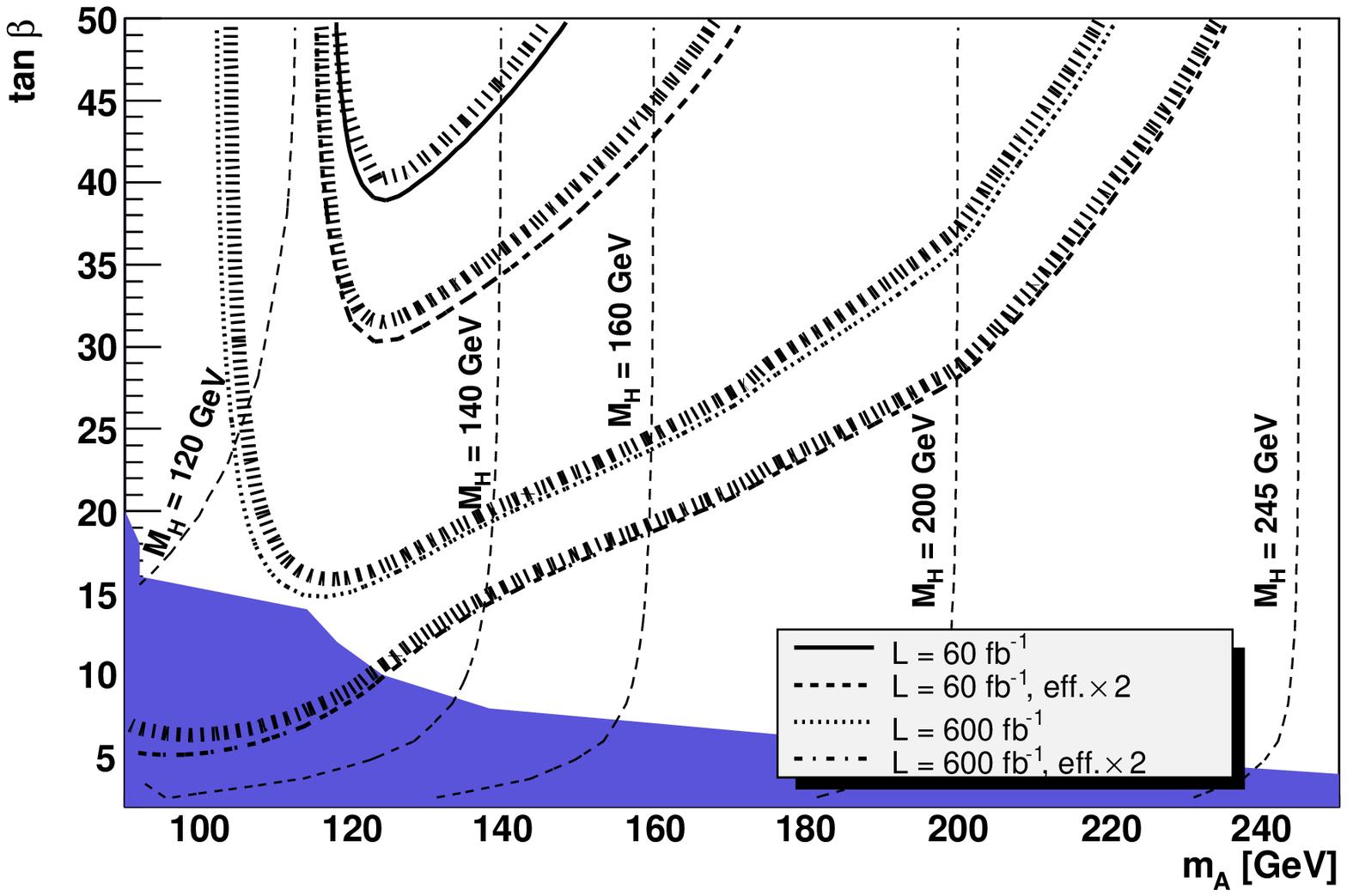}
\includegraphics[width=14cm,height=8.8cm]
                {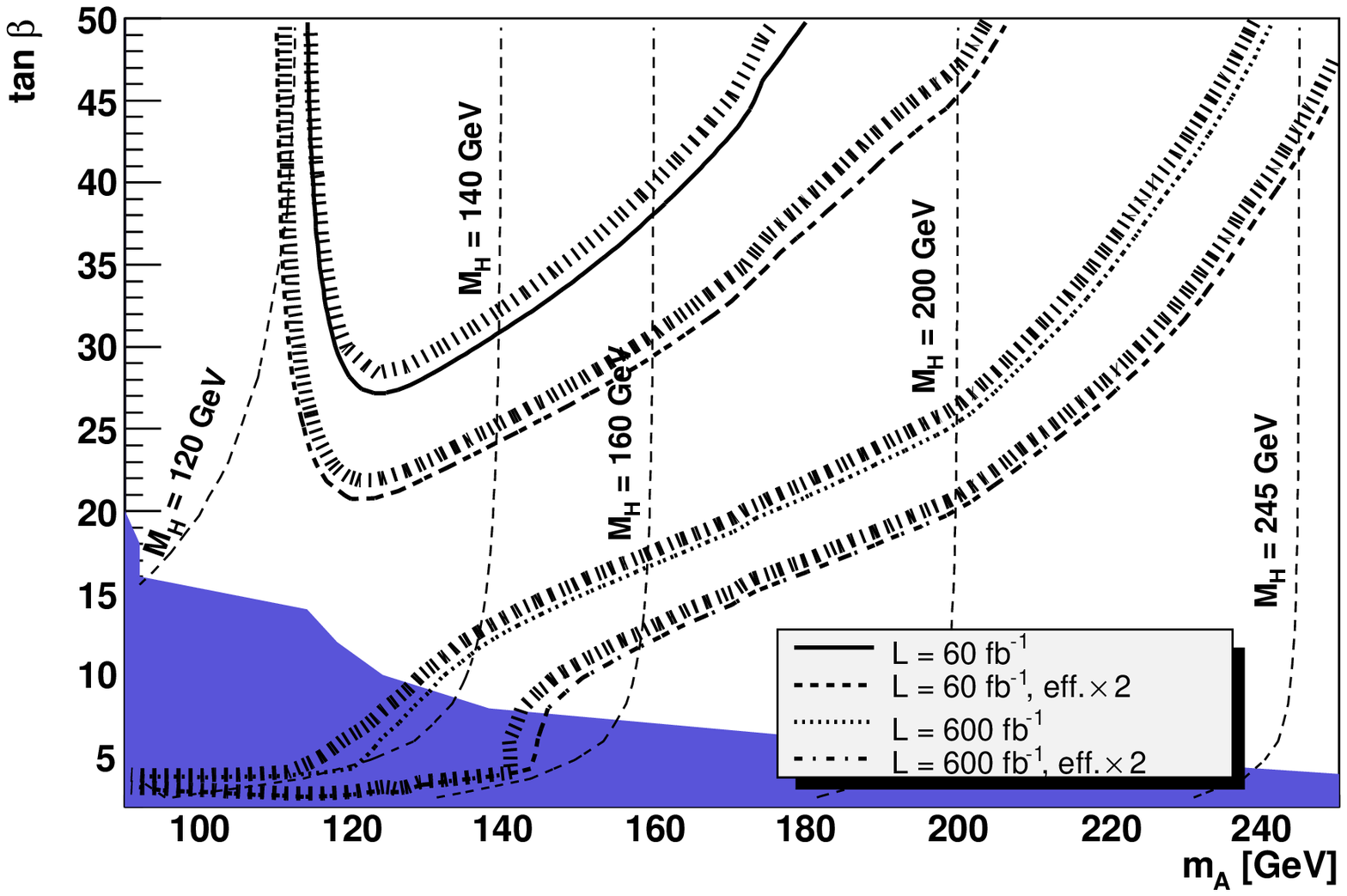}
\caption{
$5 \si$ discovery contours (upper plot) and contours of $3 \si$
statistical significance (lower plot) for the $H \to b \bar b$ channel
in CED production in the $\MA$--$\tb$ plane of the MSSM within the
no-mixing
benchmark scenario (with $\mu = +200 \gev$). The results are shown for 
assumed effective luminosities (see text, combining ATLAS and CMS) of \sixoo,
\sixooeff, \sixooo\ and \sixoooeff.
The values of the mass of the heavier $\cp$-even Higgs boson, $\MH$, are
indicated by contour lines. The dark shaded (blue)
region corresponds to the parameter region that
is excluded by the LEP Higgs
searches in the channel
$e^+e^- \to Z^* \to Z h, H$~\cite{LEPHiggsSM,LEPHiggsMSSM}.
}
\label{fig:Hbb200b}
\end{center}
\vspace{-1em}
\end{figure}
%%%%%%%%%%%%%%%%%%%%%%%%%%%%%%%% End FIGURE %%%%%%%%%%%%%%%%%%%%%%%%%%%%%%%%%%%

%%%%%%%%%%%%%%%%%%%%%%%%%%%%%%%% Begin FIGURE %%%%%%%%%%%%%%%%%%%%%%%%%%%%%%%%%
\begin{figure}[htb!]
\begin{center}
\includegraphics[width=14cm,height=8.8cm]
                {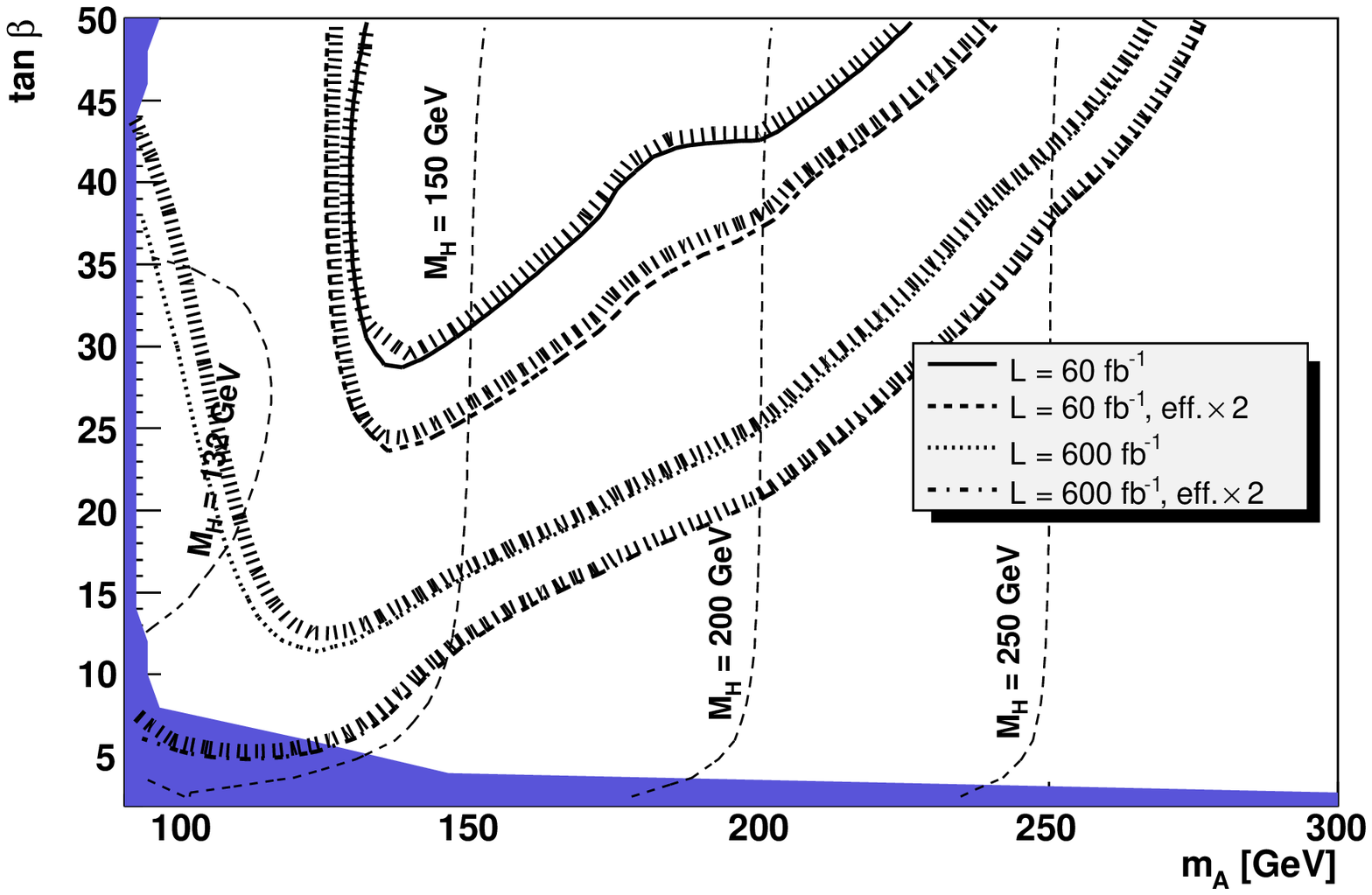} 
\includegraphics[width=14cm,height=8.8cm]
                {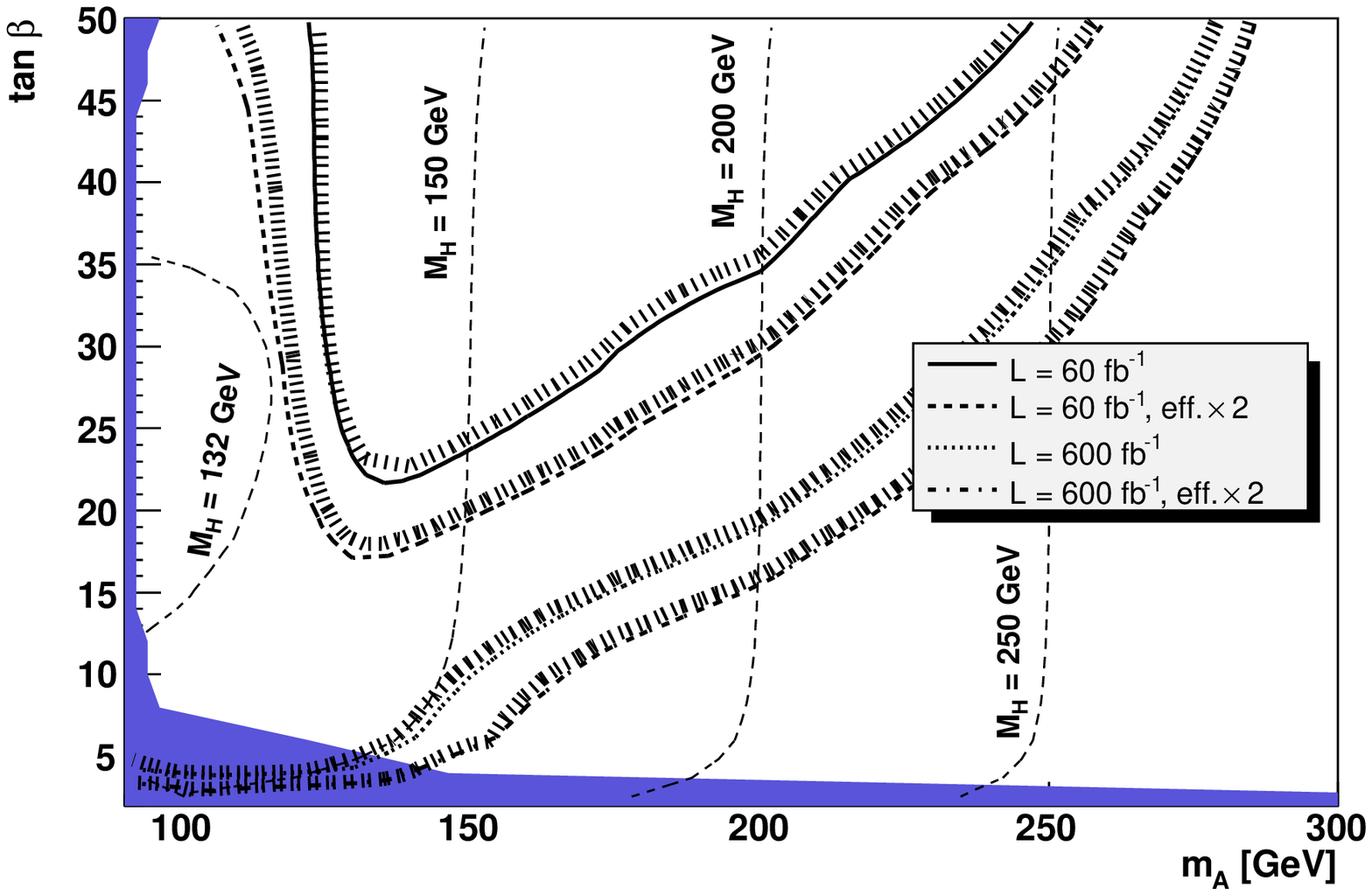}
\caption{
$5 \si$ discovery contours (upper plot) and contours of $3 \si$
statistical significance (lower plot) for the $H \to b \bar b$ channel
in CED production in the $\MA$--$\tb$ plane of the MSSM within the $\Mhmax$
benchmark scenario for the case where $\mu = -500 \gev$. The results are 
shown for 
assumed effective luminosities (see text, combining ATLAS and CMS) of \sixoo,
\sixooeff, \sixooo\ and \sixoooeff.
The values of the mass of the heavier $\cp$-even Higgs boson, $\MH$, are
indicated by contour lines. The dark shaded (blue)
region corresponds to the parameter region that
is excluded by the LEP Higgs
searches in the channel
$e^+e^- \to Z^* \to Z h, H$~\cite{LEPHiggsSM,LEPHiggsMSSM}.
}
\label{fig:Hbb-500a}
\end{center}
\vspace{-1em}
\end{figure}
%%%%%%%%%%%%%%%%%%%%%%%%%%%%%%%% End FIGURE %%%%%%%%%%%%%%%%%%%%%%%%%%%%%%%%%%%

In \reffis{fig:Hbb200a}--\ref{fig:Hbb-500a} we show the contours of $5 \si$ 
(upper plots) and $3 \si$ (lower plots) statistical significances
obtained in the four luminosity scenarios specified above for
the $H \to b \bar b$ channel in CED production within the $\Mhmax$ scenario for
$\mu = +200\gev$, the no-mixing scenario for $\mu = +200\gev$ and
the $\Mhmax$ scenario for $\mu = -500\gev$,
respectively. The pattern of the $5 \si$ and $3 \si$ contours follows from the 
discussion above: the coverage in the $\MA$--$\tb$ plane is largest in
the $\Mhmax$ scenario for $\mu = -500\gev$ (\reffi{fig:Hbb-500a}),
followed by the no-mixing scenario for $\mu = +200\gev$
(\reffi{fig:Hbb200b}) and the $\Mhmax$ scenario for $\mu = +200\gev$
(\reffi{fig:Hbb200a}). 

In the $\Mhmax$ scenario for $\mu = +200\gev$ (\reffi{fig:Hbb200a}) 
an integrated luminosity of 60~\ifb will not be sufficient to probe 
a parameter region with $\tb \leq 50$ at the $5 \si$ level. At the 
$3 \si$ level, on the other hand, the sensitivity of the ``\sixoo''
scenario extends down to about $\tb = 35$ for $\MA \approx 130 \gev$.
In the ``\sixoooeff'' scenario the discovery reach for the
heavier $\cp$-even Higgs boson goes beyond $\MH \approx 200 \gev$ in
the large $\tb$ region at the $5 \si$ level, while at the $3 \si$ level
the coverage extends to about $\MH = 250 \gev$ for $\tb \approx 50$.
In the ``\sixoooeff'' scenario the ($5 \si$ level)
discovery of a heavy $\cp$-even 
Higgs boson with a mass of about $140 \gev$ will be possible for all
values of $\tb$. This is of particular interest in view of the ``wedge
region'' left uncovered by the conventional search channels for heavy
MSSM Higgs bosons.

The search reach is somewhat extended in the no-mixing scenario for 
$\mu = +200\gev$ (\reffi{fig:Hbb200b}). In this case the ``\sixoo''
scenario can probe $\tb$ values down to about $\tb = 40$ for 
$\MA \approx 130 \gev$ at the $5 \si$ level, while the coverage extends
to $\tb \gsim 30$ at the $3 \si$ level for small $\MA$. 
As an example of anticipated event rates, for 60~\ifb one would expect
after all cuts in the no-mixing scenario for $\mu = +200\gev$,
$\MH \approx 125 \gev$ and $\tb = 50$ ($\tb = 40$) about 43 signal
events and 26 background events (28 signal events and 21 background
events).

The coverage is further enhanced in the $\Mhmax$ scenario for 
$\mu = -500\gev$ (\reffi{fig:Hbb-500a}). In this case a significant part
of the $\MA$--$\tb$ plane can be probed with 60~\ifb at the $5 \si$
level, extending down to about $\tb = 30$ for $\MA \approx 140 \gev$.
At the $3 \si$ level the reach of the ``\sixoo'' scenario extends
almost down to $\tb = 20$ for $\MA \approx 140 \gev$. 
In this scenario the expected event rates for 60~\ifb are about 124 signal
events and 65 background events (25 signal events and 14 background
events) for $\MH \approx 140 \gev$ and $\tb = 50$ ($\tb = 30$).
In the ``\sixoooeff'' scenario a
heavy $\cp$-even Higgs boson with a mass of almost $\MH = 150 \gev$
can be discovered at the $5 \si$ level for all values of $\tb$. In the
high $\tb$ region, for this luminosity
masses of $\MH \approx \MA$ in excess of $250 \gev$
can be probed at the $5 \si$ level (the coverage only slightly increases
at the $3 \si$ level). This means that CED production at the LHC may be
a unique way to access the bottom Yukawa coupling of a Higgs boson as
heavy as $250 \gev$ (which would obviously be a clear sign of physics
beyond the Standard Model).

%%%%%%%%%%%%%%%%%%%%%%%%%%%%%%%% Begin FIGURE %%%%%%%%%%%%%%%%%%%%%%%%%%%%%%%%%
\begin{figure}[htb!]
\begin{center}
\includegraphics[width=14cm,height=8.8cm]
                {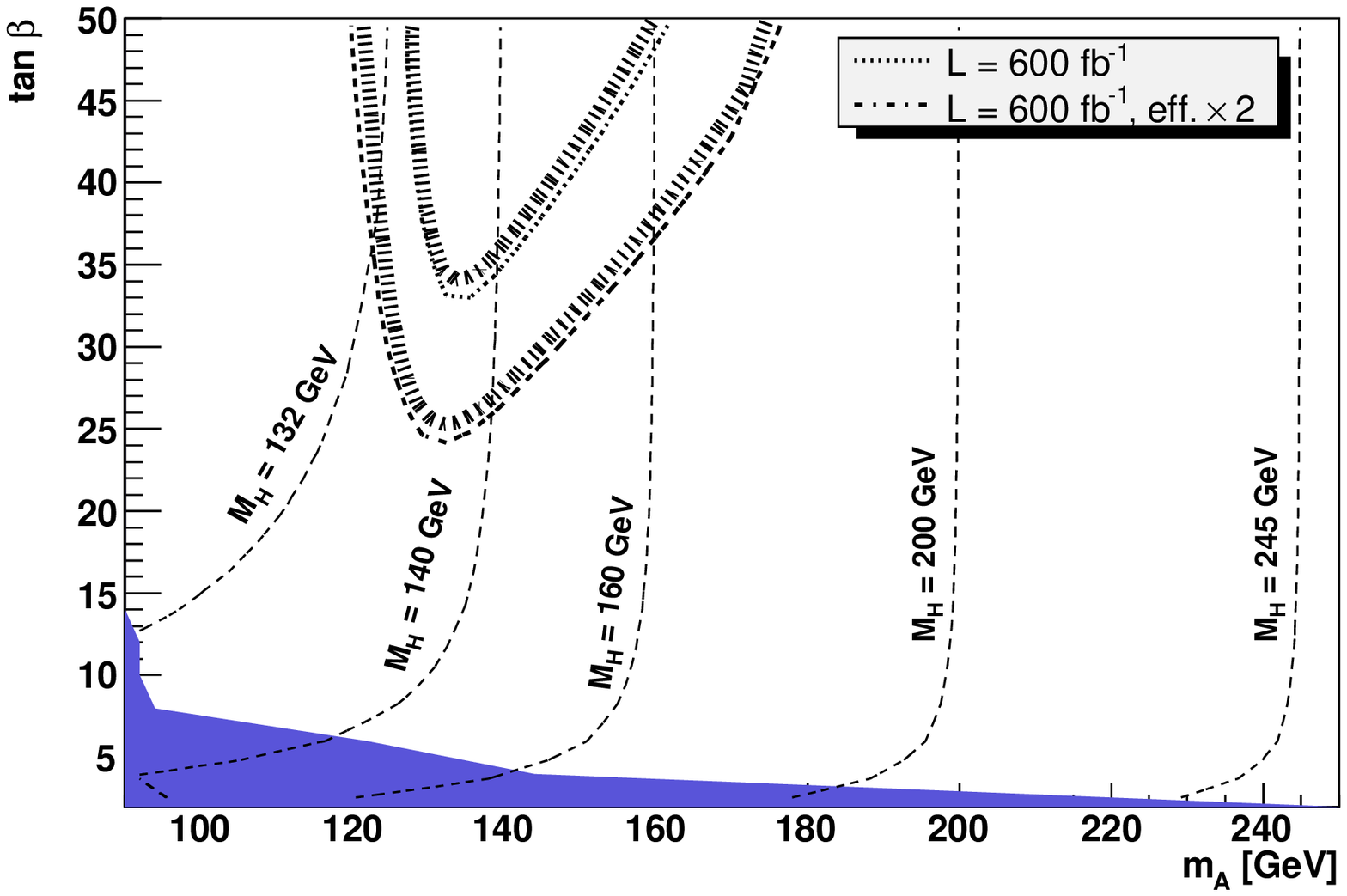}
\includegraphics[width=14cm,height=8.8cm]
                {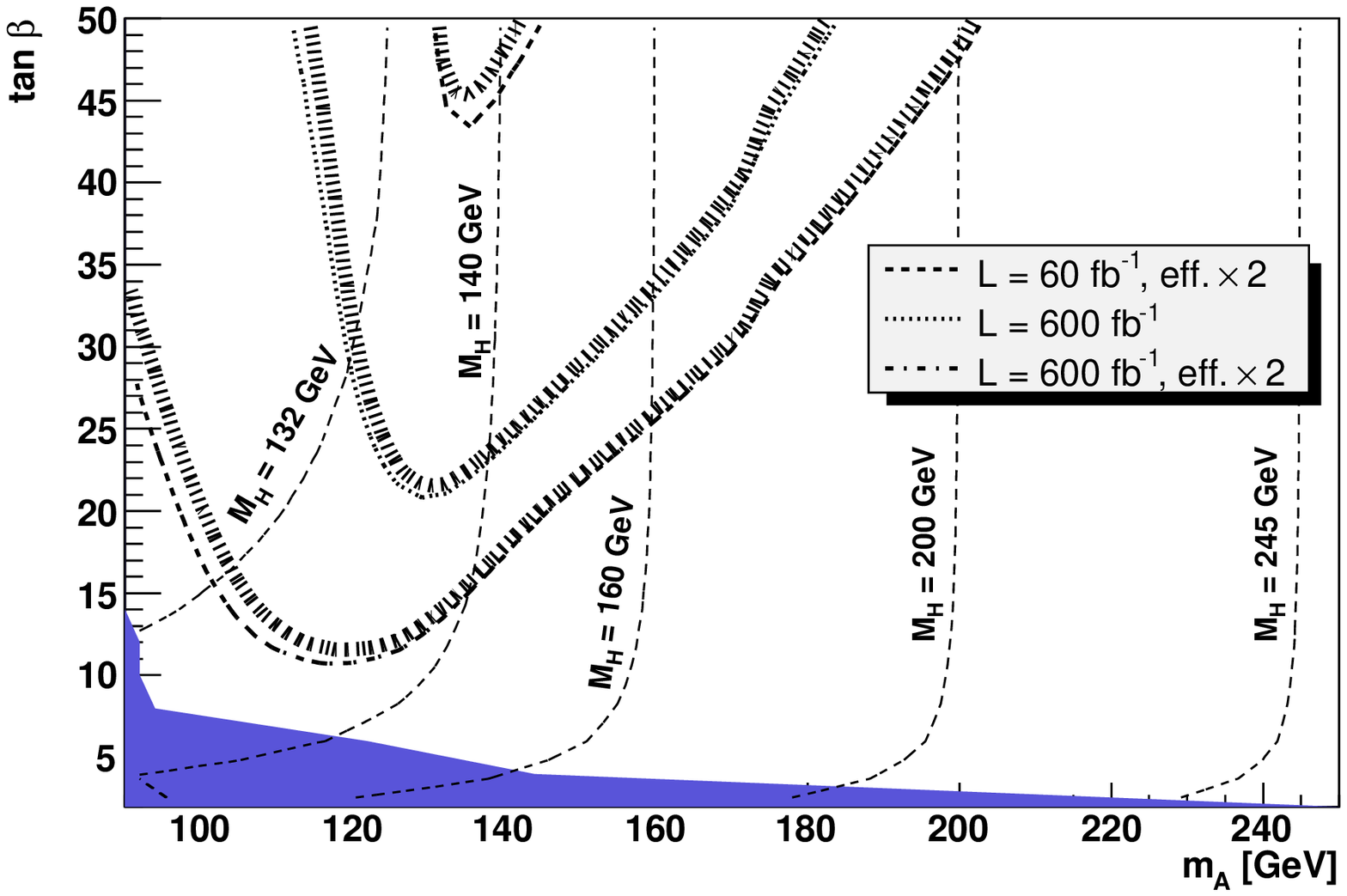}
\caption{
$5 \si$ discovery contours (upper plot) and contours of $3 \si$
statistical significance (lower plot) for the $H \to \tau^+\tau^-$ channel
in
CED production in the $\MA$--$\tb$ plane of the MSSM within the $\Mhmax$
benchmark scenario (with $\mu = +200 \gev$). The results are shown for 
assumed effective
luminosities (see text, combining ATLAS and CMS) of \sixoo,
\sixooeff, \sixooo\ and \sixoooeff.
The values of the mass of the heavier $\cp$-even Higgs boson, $\MH$, are
indicated by contour lines. The dark shaded (blue)
region corresponds to the parameter region that
is excluded by the LEP Higgs
searches in the channel
$e^+e^- \to Z^* \to Z h, H$~\cite{LEPHiggsSM,LEPHiggsMSSM}.
}
\label{fig:Htautau}
\end{center}
\vspace{-1em}
\end{figure}
%%%%%%%%%%%%%%%%%%%%%%%%%%%%%%%% End FIGURE %%%%%%%%%%%%%%%%%%%%%%%%%%%%%%%%%%%

The heavier $\cp$-even Higgs boson can also be detected via the decay 
$H \to \tau^+\tau^-$. The dependence on the parameter $\mu$ is less
pronounced in this channel compared to CED production with the decay 
$H \to b \bar b$. This is due to the fact that the $\db$ corrections
largely compensate between the production and the decay process, see
\citeres{higgscms,benchmark3} for a discussion of the analogous effect in the 
$b\bar b H/A, H/A \to \tau^+\tau^-$ channel. We therefore restrict our
discussion to the $\Mhmax$ scenario with $\mu = +200 \gev$, see 
\reffi{fig:Htautau}. The upper plot shows the $5 \si$
discovery contours, while the lower plot shows the contours of $3 \si$
statistical significance. 
Due to the suppressed branching ratio the discovery region is
significantly smaller than for the decay $H \to b \bar b$ (although the
enhancement of the signal rate compared to the SM case (not shown) is 
of similar size as for the $H \to b \bar b$ channel). 
At the $5 \si$ level an integrated luminosity of 600~\ifb is necessary
to probe parameter regions with $\tb \leq 50$.
In the ``\sixoooeff'' scenario the coverage extends to 
$\MH \approx 200 \gev$ at the $3 \si$ level for $\tb \approx 50$.

%%%%%%%%%%%%%%%%%%%%%%%%%%%%%%%%%%%%%%%%%%%%%%%%%%%%%%%%%%%%%%%%%%%%%%%%%%%%%%%
%%%%%%%%%%%%%%%%%%%%%%%%%%%%%%%%%%%%%%%%%%%%%%%%%%%%%%%%%%%%%%%%%%%%%%%%%%%%%%%

\section{Pseudoscalar Higgs-boson production in diffractive 
    processes}
\label{sec:cedA}

As is well known (see for example \citeres{atlastdr,atlasrev,cms,CMS-TDR}), 
identifying the pseudoscalar Higgs boson at the LHC
will be a challenging task. In particular, in the case of the MSSM (and many
other theories with extended Higgs sectors) the $\cp$-odd $A$~boson
decouples from 
the vector bosons, and therefore neither Higgs-boson
production via weak-boson fusion nor Higgs decay modes into gauge bosons
are of practical use.
Accordingly, it is of interest to investigate
whether the diffractive mechanism can provide some additional leverage
in the search for pseudoscalar Higgs bosons at the LHC.

Unfortunately, according to first studies performed in \citere{KKMRext},   
the situation with diffractive mechanisms for $A$~boson-production in
the MSSM looks less favourable compared to
the case of CED production of $\cp$-even Higgs bosons.
Detailed future studies (both experimental and theoretical) 
are required in order to reach a firm conclusion regarding the
prospects for detecting $A$~bosons in diffractive processes.
The discussion below reflects our current (rather incomplete)
understanding.

The main problem with the diffractive production mechanism for a heavy
$0^-$ state is the strong suppression (by about two orders of
magnitude) of the CED mode as compared to the $0^+$  case
due to the P-even selection rule, see \refse{sec:sigmaprod}.
Furthermore, for $\MA \gg \MZ$ the masses of the $\cp$-odd $A$ and the
$\cp$-even $H$ are close to each other. Consequently, 
the $H$ contribution in CED production will completely
dominate over the much smaller $A$ signal.

In order to evade the selection rule
and to have a sizable and comparable  $A$ and $H$ production rate,
it was suggested in \citere{KKMRext} to 
consider a less exclusive reaction
\BE 
pp\to X \oplus \phi \oplus Y \qquad (\phi = A,H)~,  
\label{eq:CIDP} 
\EE
where both incoming protons are allowed to dissociate, and
the Higgs bosons $\phi=A,H$ are separated from the proton remnants by 
large rapidity gaps, see \reffi{fig:A}. As shown in \citeres{KMR,KMRProsp}
such a `semi-exclusive' process has the advantage of a much
larger cross section than the CED case.
For example, for Higgs-boson masses of $120$--$140 \gev$, by
requiring rapidity gaps $\De\eta >3$, 
the effective $gg^{PP}$-luminosity is enhanced by an order of magnitude.
Moreover, the process in \refeq{eq:CIDP} can be a good testing ground for
the searches for a $\cp$-violating Higgs boson, see \citere{KMRCP}.

%%%%%%%%%%%%%%%%%%%%%%% F I G U R E %%%%%%%%%%%%%%%%%%%%%%%%%%%%%%%%%%%%%%%%%%%
\begin{figure}[htb]
\BC
\includegraphics[height=5.5cm]{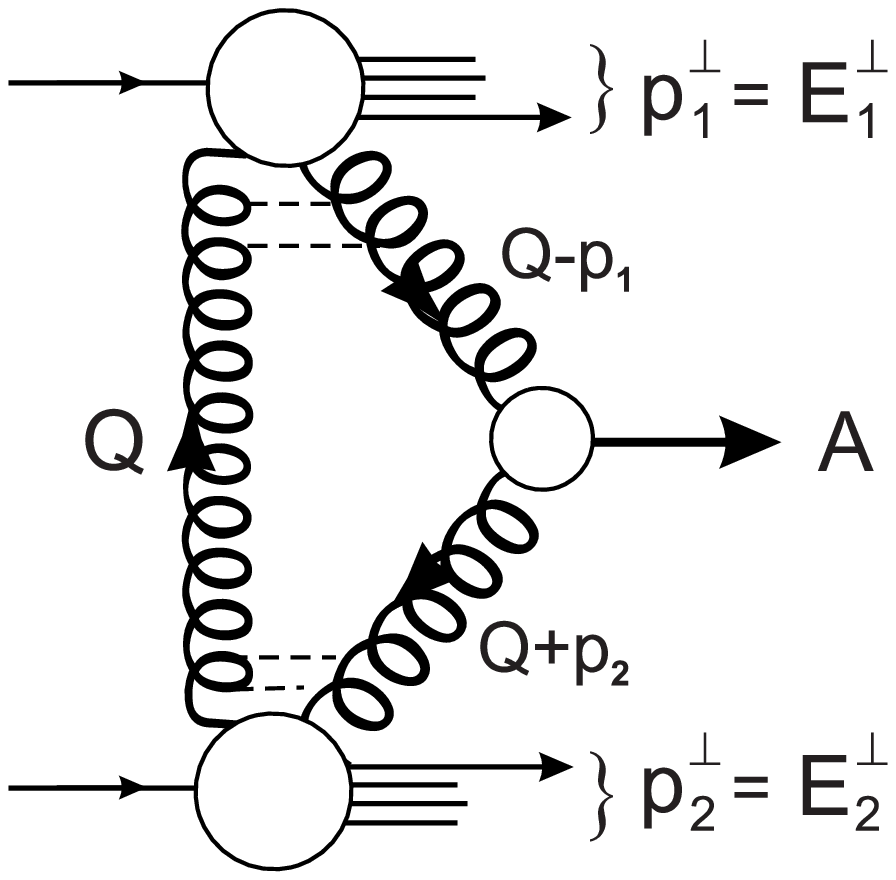}
\EC
\caption{Central production of an $A$~boson with double diffractive
dissociation (`semi-exclusive' process), in 
which the incoming protons dissociate into systems with
transverse momenta $p_i^\perp = E^\perp_i$ ($i = 1,2$).
$Q$ is the momentum of the exchanged gluon, $p_{1,2}$ are the momenta of
the incoming particles.} 
\label{fig:A}
\end{figure}
%%%%%%%%%%%%%%%%%%%%%%% F I G U R E %%%%%%%%%%%%%%%%%%%%%%%%%%%%%%%%%%%%%%%%%%%

Despite its larger cross section, this semi-exclusive
Higgs-boson production is much less advantageous for the $\cp$-even
Higgs bosons as compared to the exclusive production analysed in
\refse{sec:discovery}. 
The main reason is that one is losing the selection rule which
for the CED case gave rise to a strong suppression of the QCD backgrounds.
We consider here semi-exclusive production of 
the $\cp$-even Higgs bosons only as a
background to the semi-exclusive $A$ searches, which seems to be the
only prospective diffraction production channel for this process.  

{}From the experimental perspective, observation of the 
$A$-boson production according to \refeq{eq:CIDP} with large rapidity regions
devoid of hadron radiation appears to be very challenging.
In particular, an observation of a sizable signal rate would require a
high LHC luminosity, $\cL > 10^{33}~{\rm cm}^{-2}{\rm sec}^{-1}$,
where one faces experimental difficulties caused by the effects of pile-up 
background. As discussed in \refse{sec:cedprodhH}, it might be possible
to reduce this background to an acceptable level with the help
of fast timing detectors and other techniques.
%\mla
The Level~1 trigger can be based, for example, 
on  an observation in the forward calorimeters, at $|\eta_{1,2}| \sim 4$--6, 
of two unbalanced jets (i.e.\ one jet in the system $X$ and one in $Y$)
with $E^\perp_{1,2}> 20 \gev$.
The events with rapidity gaps can be selected subsequently
in the off-line analysis. However, no dedicated experimental study of
such a scenario has been performed so far. 
Let us recall that the signature for the Higgs production
accompanied by two forward jets is a typical characteristics
of the weak-boson fusion (WBF) mechanism~\cite{atlastdr,CMS-TDR}.
There the standard procedure
to reduce the QCD backgrounds is to use a (mini)jet veto.
In the case of the semi-exclusive process of \refeq{eq:CIDP}
the characteristic transverse momenta of the forward jets are lower
than in the WBF case (where they are typically around $10-20
\gev$). Moreover, in contrast 
to WBF Higgs production the gaps are defined as rapidity regions
devoid of any observed hadronic activity.
This definition has often been used by the HERA experiments and is similar
to that used by the CDF 
collaboration~\cite{dino}, especially in the case of the search for
exclusive diphoton events~\cite{diphot}.
Such a stronger veto on hadron activity in the rapidity gap
interval should lead to a more pronounced angular correlation between the
forward jets. 

Since the main decay mode of the $A$~boson is into bottom quarks, 
with $\br(A \to b \bar b) \approx 90\%$, the QCD 
background will be a serious obstacle for the semi-exclusive search for
$A$~bosons.
This is due in particular to the fact that in the semi-exclusive kinematics
of the process in \refeq{eq:CIDP} the $J_z$=0 selection rule can no
longer be applied to suppress the $b \bar b$  background. 
Furthermore, 
the possibility of a very good missing-mass resolution, which was
provided by the forward proton taggers in the exclusive reaction, 
does not of course exist for the semi-exclusive process.

In order to  evaluate the expected signal and background
cross sections, we use below a simplified parametrisation 
for the  `semi-exclusive' $gg^{PP}$ luminosity for production
of a system of mass $M$, which was calculated at leading order
in \citere{KMRProsp}.
In the region of interest, $100 \lsim \MA \lsim 300 \gev$,
the semi-exclusive $gg^{PP}$
luminosity at $\De\eta>3$ and  $E^\perp > 20 \gev$
can be approximated by
\BE
\frac{M^2d\cL_{\rm incl}}{dM^2}
 = 0.0024\, \exp\left(-0.416\left(\frac M{100 \gev}-2\right)\right).
\label{eq:A2}
\EE
The corresponding `semi-exclusive' $A$-signal cross section is~\cite{KMRProsp}
\BE
\hat\si^{\rm incl}=\frac{\pi^2\Ga(A\to gg)}{M^3_A} \delta(1-M^2/M^2_A).
\label{eq:A3}
\EE
Thus, to compute the cross section $\si$ for the semi-exclusive
production of the pseudoscalar $A$ the following formula
\BE
\si\cdot \br = 0.0024\, \exp\left( -0.416\left(
                               \frac{\MA}{100 \gev}-2\right)\right)
     \frac{\pi^2\Ga(A\to gg)}{M^3_A}\,  \br \cdot 0.39 \mbox{ mb} \gev^2
\label{eq:A4}
\EE
can be used.
Here $\br$ is the corresponding branching fraction for 
$A\to b\bar b$ or $A\to\tau^+\tau^-$ (as calculated with 
{\tt FeynHiggs}~\cite{feynhiggs,mhiggslong,mhiggsAEC,feynhiggs2.5}).

To evaluate the $b \bar b$ QCD background, 
we  simply convolute the effective luminosity 
$\cL_{\rm incl}$ with the spin-summed leading-order cross section for
the hard subprocess 
$gg\to b\bar b$, given in \citere{KMRProsp} (see also \citere{KRSW}).
Thus, the signal-to-background ratio in the $b \bar b$ channel can
be expressed as 
\BE
S/B \simeq \frac{\Ga(A\to gg)\cdot \br(A\to b\bar b)}
                {0.25 \alpha_s^2 \De M_{bb}}~. 
\label{eq:SB}
\EE
Here $\De M_{bb}$ denotes, as before, the $b \bar b$ mass window.
To illustrate the typical expectation for the signal-to-background ratio
in the $A\to b\bar b$ channel we first consider a parameter point in the 
$\Mhmax$ scenario with $\mu=\pm 200 \gev$ at $\tb=30$ and $\MA=140 \gev$.
Unfortunately, the result for the signal-to-background ratio
appears to be not encouraging.
For a mass window $\Delta M_{bb} \simeq 24 \gev$ 
(taken as  twice the mass resolution in the central detector) and
the angular cut $60^\circ<\theta<120^\circ$ to allow for the suppression of
the collinear singularity in the background $gg\to b\bar b$ subprocess,
we arrive at $S/B\sim 1/100$, with the signal being at the fb level.
To gain insight into what one might expect at best in the $b \bar b$
channel, we focus on a `most optimistic' $\Mhmax$~scenario 
with $\mu=-700 \gev$, $\tb=50$ and $\mgl = 1000 \gev$. As discussed in
\refse{subsec:HO}, a large negative value of $\mu$ (together with large
$\tb$ and $\mgl$) leads to a significant enhancement of the bottom Yukawa
coupling of the $A$~boson.
For this `most optimistic' scenario we find at $\MA = 160$--$200 \gev$
\BE 
\Ga(A\to gg) \cdot \br(A\to b \bar b) \approx 22-24 \mev ,
\label{eq:gam} 
\EE
and $\si \cdot {\rm BR}(A\to bb)$ is decreasing from 65~fb
for $\MA = 160 \gev$ to 25~fb for $\MA = 200 \gev$. 

This is illustrated in 
\reffi{fig:Abb}, where we show for the $\Mhmax$ scenario with 
$\mu = -700 \gev$ and $\mgl = 1000 \gev$ the production cross section for
the $A$~boson 
(upper plot) and the ratio of signal events in the MSSM to those in the SM with
$\MHSM = \MA$ (lower plot). The difference in the region exluded by the 
LEP Higgs searches in the channel
$e^+e^- \to Z^* \to Z h, H$~\cite{LEPHiggsSM,LEPHiggsMSSM} compared to
the other plots in this paper results from a downward shift in $\Mh$
caused by the enhanced bottom Yukawa coupling in the region of large
$\tb$. It should be noted that the large-$\tb$ region for 
$\mu = -700\gev$ is also affected by the limits from the MSSM
Higgs-boson searches at the Tevatron, in particular in the 
$b \bar b +(H, A)$, $H, A \to b \bar b$
channel~\cite{D0bounds}.
The cross section in \reffi{fig:Abb} ranges from 1~fb to
100~fb for large $\tb$. The ratio of $A$~signal events in the MSSM to those
in the SM case (with $\MHSM = \MA$) reaches up to $R = 100$ for 
$\tb \lsim 15$ or $\MA \lsim 150 \gev$. For large $\tb$ and $\MA$ a
ratio of up to $R = 5000$ is possible. Accordingly, \reffi{fig:Abb}
gives an idea in which
part of the MSSM parameter space there may be a chance for a successful $A$
search in the semi-exclusive process.
In the discussed `most optimistic' case, the signal-to-background ratio
in the $b \bar b$ channel, using \refeq{eq:SB}, 
gives $S/B\approx 5.5\gev/\,\De M_{bb}$.
Therefore, since the production cross section and branching ratio are
already chosen at optimistic values, 
an immediate conclusion is that the prospects for
hunting a $0^-$ Higgs boson depend strongly on progress in
improving $\Delta M_{bb}$.

%%%%%%%%%%%%%%%%%%%%%%% F I G U R E %%%%%%%%%%%%%%%%%%%%%%%%%%%%%%%%%%%%%%%%%%%
\begin{figure}[htb!]
\begin{center}
\includegraphics[width=14cm,height=8.8cm]
                {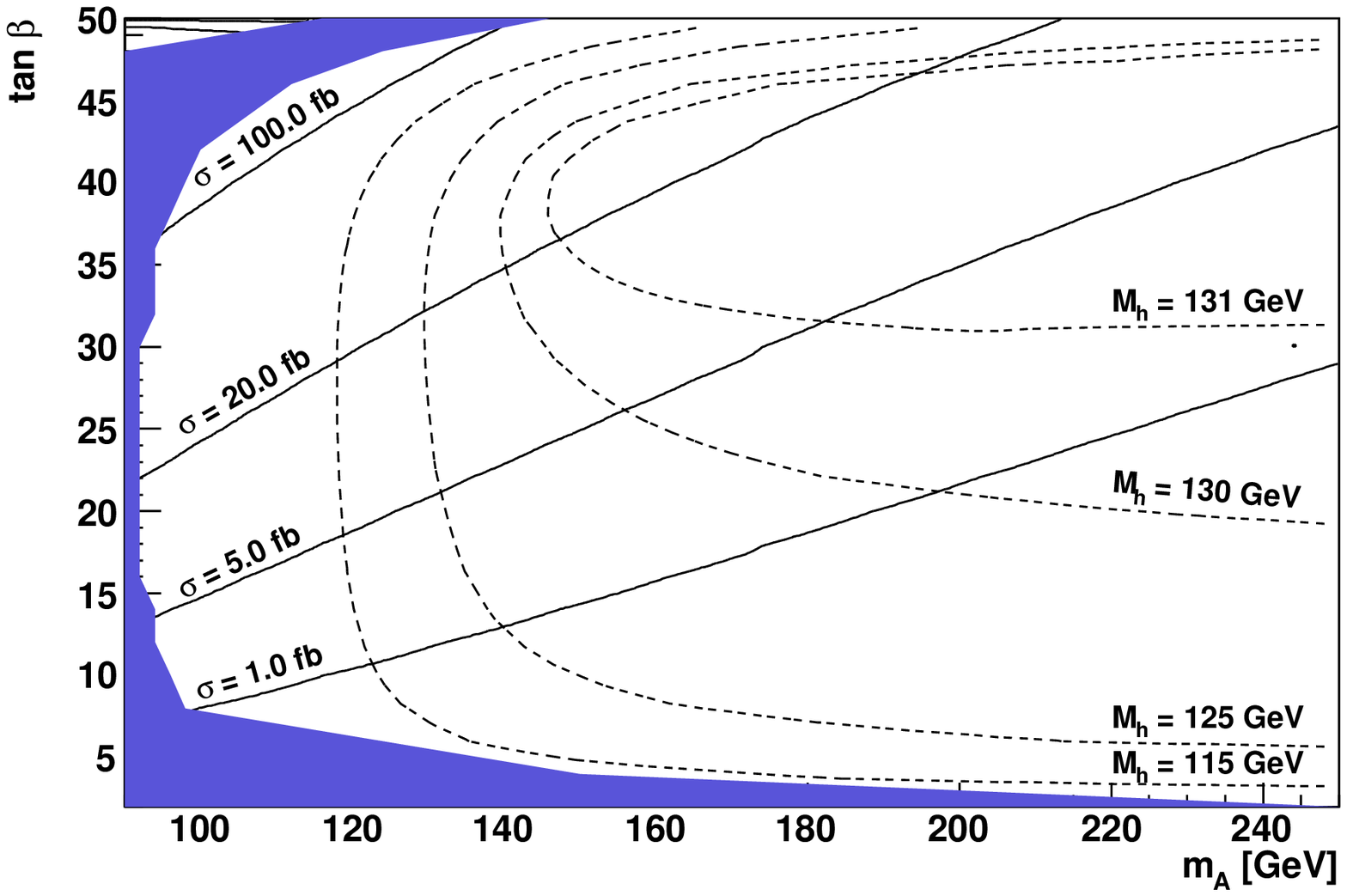}
\includegraphics[width=14cm,height=8.8cm]
                {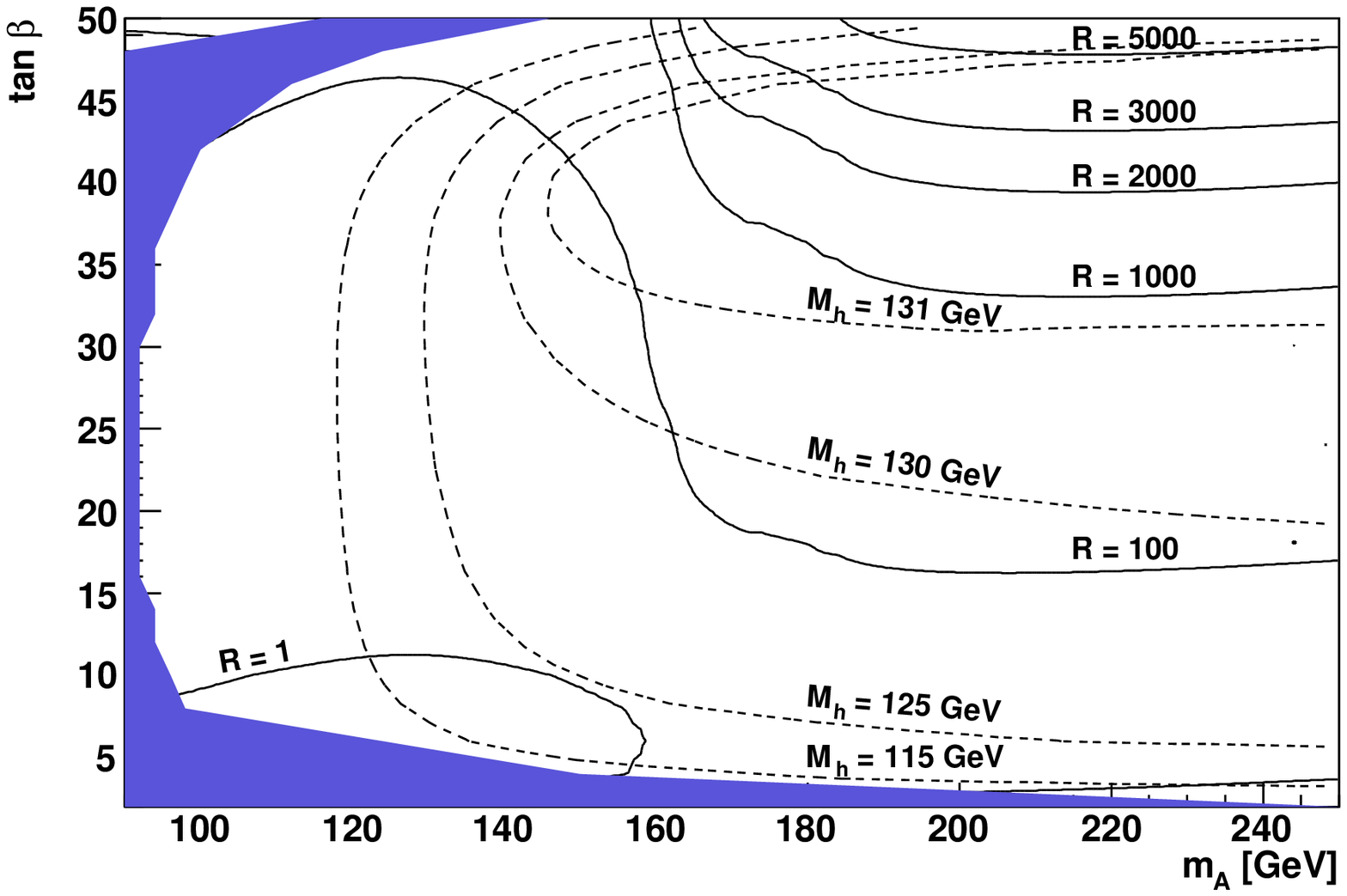}
\end{center}
\caption{Upper plot: The cross section (in fb) for
central production of the $A$ boson with double diffractive dissociation
(`semi-exclusive' process)
in the $\MA$--$\tb$ plane for the $\Mhmax$ scenario with $\mu = -700 \gev$.
The values of the mass of the light $\cp$-even Higgs boson, $\Mh$, are
indicated by contour lines. The dark shaded (blue)
region corresponds to the parameter region that
is excluded by the LEP Higgs
searches in the channel
$e^+e^- \to Z^* \to Z h, H$~\cite{LEPHiggsSM,LEPHiggsMSSM}.
Lower plot: Contours for the ratio of $A$~signal events in the MSSM over 
the SM case (with $\MHSM = \MA$) for the same parameters.
\label{fig:Abb}
}
\end{figure}
%%%%%%%%%%%%%%%%%%%%%%% F I G U R E %%%%%%%%%%%%%%%%%%%%%%%%%%%%%%%%%%%%%%%%%%%

Even with such an overwhelming QCD background there may
still be a chance to identify the $A\to b\bar b$ signal
by studying the azimuthal angular distributions 
of the  transverse energy flows in the forward (and backward)
regions of proton dissociation, i.e.\ of the total transverse momenta
of the $X$ and $Y$ systems, see \reffi{fig:A}.%
\footnote{
It should be noted that at $E^\perp > 3$--$5 \gev$ the
energy flow is dominated by one (gluon) jet with the lowest rapidity.
}
The dependence on the azimuthal angle $\varphi$ between the total
transverse momenta of the~$X$ and~$Y$ systems has the form
\BEA 
&\cos^2 \varphi &\quad {\rm for}\ H(0^+)
\nonumber \\
{\rm and} & \sin^2 \varphi &\quad {\rm for}\ A(0^-),
\label{eq:A1}
\EEA
whereas the $\varphi$-dependence of the backgrounds is practically flat.
The different behaviour of the azimuthal angular distributions of signal
and background may therefore allow for an improvement in the statistical
significance of the signal.

The difference in the azimuthal angular
dependence may even provide a way to discriminate between $0^+$ and $0^-$
Higgs-boson production~\cite{KKMRext}. 
By selecting  events with rather large
$E^\perp_{1,2} \gsim 10$--$20 \gev$, where the transverse energy flows 
in the forward and backward hemispheres are orthogonal to each other,
the $H$ signal can be suppressed.
In the region of $E^\perp_{1,2}> 20 \gev$ and $|\eta_{1,2}|= 3$--$6$,
for example, the T2 detector of TOTEM~\cite{totem} and/or
forward hadronic calorimeters (HF (CMS) or FCAL (ATLAS) and ZDC (CMS and
ATLAS)) could be used for such a discrimination between $A$ and $H$.
However, at the moment it is still unclear whether this is in fact a viable
option. Before drawing a firm conclusion, the whole issue of experimental
studies related to proton dissociation needs further detailed
investigation.

We also briefly investigate diffractive $A$~production with
subsequent decay to $\tau^+\tau^-$. 
Here the dominant non-pile-up backgrounds, arising from the 
QED production process $pp \to X\ +\ (\tau^+\tau^-)\ +\ Y$ (see \citere{KMRCP}) 
and from misidentification of a gluon dijet as a $\tau^+\tau^-$ system, 
are small, while the branching fraction 
$\br(A\to \tau^+\tau^-)$ is about 10\% .
Using the cut $E_{\rm T} >20 \gev$ the cross section for the
$A\to \tau^+\tau^-$ channel in the mass range 100--150~GeV and 
$\tb > 20$ ranges  
between 1--2~fb. If the overall efficiency were about 4--10\% as  
in the case of the $\tau^+\tau^-$ channel in the exclusive processes
(see \refta{expeff}), 
an integrated LHC luminosity of $\cL = 600$~\ifb could be sufficient 
for the observation of the semi-exclusive Higgs-boson signal. Moreover,
in such a case one could be even able to discriminate between the $A$~and 
$H$~bosons by studying the azimuthal angular distribution.
However, this simple estimate may turn out to be too optimistic since various 
experimental issues affecting the semi-exclusive process
at such high values of the instantaneous luminosity, in particular the
pile-up effect (see \refse{sec:cedprodhH}), are currently
unclear and need further dedicated experimental studies.

\smallskip
Our current {\em tentative}\ conclusion is that 
the prospects for the observation of the $\cp$-odd Higgs boson
in diffractive processes at the LHC look
borderline. Detailed experimental simulation studies would be helpful in
order to arrive at a more definite conclusion on the prospects for $A$-boson
production in this channel.

%%%%%%%%%%%%%%%%%%%%%%%%%%%%%%%%%%%%%%%%%%%%%%%%%%%%%%%%%%%%%%%%%%%%%%%%%%%%%%%
%%%%%%%%%%%%%%%%%%%%%%%%%%%%%%%%%%%%%%%%%%%%%%%%%%%%%%%%%%%%%%%%%%%%%%%%%%%%%%%

\section{Conclusions}
\label{sec:conclusions}

We have analysed in this paper the prospects for probing the Higgs
sector of the MSSM with central exclusive Higgs-boson production
processes at the LHC, utilising forward proton detectors installed
at 220~m and 420~m distance around ATLAS and CMS.
We have studied CED production of the neutral $\cp$-even Higgs bosons
$h$ and $H$ and their decays into bottom quarks, $\tau$ leptons and $W$
bosons, accounting in each case for the relevant background processes.
The experimental efficiencies for the various CED channels are based on 
existing dedicated studies.
The impact of pile-up
backgrounds, which are important in particular at high instantaneous LHC
luminosity, has been discussed, and various approaches for reducing this
background source have been summarised. In order to illustrate the
physics potential of the CED processes, four luminosity scenarios have
been investigated, corresponding to different assumptions on the
achievable overall exerimental efficiencies and the integrated
luminosity that can be utilised for the CED channels.

A striking feature of CED Higgs-boson production is that this channel 
provides good prospects for detecting Higgs-boson decays into bottom
quarks, $\tau$ leptons and $W$ bosons. Although the decay into bottom
quarks is the dominant decay mode for a light SM-like Higgs boson, this
decay channel is very difficult to access in the conventional search
channels at the LHC. In the MSSM the $b \bar b$ and $\tau^+\tau^-$ decay
channels are of particular importance, since they are in general 
dominant even for heavy MSSM Higgs bosons, whereas a SM Higgs boson of the
same mass would have a negligible branching ratio into $b \bar b$ and
$\tau^+\tau^-$.
It should be noted that 
heavy Higgs bosons that decouple from gauge bosons and therefore
predominantly decay into heavy SM fermions are a quite generic feature
of extended Higgs-boson sectors. 

We have analysed the $5 \si$ discovery contours and contours of $3 \si$
statistical significances for the CED channels with 
$h, H \to b \bar b$ and $h, H \to \tau^+\tau^-$ 
in the $\MA$--$\tb$ parameter plane of the MSSM for various benchmark
settings of the other parameters. Concerning the search for the 
light $\cp$-even Higgs boson, $h$, we find that for the
(most optimistic) 
``\sixoooeff'' scenario the whole $\MA$--$\tb$ plane of the
MSSM, with the exception of a small parameter region for small $\MA$,
can be covered with the CED process at the $3 \si$ level.
The case  of a light SM-like Higgs can also be covered at the $3 \si$
level in this luminosity scenario. Thus, if the CED channel can be
utilised at high instantaneous
luminosity (which requires in particular that pile-up background is brought
under control) it can contribute very important
information on the Higgs sector of the MSSM. Besides giving access to
the bottom Yukawa coupling, which is a crucial input for determining all
other Higgs-boson couplings, observation of a Higgs
boson in CED production with subsequent decay into bottom quarks would
provide information on the $\cp$ quantum numbers of the new state, yield
an (additional)
precise mass measurement, and may even allow a direct measurement of the
Higgs-boson width. We have furthermore shown that even if only an
integrated luminosity of 60~\ifb\ can be utilised for the CED production
process, the MSSM parameter region with large $\tb$ and relatively small
$\MA$ can still be probed. For the $h \to \tau^+\tau^-$ channel we find a
slightly weaker coverage compared to the $h \to b \bar b$ channel, based
however on conservatively assuming the same selection efficiencies for
this channel as for the $h \to b \bar b$ channel. An improved selection
procedure could yield a significant gain for the $\tau^+\tau^-$ channel.

We have shown that CED production of the heavier
$\cp$-even Higgs boson of the MSSM with subsequent decay into bottom
quarks provides a unique opportunity for accessing its bottom Yukawa
coupling in a mass range where for a SM Higgs boson the decay rate into
bottom quarks would be negligibly small. 
In the ``\sixoooeff'' scenario the discovery of a heavy $\cp$-even 
Higgs boson with a mass of about $140 \gev$ will be possible for all
values of $\tb$. This is of particular interest in view of the ``wedge
region'' left uncovered by the conventional search channels for heavy
MSSM Higgs bosons. With an effective integrated luminosity of only 
``\sixooeff'',
Higgs masses of up to $\MH \approx 200 \gev$ can be probed in the 
high-$\tb$ region at the $3 \si$ level. If the bottom Yukawa coupling is
enhanced by higher-order corrections, this sensitivity extends beyond
$\MH = 250\gev$. For the $H \to  \tau^+\tau^-$ channel, as a consequence
of the reduced branching ratio, an effective integrated luminosity of at
least 600~\ifb\ will be necessary to probe significant parts of the MSSM
parameter space. 

We have furthermore analysed the channels $h, H \toWW$ in CED production
and compared them with the SM case. Since an enhancement of the bottom
Yukawa coupling happens at the expense of the branching ratio into 
$W$ bosons, the $h, H \toWW$ channel is less favourable compared to the
SM case in significant parts of the MSSM parameter space. However,
the opposite effect is also possible, giving rise to an enhancement of the 
CED Higgs-boson production with subsequent decay into $W$ bosons
compared to the SM case. We have shown that for
$140 \gev \lsim \MA \lsim 170 \gev$ and intermediate $\tb$ an
enhancement of the MSSM rate of the CED production with $h \toWW$ 
of up to a factor of four is possible compared to the SM case.
Since the irreducible background to this
channel has not been fully investigated yet, we have not presented 
$5\,\si$~discovery regions for this channel. 
Clearly, more detailed experimental studies of the $WW$ modes would be
very desirable to assess the physics potential of this interesting
channel.

We have furthermore discussed the 
prospects for identifying the $\cp$-odd Higgs boson, $A$, in diffractive
processes at the LHC. Since the 
CED production of the $\cp$-odd Higgs boson is less promising than
production of the $\cp$-even state because of a strong suppression of
this mode caused by the $P$-even selection rule, we have 
investigated $A$~boson production in a less exclusive
reaction. The experimental analysis of the double diffractive
dissociation of the incoming protons in the presence of severe QCD
backgrounds appears to be challenging at the
present time. However, further detailed experimental and theoretical
studies will be required to reach a firm conclusion on the prospects for
this channel.

For clarity of presentation, the analyses in this paper have been
performed in the $\cp$-conserving MSSM, i.e.\ in terms of the $\cp$
eigenstates $h, H, A$. Our analysis could easily be extended to the case
where $\cp$-violating complex phases are present, giving rise to a 
mixing between the three neutral Higgs bosons. In fact, the CED
production channels may provide crucial information on the $\cp$
properties of Higgs-like states detected at the LHC.

%%%%%%%%%%%%%%%%%%%%%%%%%%%%%%%%%%%%%%%%%%%%%%%%%%%%%%%%%%%%%%%%%%%%
%%%%%%%%%%%%%%%%%%%%%%%%%%%%%%%%%%%%%%%%%%%%%%%%%%%%%%%%%%%%%%%%%%%%

\subsection*{Acknowledgements}

We thank Michele Arneodo, 
Brian Cox, Albert De Roeck, Nigel Glover, Monika Grothe,
Alan Martin, Alexandre Nikitenko, Risto Orava and Andrew Pilkington 
for useful discussions.
MGR thanks the IPPP at the University of
Durham for hospitality. This work was supported by
INTAS grant 05-103-7515, by grant RFBR 07-02-00023,
by the Federal Program of Russian Ministry of Industry,
Science and Technology RSGSS-5788.2006.02, by the Ministry of Education of 
the Czech Republic under the project LC527 and by Interuniversity Attraction 
Poles Programme - Belgian Science Policy.  
The work of SH was partially supported by CICYT (grant FPA2006--02315).
This work is also supported in part by the European Community's
Marie-Curie Research Training Network under contract MRTN-CT-2006-035505
`Tools and Precision Calculations for Physics Discoveries at Colliders'.

%%%%%%%%%%%%%%%%%%%%%%%%%%%%%%%%%%%%%%%%%%%%%%%%%%%%%%%%%%%%%%%%%%%%
%%%%%%%%%%%%%%%%%%%%%%%%%%%%%%%%%%%%%%%%%%%%%%%%%%%%%%%%%%%%%%%%%%%%

%%%%%%%%%%%%%%%%%%%%%%%%%%%%%%%%%%%%%%%%%%%%%%%%%%%%%%%%%%%%%%%%%%%%%%%%%%%%%%%
%%%%%%%%%%%%%%%%%%%%%%%%%%%%%%%%%%%%%%%%%%%%%%%%%%%%%%%%%%%%%%%%%%%%%%%%%%%%%%%

%\newpage

\end{document}

%%%%%%%%%%%%%%%%%%%%%%%%%%%%%%%%%%%%%%%%%%%%%%%%%%%%%%%%%%%%%%%%%%%%%%%%%%%%%%%
%%%%%%%%%%%%%%%%%%%%%%%%%%%%%%%%%%%%%%%%%%%%%%%%%%%%%%%%%%%%%%%%%%%%%%%%%%%%%%%